\documentclass[11pt,twoside]{article}
\pdfoutput=1

\usepackage{pslatex}    % fit more on page using PS-fonts 
\usepackage{amsfonts}
\usepackage{amsmath}
\usepackage{amsthm}
\usepackage{amscd}
\usepackage{cite}
\usepackage{eucal}
\usepackage{graphicx}
\usepackage{a4}
\usepackage{booktabs}
\usepackage{microtype}
\usepackage{appendix}
\def\beq{\begin{equation}}
\def\eeq{\end{equation}}
\def\bea{\begin{eqnarray}}
\def\eea{\end{eqnarray}}

\def\beann{\begin{eqnarray*}}
\def\eeann{\end{eqnarray*}}

\let\a=\alpha \let\be=\beta \let\g=\gamma \let\de=\delta
\let\e=\varepsilon  \let\h=\eta \let\th=\theta

 \let\k=\kappa \let\la=\lambda \let\m=\mu
\let\n=\nu \let\x=\xi \let\p=\pi \let\r=\rho \let\s=\sigma
 
 \let\Ph=\phi  \let\Ps=\Psi
  
\let\La=\Lambda \let\G=\Gamma \let\D=\Delta

\let\qd=\quad  

\def\epp{\, .}
\def\epc{\, ,}

\def\tst#1{{\textstyle #1}}
\def\dst#1{{\displaystyle #1}}

\theoremstyle{plain}

\newtheorem{lemma}{Lemma}

\newtheorem{corollary}{Corollary}
\newtheorem*{corollary*}{Corollary}

\newtheorem*{conjecture*}{Conjecture}

\theoremstyle{definition}

\def\2{\frac{1}{2}} \def\4{\frac{1}{4}}

\def\6{\partial}

\def\+{\dagger}

\def\<{\langle} \def\>{\rangle}

\def\CO{{\cal O}}

\def\i{{\rm i}}

\def\rd{{\rm d}}
\def\re{{\rm e}}
\def\rp{{\rm p}}

\DeclareMathOperator{\sh}{sh}
\DeclareMathOperator{\ch}{ch}
\DeclareMathOperator{\tgh}{th}

\DeclareMathOperator{\cth}{cth}

\DeclareMathOperator{\tr}{tr}
\DeclareMathOperator{\Tr}{Tr}

\DeclareMathOperator{\sign}{sign}

\DeclareMathOperator{\res}{res}

\def\Re{{\rm Re\,}} \def\Im{{\rm Im\,}}

\def\av{\mathbf{a}}

\def\ev{\mathbf{e}}

\def\uv{\mathbf{u}}
\def\vv{\mathbf{v}}
\def\wv{\mathbf{w}}
\def\xv{\mathbf{x}}
\def\yv{\mathbf{y}}
\def\zv{\mathbf{z}}

\def\fa{\mathfrak{a}}

\def\faq{\overline{\mathfrak{a}}}

\newcommand\uq{\overline{u}}

\newcommand{\wts}{\widetilde{s}}
\newcommand{\whk}{\widehat{K}}

\renewcommand{\appendix}{%
   \renewcommand{\section}{%\newpage%
        \secdef\Appendix\sAppendix}%
   \setcounter{section}{0}%
   \renewcommand{\thesection}{\Alph{section}}%
   \renewcommand{\theequation}{\thesection.\arabic{equation}}%
}

\newcommand{\Appendix}[2][?]{%
     \refstepcounter{section}%
     \setcounter{equation}{0}%
     \addcontentsline{toc}{appendix}%
          {\protect\numberline{\appendixname~\thesection} #1}%
     \vspace{\baselineskip}%
     {\noindent\large\bfseries\appendixname\ \thesection: #2\par}%
     \sectionmark{#1}\vspace{\baselineskip}}

\newcommand{\sAppendix}[1]{%
     {\noindent\large\bfseries\appendixname\:: #1\par}%
     \sectionmark{#1}\vspace{\baselineskip}}

\begin{document}

\thispagestyle{empty}

\begin{center}

{\Large {\bf Thermal form factors of the XXZ chain and the
large-distance asymptotics of its temperature dependent correlation functions
\\}}

\vspace{7mm}

{\large
Maxime Dugave\footnote{e-mail: dugave@uni-wuppertal.de},
Frank G\"{o}hmann\footnote{e-mail: goehmann@uni-wuppertal.de}}%
\\[1ex]
Fachbereich C -- Physik, Bergische Universit\"at Wuppertal,\\
42097 Wuppertal, Germany\\[2.5ex]
{\large Karol K. Kozlowski\footnote{e-mail: karol.kozlowski@u-bourgogne.fr}}%
\\[1ex]
IMB, UMR 5584 du CNRS,
Universit\'e de Bourgogne, France

\vspace{40mm}

{\large {\bf Abstract}}

\end{center}

\begin{list}{}{\addtolength{\rightmargin}{9mm}
               \addtolength{\topsep}{-5mm}}
\item
We derive expressions for the form factors of the
quantum transfer matrix of the spin-$\2$ XXZ chain
which are suitable for taking the infinite Trotter number
limit. These form factors determine the finitely many
amplitudes in the leading asymptotics of the finite-%
temperature correlation functions of the model.
We consider form-factor expansions of the longitudinal
and transversal two-point functions. Remarkably, the
formulae for the amplitudes are in both cases of the
same form. We also explain how to adapt our formulae to
the description of ground state correlation functions of
the finite chain. The usefulness of our novel formulae is
demonstrated by working out explicit results in the
high- and low-temperature limits. We obtain, in particular,
the large-distance asymptotics of the longitudinal
two-point functions for small temperatures by summing
up the asymptotically most relevant terms in the form
factor expansion of a generating function of the
longitudinal correlation functions. As expected the
leading term in the expansion of the corresponding
two-point functions is in accordance with conformal field
theory predictions. Here it is obtained for the first
time by a direct calculation.
\\[2ex]
\end{list}

\clearpage

\section{Introduction}
For a long time form factors have been the main tool for studying the
correlation functions of integrable massive quantum field theories
through their Lehmann representation \cite{KaWe78,Smirnov86b}.
For these theories they can be determined as solutions of appropriate
functional equations \cite{Smirnov92,Smirnov92b}.

Form factors of integrable non-relativistic models can be studied by
means of the vertex operator approach \cite{JiMi95} and by means of
the algebraic Bethe ansatz \cite{Slavnov89, KMT99a,KMST05a,KKMST09b,%
KKMST11a,KKMST11b}. In this work we apply the latter approach in order
to derive expressions for the form factors of the quantum transfer
matrix \cite{Suzuki85,SuIn87} that are suitable for taking the limit of
an infinite Trotter number. Similar formulae hold for the form factors
of the usual row-to-row transfer matrix for finite chains. In both
cases the form factors will be described in terms of solutions of
linear and non-linear integral equations, similar to those appearing in
the description of the thermodynamics \cite{Kluemper92,Kluemper93} and
correlation functions \cite{GKS04a,BoGo09} of the model. We shall call
such form factors thermal form factors.

At finite temperature the correlation functions of the XXZ chain decay
exponentially. Their asymptotic behaviour is described by correlation
lengths, determined by the ratio of one of the sub-dominant eigenvalues of
the quantum transfer matrix and its dominant eigenvalue, and by amplitudes
given as products of two corresponding thermal form factors. While the
correlations lengths were studied thoroughly in the literature
\cite{Kluemper92,Kluemper93,SSSU99,KMSS01}, so far rather little attention
has been paid to the amplitudes.

Our expressions for the form factors involve, except for integrals over
functions obtained as solutions of integral equations, also Fredholm
determinants of the corresponding integral operators. We consider form
factors arising in expansions of a generating function of the longitudinal
correlation functions and of the transversal two-point correlation
functions. The usefulness of our formulae is demonstrated in several
limiting cases for high and low temperatures.

The analysis of the low-temperature limit is delicate and cumbersome.
In this work we concentrate on the longitudinal correlation functions
in the critical regime. The low-temperature analysis of the transversal
correlation functions in the critical regime is similar but requires
a separate calculation which we leave for future work. Quite generally,
critical models at low temperature are characterized by the fact
that infinitely many of their correlation lengths diverge \cite{Cardy84,%
Cardy86}. We shall see that the corresponding amplitudes vanish
algebraically as functions of the temperature with the same critical
exponents that govern the spatial decay of the ground state correlation
functions. The situation is very much like for the finite-size
dependence of the `critical amplitudes' occurring in the form factor
expansion of the ground state correlation functions \cite{KKMST09b,KKMST11a}.
A similar temperature dependence was also reported for the amplitudes
in a series expansion of the density-density correlation functions of
the Bose gas with delta-function interactions \cite{KMS11a,KMS11b}
obtained from a multiple-integral representation of the generating
function.

In fact, we shall closely follow the work \cite{KMS11b}
in our low-temperature analysis. This allows us to keep the presentation
short, as many details can be proved along the lines of \cite{KMS11b}.
As in \cite{KMS11b} we can also sum up the critical particle-hole
contributions to the form factor expansion by means of a summation
formula proved in \cite{KOV93,Olshanski03,KKMST11b}. As a result we obtain the
low-temperature large-distance asymptotics of the longitudinal
correlation functions. The form of the exponential decay is in accordance
with conformal field theory predictions \cite{Cardy84,Cardy86}. The
amplitude of the leading oscillating term so far could not be obtained
by approximate field theoretical calculations. It is the same as
the ground state's leading oscillatory amplitude which was obtained
by a direct asymptotic analysis of multiple integrals in \cite{KKMST09a}.

\section{Hamiltonian and correlation functions}
The XXZ-chain in a longitudinal magnetic field $h$ is defined by the Hamiltonian
\begin{equation} \label{ham}
     H = J \sum_{j = - L + 1}^L \Bigl( \s_{j-1}^x \s_j^x + \s_{j-1}^y \s_j^y
                               + \D \bigl( \s_{j-1}^z \s_j^z - 1 \bigr) \Bigr)
         - \frac{h}{2} \sum_{j = - L + 1}^L \s_j^z \epc
\end{equation}
acting on $({\mathbb C}^2)^{\otimes 2L}$. Here the $\s_j^\a$, $\a = x, y, z$,
are the Pauli matrices and periodic boundary conditions, $\s_{-L}^\a = \s_L^\a$,
are implied. $J$ and $\D$ are  the exchange energy and the anisotropy parameter.
We shall employ the standard parameterization $\D = (q + q^{-1})/2$, $q = \re^\h$.

The Hamiltonian (\ref{ham}) is integrable with $R$-matrix
\begin{subequations}
\begin{align} \label{rxxz}
     R(\la,\m) & = \begin{pmatrix}
                    1 & 0 & 0 & 0 \\
		    0 & b(\la, \m) & c(\la, \m) & 0 \\
		    0 & c(\la, \m) & b(\la, \m) & 0 \\
		    0 & 0 & 0 & 1
		   \end{pmatrix} \epc \\[2ex] \label{defbc}
     b(\la, \m) & = \frac{\sh(\la - \m)}{\sh(\la - \m + \h)} \epc \qd
     c(\la, \m) = \frac{\sh(\h)}{\sh(\la - \m + \h)} \epp
\end{align}
\end{subequations}

For the exact calculation of thermal averages \cite{GKS04a} at temperature
$T$ we have to express the statistical operator of the grand-canonical ensemble
\begin{equation} \label{statop}
      \r_L = \re^{- H/T}
\end{equation}
in terms of the $R$-matrix (\ref{rxxz}). For this purpose we introduce
an auxiliary vertex model with monodromy matrix
\begin{equation} \label{monoqtm}
     T_j (\la ) =% \begin{pmatrix} A(\la) & B(\la) \\ C(\la) & D(\la)
                 % \end{pmatrix}_j =
        q^{\k \s_j^z}
        R_{j \overline{N}} \bigl(\la, \tst{\frac{\be}{N}} \bigr)
	R_{\overline{N-1} j}^{t_1}
	   \bigl(- \tst{\frac{\be}{N}}, \la \bigr) \dots
        R_{j \overline{2}} \bigl(\la, \tst{\frac{\be}{N}} \bigr)
	R_{\bar 1 j}^{t_1} \bigl(- \tst{\frac{\be}{N}}, \la \bigr) \epp
\end{equation}
It acts non-trivially in the tensor product of an `auxiliary space'
$j = - L + 1, \dots, L$ and $N \in 2 {\mathbb N}$ `quantum spaces'
$\bar 1, \dots, \overline{N}$ which are all isomorphic to ${\mathbb C}^2$.
The superscript $t_1$ denotes the transposition with respect to the
first of the two spaces $R(\la,\m)$ is acting on. The parameters $\be$
and $\k$ are related to the temperature $T$ and to the magnetic field $h$,
\begin{equation} \label{defbeta}
     \be = \frac{2J \sh(\h)}{T} \epc \qd \k = \frac{h}{2 \h T} \epp
\end{equation}
$T_j (\la)$ is a $2 \times 2$ matrix in auxiliary space $j$. Its
matrix elements satisfy the Yang-Baxter algebra relations
\begin{equation} \label{yba}
    R_{jk} (\la,\m) T_j (\la) T_k (\m)
       = T_k (\m) T_j (\la) R_{jk} (\la,\m)
\end{equation}
by construction.

Using the monodromy matrix (\ref{monoqtm}) we can express the statistical
operator (\ref{statop}) as a limit,
\begin{equation} \label{staatsoper}
     \r_L = \lim_{N \rightarrow \infty}
            \Tr_{\bar 1 \dots \overline N} \{T_{- L + 1} (0) \dots T_L (0)\} \epp
\end{equation}
We call this limit the Trotter limit and the number $N$ the Trotter number.
For finite Trotter number the operator
\begin{equation}
     \r_{N, L} = \Tr_{\bar 1 \dots \overline N} \{T_{- L + 1} (0) \dots T_L (0)\}
\end{equation}
approximates the statistical operator. It can be used to define finite
Trotter number approximants of correlation functions of $m$ local operators
$\CO^{(1)}, \dots, \CO^{(m)}$ acting on sites $1, \dots, m$ of the infinite
chain
\begin{multline} \label{finitecor}
     \bigl\< \CO_1^{(1)} \dots \CO_m^{(m)} \bigr\>_N
        = \lim_{L \rightarrow \infty}
	  \frac{\Tr_{- L + 1 \dots L} \{\r_{N, L} \CO_1^{(1)} \dots \CO_m^{(m)}\}}
               {\Tr_{- L + 1 \dots L} \{\r_{N, L}\}} \\[1ex]
        = \lim_{L \rightarrow \infty}
          \frac{\Tr_{\bar 1 \dots \overline N} \bigl\{
	        \Tr^L \{ T(0) \}
	        \Tr \{ \CO^{(1)} T(0) \} \dots \Tr \{ \CO^{(m)} T(0)\}
	        \Tr^{L-m} \{ T(0) \}\bigr\}}
               {\Tr_{\bar 1 \dots \overline N} \bigl\{ \Tr^{2L}\{ T(0) \} \bigr\}} \epp
\end{multline}

Here the expression
\begin{equation} \label{finitepart}
     Z_{N, L} = \Tr_{\bar 1 \dots \overline N} \bigl\{ \Tr^{2L} \{T(0)\} \bigr\}
\end{equation}
in the denominator is the finite Trotter number approximant to the
partition function $Z_L = \Tr_{- L + 1 \dots L} \{\exp (- H/T)\}$ of the
XXZ chain of length $2L$. The operator occurring on the right hand
side of equation (\ref{finitepart}),
\begin{equation} \label{tqtm}
     t(\la) = \Tr \{ T (\la) \}\epc
\end{equation}
is called the quantum transfer matrix \cite{Suzuki85}. It has the peculiar
property that the spectrum of $t(0)$ contains a unique real eigenvalue
$\La_0 (0)$, whose modulus is strictly larger than those of all other
eigenvalues, even if the Hamiltonian (\ref{ham}) is in the critical regime
$|\D| < 1$. This implies that $t(0)^L$ in (\ref{finitecor}) acts as
a projector onto the corresponding eigenstate $|\Ps_0\>$, when $L$ becomes
large. It follows that
\begin{equation} \label{corn}
     \bigl\< \CO_1^{(1)} \dots \CO_m^{(m)} \bigr\>_N
        = \frac{\<\Ps_0| \Tr \{ \CO^{(1)} T(0) \} \dots \Tr \{ \CO^{(m)} T(0) \}|\Ps_0\>}
               {\<\Ps_0|\Ps_0\> \La_0^m (0)} \epp
\end{equation}
We shall call $|\Ps_0\>$ the dominant eigenstate and $\La_0 (\la)$ the
dominant eigenvalue. We see that the dominant eigenstate determines all
thermal correlation functions of the system in the thermodynamic limit,
or, to phrase it differently, that it determines the state of thermal
equilibrium completely. The physical correlation functions follow from
(\ref{corn}) in the Trotter limit $N \rightarrow \infty$. For more details
see \cite{GKS04a}.

\section{Bethe ansatz and auxiliary functions}
Eigenvectors and eigenvalues of the quantum transfer matrix can be
constructed by means of the algebraic Bethe ansatz. This follows since
the monodromy matrix elements, for which we use the standard notation
\begin{equation}
     T (\la) = \begin{pmatrix}
                 A(\la) & B(\la) \\ C(\la) & D(\la) \end{pmatrix} \epc
\end{equation}
satisfy the Yang-Baxter algebra relation (\ref{yba}) and since
a pseudo vacuum $|0\> = \bigl[\bigl( \begin{smallmatrix} 0 \\ 1
\end{smallmatrix} \bigr) \otimes \bigl( \begin{smallmatrix} 1 \\ 0
\end{smallmatrix} \bigr)\bigr]^{\otimes N/2}$ exists on which the
monodromy matrix elements act as
\begin{equation}
\begin{split}
     C(\la) |0\> & = 0 \epc \\
     A(\la) |0\> & = q^\k b^\frac{N}{2}
                     \bigl(- \tst{\frac{\be}{N}}, \la \bigr)
		     |0\> = a(\la) |0\> \epc \\
     D(\la) |0\> & = q^{- \k} b^\frac{N}{2} \bigl(\la, \tst{\frac{\be}{N}} \bigr)
                     |0\> = d(\la) |0\> \epp
\end{split}
\end{equation}
Then (see e.g.\ \cite{KBIBo})
\begin{equation} \label{abaevec}
     |\Ps_n \> = B(\la_M) \dots B(\la_1) |0\>
\end{equation}
is an eigenvector of the quantum transfer matrix $t (\la)$ if the Bethe roots
$\la_j$, $j = 1, \dots, M$, satisfy the system
\begin{equation} \label{bae}
     \frac{a(\la_j)}{d(\la_j)}
        = \prod_{\substack{k = 1 \\k \ne j}}^M
	  \frac{\sh(\la_j - \la_k + \h)}{\sh(\la_j - \la_k - \h)}
\end{equation}
of Bethe ansatz equations. We note that (\ref{bae}) has many solutions,
for $M = 0, \dots, N$, and that every eigenstate $|\Ps_n \>$ is
parametrized by its own characteristic set of Bethe roots, which
therefore depends on $n$. We could exhibit this dependence writing
$\la_j^{(n)}$ and $M_n$ instead of $\la_j$ and $M$, but, for the
sake of brevity, we follow the tradition and refrain from such explicit
notation.

The eigenvalue corresponding to $|\Ps_n \>$ is
\begin{equation} \label{abaeval}
     \La_n (\la) = a(\la) \prod_{j=1}^M \frac{\sh(\la - \la_j - \h)}
                                             {\sh(\la - \la_j)}
	           + d(\la) \prod_{j=1}^M \frac{\sh(\la - \la_j + \h)}
		                               {\sh(\la - \la_j)}
\end{equation}
and a left eigenvector with the same eigenvalue can be written as
\begin{equation} \label{leftabaevec}
     \<\Ps_n| = \<0|C (\la_1) \dots C(\la_M) \epc
\end{equation}
where $\<0| = |0\>^t$ is the left pseudo vacuum satisfying $\<0| B(\la) = 0$.

We have used $n = 0, 1, \dots, 2^N - 1$ to enumerate the eigenstates. This
means we assume implicitly that, for generic $q$ and non-zero $\k$, the
sets of right and left eigenvectors defined by (\ref{abaevec}) and
(\ref{leftabaevec}) form bases of $({\mathbb C}^2)^{\otimes N}$ and of its
dual. We would like to emphasize that this point is important for the
interpretation of our results below, but not so much for the results
themselves. The quantum transfer matrix of the XXZ chain can be seen as
a special inhomogeneous case of the usual transfer matrix of the model
(see e.g.\ \cite{BoGo09}). For a discussion of the issue of completeness
of the Bethe states of the latter, as far as it is relevant in the present
context, the reader is referred to \cite{KMST05a}. In the following we shall
assume that the eigenvalues are ordered by size in such a way that
$\La_0 (0) > |\La_1 (0)| \ge |\La_2 (0)| \ge \dots$.

Along with the quantum transfer matrix $t(\la)$ we consider a `twisted version'
$t(\la|\a)$ obtained from $t(\la)$ by shifting $\k \rightarrow \k + \a$. Then
$a(\la) \rightarrow q^\a a(\la)$,  $d(\la) \rightarrow q^{- \a} d(\la)$ and
the Bethe roots are altered as well. For the corresponding eigenvalues and
eigenvectors we use the notation $\La_n (\la|\a)$ and $|\Ps_n^\a\>$.

Our goal is to write the form factors of the quantum transfer matrix in a
form that admits the Trotter limit. For this purpose we introduce certain
auxiliary functions generalizing some of those functions that proved to be
useful in the description of the finite-temperature correlation functions
of the model \cite{GKS04a,BoGo09}. For every regular solution\footnote{We call
a solution of the Bethe equations regular if it corresponds to an eigenvector
of the form (\ref{abaevec}).} of the Bethe equations (\ref{bae}) we define an
auxiliary function
\begin{equation} \label{auxa}
     \fa_n (\la) = \frac{d(\la)}{a(\la)} \prod_{j=1}^M
		   \frac{\sh(\la - \la_j + \h)}{\sh(\la - \la_j - \h)} \epp
\end{equation}
In the twisted case we write $\fa_n (\la|\a)$. We also define
\begin{equation} \label{rho}
     \r_n (\la|\a) = \frac{\La_n (\la|\a)}{\La_0 (\la)} \epp
\end{equation}

In the Trotter limit the functions $\fa_n (\la)$ are uniquely determined by
the integral equation
\begin{equation} \label{nlie}
     \ln (\fa_n (\la)) =  - (2 \k + N - 2M) \h - \be \re (\la)
        - \int_{{\cal C}_n} \frac{\rd \m}{2\p \i} \:
	  K(\la - \m) \ln \bigl(1 + \fa_n (\m)\bigr) \epp
\end{equation}
Here we have introduced the kernel
\begin{equation} \label{kernel}
     K(\la) = K_0 (\la) \epc \qd
     K_\a (\la) = q^{- \a} \cth (\la - \h) - q^\a \cth (\la + \h) \epc
\end{equation}
which is the derivative of the bare two-particle scattering phase, and
the bare energy
\begin{equation} \label{baree}
     \re (\la) = \cth (\la) - \cth(\la + \h) \epp
\end{equation}

In equation (\ref{nlie}) the information about the specific excitation
is hidden in the choice of the integration contours ${\cal C}_n$.
These are deformations of the contour ${\cal C}_0$ pertaining to
the dominant state $|\Ps_0\>$. We have to distinguish between the
massive and the massless case $\D > 1$ and $|\D| < 1$, respectively.
In the massive case $\h$ is real and the contour ${\cal C}_0$ is a
finite rectangle oriented parallel to the real and imaginary axis
and cutting the axes at $\pm \h/2$ and $\pm \i \p/2$ (for a picture
see \cite{GKS04a}). In the massless case $\h = - \i \g$, $\g \in (0, \p)$,
and the contour ${\cal C}_0$ consists of two straight lines parallel
to the real axis at $\pm \i \min \{\g /2, \p/2 - \g/2\}$, encircling
the real axis in positive direction.

\begin{figure}
\begin{center}
\includegraphics[height=15em]{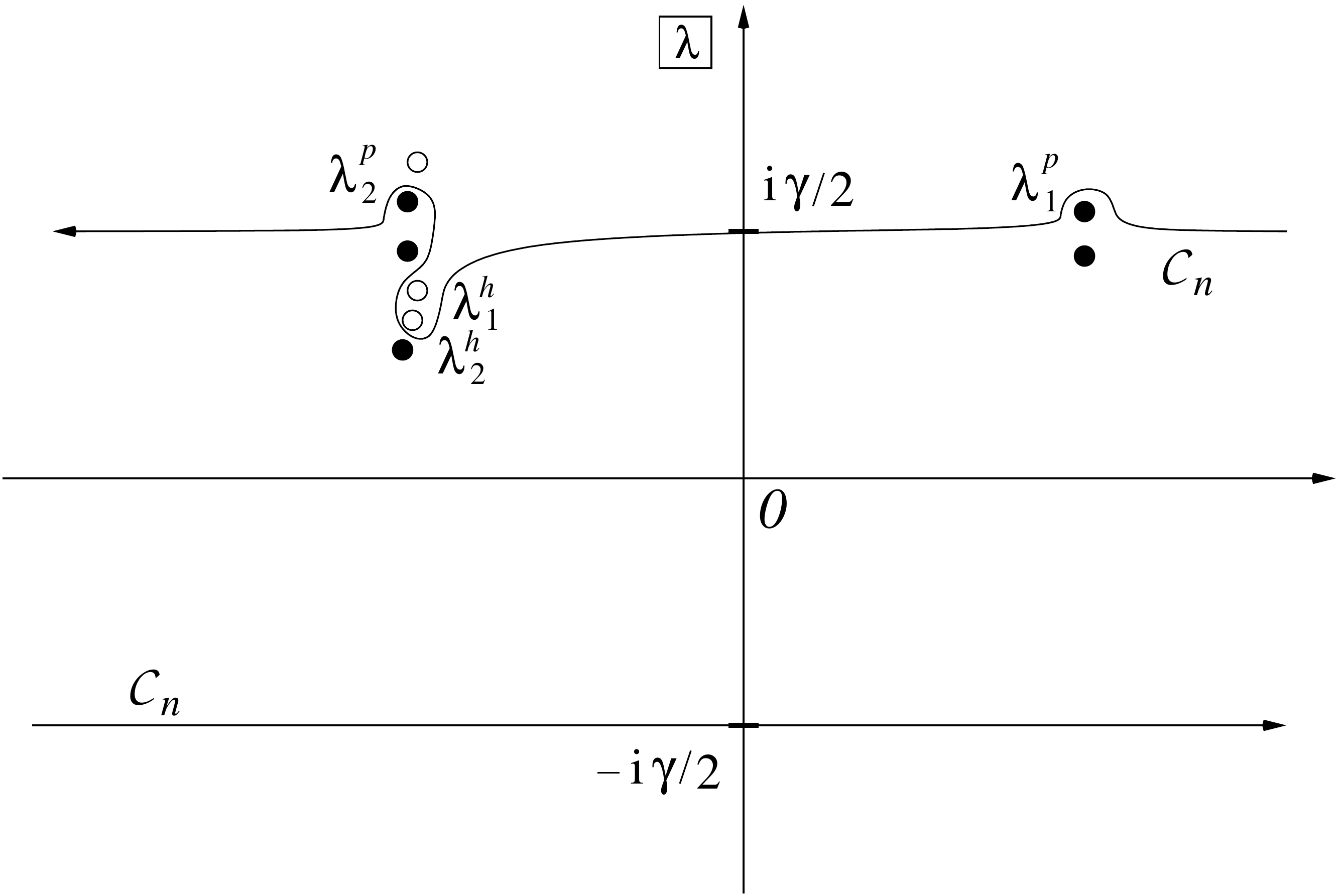}
\caption{\label{fig:cn} Sketch of a contour ${\cal C}_n$ for $0 < \D < 1$.
Hole parameters $\la_j^h$ inside the `physical strip' between $- \i \g/2$ and
$\i \g/2$ are excluded from the contour, and particle parameters (Bethe roots)
$\la_k^p$ outside the physical strip are included into the contour.
}
\end{center}
\end{figure}

All Bethe roots of the dominant state are located inside ${\cal C}_0$.
All zeros of $\La_0 (\la|\a)$, which are zeros of $1 + \fa_0 (\la|\a)$ as
well, are located outside ${\cal C}_0$. These zeros are called holes. 
If $\la$ is on the contour we infinitesimally deform it in such a way that
the point at the opposite side of the contour ($\la + \h$ or $\la - \h$)
is outside. Excitations are characterized by a finite number of holes
$\la_j^h$ inside ${\cal C}_0$ and a finite number of Bethe roots $\la_k^p$
outside. The contours ${\cal C}_n$ are obtained by deforming ${\cal C}_0$ in
such a way that the $\la_j^h$ are excluded from ${\cal C}_n$ but the
$\la_k^p$ are included (see figure~\ref{fig:cn}). For our convenience we
shall also assume that all Bethe roots of the dominant state are located
inside ${\cal C}_n$ while the corresponding holes remain outside.%
\footnote{This implies that we exclude the free Fermion case $\g = \p/2$
from our considerations.}

The contours ${\cal C}_n$ can be straightened. Then modified driving terms
depending explicitly on the particle and hole parameters $\la_k^p$ and
$\la_j^h$ appear, which can be determined by a set of subsidiary conditions
of the form $\fa_n (\la_k^p) = - 1$, $\fa_n (\la_j^h) = - 1$ (see
\cite{Kluemper93}). Such type of formulation is useful for numerical
calculations and for the low-temperature analysis of section~\ref{sec:lot}.
For the derivation of form-factor formulae, however, the deformed contour
formulation seems to be more convenient.

In order to keep our exposition simple we shall restrict the range
of $\D$ in the following to $0 < \D < 1$. The range $- 1 < \D < 0$
can be accessed by a canonical transformation \cite{YaYa66a}. Also
note that our results, except for those in section~\ref{sec:lot} on the
low-temperature limit, are valid in the massive case as well.\footnote{Possibly
certain phase factors stemming from integration boundaries have to be adapted.}

The functions $\r_n (\la|\a)$ can be represented as integrals over the
auxiliary functions,
\begin{equation} \label{rhoint}
     \r_n (\la|\a) = q^{\a + \frac{N}{2} - M}
        \exp \biggl\{\int_{{\cal C}_n} \frac{\rd \m}{2\p \i} \: \re (\m - \la)
	\ln \biggl( \frac{1 + \fa_n (\m|\a)}{1 + \fa_0 (\m)} \biggr) \biggr\} \epc
\end{equation}
where $\la$ is located inside the contour ${\cal C}_n$. The number $N/2 - M$
is the eigenvalue of the conserved $z$-component of the pseudo spin,
$\h^z = \2 \sum_{j=1}^N (-1)^j \s_j^z$. In our discussion of the form factors
below we have to distinguish two cases: (i) $N/2 - M = 0$, which is relevant
for the longitudinal correlation functions, and (ii) $N/2 - M = 1$, which
applies to the transversal correlation functions. In both cases the right
hand side of (\ref{rhoint}) is well defined and non-trivial in the Trotter limit.

\section{Form factor expansion of correlation functions}
We shall consider form factor expansions of the longitudinal and transversal
two-point functions $\<\s_1^z \s_{m+1}^z\>$ and $\<\s_1^- \s_{m+1}^+\>$.
For the longitudinal two-point functions we start with a generating function
\cite{IzKo84} which is closely related to the twisted transfer matrix.
Setting
\begin{equation} \label{sm}
     S(m) = \2 \sum_{j=1}^m \s_j^z
\end{equation}
the generating function is defined as $\bigl\< q^{2 \a S(m)}\bigr\>$, where
$\a$ is a complex parameter. In fact,
\begin{equation} \label{genzz}
     \<\s_1^z \s_{m+1}^z\> =
        \2 D_m^2 \6_{\h \a}^2 \bigl\<q^{2 \a S(m+1)}\bigr\>\Bigr|_{\a = 0} \epc
\end{equation}
where $D_m$ is the difference operator defined by $D_m f(m) = f(m) - f(m-1)$.
Note that the generating function provides more information. The one-point
functions are given by
\begin{equation} \label{genonep}
     \<\s_1^z\> = D_m \6_{\h \a} \bigl\<q^{2 \a S(m+1)}\bigr\>\Bigr|_{\a = 0} \epc
\end{equation}
and the emptiness formation probability can be obtained as the limit
\begin{equation}
     \bigl\< {e_1}_+^+ \dots {e_m}_+^+\bigr\>
        = \lim_{\Re \h \a \rightarrow + \infty} q^{- m \a} \bigl\<q^{2 \a S(m)}\bigr\>
	  \epp
\end{equation}
Here ${e_j}_+^+$ is a projector onto the local spin-up state.

It follows from (\ref{corn}) that
\begin{equation}
     \bigl\<q^{2 \a S(m)}\bigr\>_N
        = \frac{\<\Ps_0| t^m (0|\a) |\Ps_0\>}{\<\Ps_0|\Ps_0\> \La_0^m (0)}
\end{equation}
is the finite Trotter number approximant to the generating function.
Expanding $t^m (0|\a) |\Ps_0\>$ in the basis of Bethe states $\{ |\Ps_n^\a\> \}$
we obtain the form factor expansion
\begin{equation} \label{genfunexp}
     \bigl\<q^{2 \a S(m)}\bigr\>_N
        = \sum_{n=0}^{N_M - 1} A_n (\a) \r_n^m (0|\a) \epc \qd
     A_n (\a) = \frac{\<\Ps_0|\Ps_n^\a\> \<\Ps_n^\a|\Ps_0\>}
                     {\<\Ps_0|\Ps_0\> \<\Ps_n^\a|\Ps_n^\a\>} \epp
\end{equation}
Here we have adapted the numbering of the states. We count only those 
$N_M = \binom{N}{M}$, $M = N/2$, states which have non-zero overlap
with the dominant state.

Inserting now (\ref{genfunexp}) into (\ref{genzz}) we obtain the form
factor expansion of the connected longitudinal correlation functions,
\begin{equation} \label{zzexp}
     \<\s_1^z \s_{m+1}^z\>_N - \<\s_1^z\>_N \<\s_{m+1}^z\>_N
        = \sum_{n=1}^{N_M - 1} A_n^{zz} \,
	  \Bigl(\r_n^\2 - \r_n^{- \2}\Bigr)^2 \r_n^m \epp
\end{equation}
In this expression we introduced the abbreviations
\begin{equation} \label{ampzz}
     \r_n = \r_n (0|0) \epc \qd
     A_n^{zz} = \2 \6_{\h \a}^2 A_n (\a) \bigr|_{\a = 0}
           = \frac{\<\Ps_0|\Ps_n'\> \<\Ps_n'|\Ps_0\>}
                  {\<\Ps_0|\Ps_0\> \<\Ps_n|\Ps_n\>} \epp
\end{equation}
By $|\Ps_n'\>$ and $\<\Ps_n'|$ we mean the derivatives of the corresponding
$\a$-dependent vectors with respect to $\h \a$ at $\a = 0$.

Note that $1 > |\r_1| \ge |\r_2| \ge \dots$ by construction. This means
that (\ref{zzexp}) is the finite-temperature asymptotic expansion of the
longitudinal correlation functions. The leading asymptotics is determined by
the first few terms. The correlation length of the longitudinal correlation
functions is
\begin{equation} \label{corrl}
     \x_1 = - \frac{1}{\ln |\r_1|} \epp
\end{equation}
This correlation length was studied in \cite{Takahashi91,Kluemper93,FKM99,%
SSSU99,KMSS01}. So far the amplitudes in (\ref{zzexp}) were only studied
numerically \cite{FKM99} for finite Trotter numbers. Below in section
\ref{sec:longamp} we shall calculate them analytically in the Trotter limit.

Using once more (\ref{corn}) we obtain a finite-temperature asymptotic
expansion for the transversal correlation functions,
\begin{equation} \label{pmexp}
     \<\s_1^- \s_{m+1}^+\>_N = \sum_{n=1}^{N_M} A_n^{-+} \r_n^m \epc
\end{equation}
where
\begin{equation} \label{amppm}
     A_n^{-+} = \frac{\<\Ps_0|B(0)|\Ps_n\>}{\La_n (0) \<\Ps_0|\Ps_0\>} \,
                \frac{\<\Ps_n|C(0)|\Ps_0\>}{\La_0 (0) \<\Ps_n|\Ps_n\>} \epp
\end{equation}
Again the sum is restricted to the non-vanishing amplitudes (for which
necessarily $M = N/2 -1$) and the numbering is adapted (in this case we
start with `1', since we have reserved the label `0' for the dominant
eigenstate). Accordingly the $\r_n$ in the above formula (\ref{pmexp})
belong to excitations with $M = N/2 - 1$ Bethe roots and are different
from the $\r_n$ in equation (\ref{zzexp}) belonging to excitations with
$M = N/2$ Bethe roots. As before the correlation length is defined by
(\ref{corrl}) with the appropriate $\r_1$ inserted. The amplitudes in
the Trotter limit were heretofore unknown and will be calculated in
section \ref{sec:transamp}.

\section{Amplitudes in the asymptotic expansion of the longitudinal
correlation functions} \label{sec:longamp}
The basic equation underlying the algebraic Bethe ansatz approach to the
correlation functions of the XXZ chain is the following scalar product formula
due to Nikita Slavnov \cite{Slavnov89}:
\begin{multline} \label{slavnov}
     \<0|C(\m_1) \dots C(\m_M) B(\la_M) \dots B(\la_1)|0\> =
     \<0|C(\la_1) \dots C(\la_M) B(\m_M) \dots B(\m_1)|0\> \\ =
     \biggl[ \prod_{j=1}^M a(\m_j) d(\la_j) \prod_{k=1}^M \frac{1}{b(\la_j, \m_k)} \biggr]
     \frac{\det_M \bigl( \re(\la_j - \m_k) - \re(\m_k - \la_j) \fa_n (\m_k) \bigr)}
          {\det_M \Bigl( \frac{1}{\sh(\la_j - \m_k)} \Bigr)} \epp
\end{multline}
Here the $\la_j$, $j = 1, \dots, M$, are a solution of the Bethe equations
(\ref{bae}) and $\fa_n$ is the associated auxiliary function. The $\m_j$
in this formula are free. They may or may not be a solution of the Bethe
equations. In particular, we can take the limit $\m_j \rightarrow \la_j$ in
(\ref{slavnov}) and obtain the `norm formula' for Bethe states \cite{Korepin82},
\begin{equation} \label{norm}
     \<\Ps_n|\Ps_n\> =
        \biggl[ \prod_{j=1}^M d(\la_j) \La_n (\la_j) \biggr]
	\det_M \biggl\{ \de^j_k - \frac{K(\la_j - \la_k)}{\fa_n' (\la_j)} \biggr\} \epp
\end{equation}

Using (\ref{slavnov}) and (\ref{norm}) in (\ref{ampzz}) or (\ref{amppm})
we obtain expressions for the amplitudes in terms of Bethe roots (compare
\cite{KMT99a}). These are not directly suitable for taking the Trotter
limit. Still, after a sequence of elementary manipulations (see appendix
\ref{app:zz}) we obtain the following result,
\begin{multline} \label{aalphan}
     A_n (\a) = \biggl[ \prod_{j=1}^M \frac{\r_n (\la_j|\a)}{\r_n (\m_j|\a)} \biggr]
                \frac{\det_M \Bigl\{ \de^j_k - \frac{\r_n (\m_j|\a)}{\fa_n' (\m_j|\a)}
		      {\cal K}_{- \a} (\m_j - \m_k) \Bigr\}}
		     {\det_M \Bigl\{\de^j_k - \frac{1}{\fa_n' (\m_j|\a)}
		      {\cal K} (\m_j - \m_k)\Bigr\}} \\ \times
                \frac{\det_M \Bigl\{ \de^j_k - \frac{\r_n^{-1} (\la_j|\a)}{\fa_0' (\la_j)}
		      {\cal K}_\a (\la_j - \la_k) \Bigr\}}
		     {\det_M \Bigl\{\de^j_k - \frac{1}{\fa_0' (\la_j)}
		      {\cal K} (\la_j - \la_k)\Bigr\}} \epp
\end{multline}
In this equation $M = N/2$, the $\la_j$, $j = 1, \dots, M$, are the Bethe roots
of the dominant state, while the $\m_j$, $j = 1, \dots, M$, are the Bethe roots
of an excited state of the twisted transfer matrix $t(\la|\a)$. The kernel functions
in the determinants are defined as
\begin{equation} \label{kernela}
     {\cal K}(\la) = {\cal K}_0 (\la) \epc \qd
     {\cal K}_\a (\la) = \frac{\re^{(\a - 1)(\la - \h)}}{\sh(\la - \h)}
                         - \frac{\re^{(\a - 1)(\la + \h)}}{\sh(\la + \h)} \epp
\end{equation}

The right hand side of (\ref{aalphan}) is suitable for taking the Trotter
limit. For the product this is obvious from the representation
\begin{equation}
     \prod_{j=1}^M \frac{\r_n (\la_j|\a)}{\r_n (\m_j|\a)}
        = \exp \biggl\{ \int_{{\cal C}_n} \frac{\rd \la}{2 \p \i} \:
	                \frac{\r_n' (\la|\a)}{\r_n (\la|\a)}
			\ln \biggl( \frac{1 + \fa_n (\la|\a)}{1 + \fa_0 (\la)} \biggr)
			\biggr\} \epp
\end{equation}
The determinants are all of the same structure. Expanding e.g.\ the first determinant
in the numerator we obtain
\begin{multline} \label{detalphan}
     \det_M \Bigl\{ \de^j_k - \frac{\r_n (\m_j|\a)}{\fa_n' (\m_j|\a)}
                    {\cal K}_{- \a} (\m_j - \m_k) \Bigr\}
        = 1 - \sum_{j=1}^M \frac{\r_n (\m_j|\a)}{\fa_n' (\m_j|\a)} {\cal K}_{- \a} (0) \\
	    + \sum_{1 \le j < k \le M} \frac{\r_n (\m_j|\a)}{\fa_n' (\m_j|\a)}
	         \frac{\r_n (\m_k|\a)}{\fa_n' (\m_k|\a)}
		 \det \left| \begin{array}{cc}
		      {\cal K}_{- \a} (0) & {\cal K}_{- \a} (\m_j - \m_k)  \\
		      {\cal K}_{- \a} (\m_k - \m_j) & {\cal K}_{- \a} (0)
		      \end{array} \right| - \dots \\
        = 1 + \sum_{k=1}^M \frac{(-1)^k}{k!}
	         \biggl[ \prod_{j=1}^k \int_{{\cal C}_n} \rd m_+^\a (\n_j) \biggr]
		 \det_k \bigl\{ {\cal K}_{- \a} (\n_l - \n_m) \bigr\} \epc
\end{multline}
where we have introduced the `measure'
\begin{equation}
     \rd m_+^\a (\la)
        = \frac{\rd \la \: \r_n (\la|\a)}{2 \p \i (1 + \fa_n (\la|\a))} \epp
\end{equation}
In the Trotter limit $N = 2M \rightarrow \infty$ the right hand side of
(\ref{detalphan}) converges to the Fredholm determinant of the integral
operator $\widehat{{\cal K}}_{-\a}$ defined by the kernel ${\cal K}_{-\a}$,
the measure $\rd m^\a_+$ and the contour~${\cal C}_n$,
\begin{multline} \label{detalpha}
     \lim_{N \rightarrow \infty}
     \det_M \Bigl\{ \de^j_k - \frac{\r_n (\m_j|\a)}{\fa_n' (\m_j|\a)}
                    {\cal K}_{- \a} (\m_j - \m_k) \Bigr\} \\
        = 1 + \sum_{k=1}^\infty \frac{(-1)^k}{k!}
	         \biggl[ \prod_{j=1}^k \int_{{\cal C}_n} \rd m_+^\a (\n_j) \biggr]
		 \det_k \bigl\{ {\cal K}_{- \a} (\n_l - \n_m) \bigr\}
        = \det_{\rd m^\a_+, {\cal C}_n} \bigl\{ 1 - \widehat{\cal K}_{-\a} \bigr\} \epp
\end{multline}

The other determinants in (\ref{aalphan}) can be treated in a similar way.
Introducing the measures
\begin{subequations}
\begin{align}
     \rd m (\la)
        & = \frac{\rd \la}{2 \p \i (1 + \fa_0 (\la))} \epc \\
     \rd m_0^\a (\la)
        & = \frac{\rd \la}{2 \p \i (1 + \fa_n (\la|\a))} \epc \\
     \rd m_-^\a (\la)
        & = \frac{\rd \la \: \r_n^{-1} (\la|\a)}{2 \p \i (1 + \fa_0 (\la))}
\end{align}
\end{subequations}
we obtain
\begin{multline} \label{ancalform}
     A_n (\a) =  \exp \biggl\{
                   \int_{{\cal C}_n} \frac{\rd \la}{2 \p \i} \:
	           \frac{\r_n' (\la|\a)}{\r_n (\la|\a)}
		   \ln \biggl( \frac{1 + \fa_n (\la|\a)}{1 + \fa_0 (\la)} \biggr) \biggr\}\\
		   \times
		   \frac{\det_{\rd m^\a_+, {\cal C}_n}
		         \bigl\{ 1 - \widehat{\cal K}_{-\a} \bigr\}
		         \det_{\rd m^\a_-, {\cal C}_n}
		         \bigl\{ 1 - \widehat{\cal K}_{\a} \bigr\}}
			{\det_{\rd m^\a_0, {\cal C}_n}
			 \bigl\{ 1 - \widehat{\cal K} \bigr\}
			 \det_{\rd m, {\cal C}_n}
			 \bigl\{ 1 - \widehat{\cal K} \bigr\}} \epp
\end{multline}
for the amplitudes in the Trotter limit. At least at first sight this
looks rather appealing. The amplitudes are entirely described in terms
of functions which appeared earlier in the description of the
thermodynamic properties, the correlation lengths and the correlation
functions of the model. These are the auxiliary functions $\fa_n$
(see \cite{Kluemper93}), the eigenvalue ratios $\r_n$ (see \cite{BoGo09,
BJMS10}) and the deformed kernel ${\cal K}_\a$.

The latter is a close relative of the deformed kernel $K_\a$, equation
(\ref{kernel}),  which first appeared in the context of the description
of the correlation functions of the XXZ chain in \cite{BGKS07,BoGo09}.
It seems that the kernel ${\cal K}_\a$ is more suitable in the analysis
of the scaling limit towards conformal field theory \cite{BJMS10,Boos11},
while $K_\a$ seems to be more convenient for the high-temperature analysis.
This is also reflected in the different symmetry properties of the two
kernels,
\begin{equation}
     {\cal K}_\a (- \la) = {\cal K}_{2 - \a} (\la) \epc \qd
     K_\a (- \la) = K_{- \a} (\la) \epp
\end{equation}

Since the two kernels are related by
\begin{equation} \label{kernrel}
     \re^{- \a \la} {\cal K}_\a (\la) = K_\a (\la) - q^{- \a} + q^\a \epc
\end{equation}
the integral operators in the determinants in (\ref{ancalform}) differ
from the corresponding integral operators with kernel $K_\a$ only by
a one-dimensional projector. Hence, using also the fact that ${\cal K}_0
= K_ 0 = K$, we can rewrite (\ref{ancalform}) as
\begin{multline} \label{an}
     A_n (\a) = \overline{\s}_+^- \s_-^- \exp \biggl\{
                   \int_{{\cal C}_n} \frac{\rd \la}{2 \p \i} \:
	           \frac{\r_n' (\la|\a)}{\r_n (\la|\a)}
		   \ln \biggl( \frac{1 + \fa_n (\la|\a)}{1 + \fa_0 (\la)} \biggr) \biggr\}\\
		   \times
		   \frac{\det_{\rd m^\a_+, {\cal C}_n}
		         \bigl\{ 1 - \widehat{K}_{-\a} \bigr\}
		         \det_{\rd m^\a_-, {\cal C}_n}
		         \bigl\{ 1 - \widehat{K}_\a \bigr\}}
			{\det_{\rd m^\a_0, {\cal C}_n}
			 \bigl\{ 1 - \widehat{K} \bigr\}
			 \det_{\rd m, {\cal C}_n}
			 \bigl\{ 1 - \widehat{K} \bigr\}} \epc
\end{multline}
where
\begin{equation}
     \overline{\s}_+^- = \lim_{\Re \la \rightarrow - \infty} \overline{\s}_+ (\la) \epc \qd
     \s_-^- = \lim_{\Re \la \rightarrow - \infty} \s_- (\la)
\end{equation}
and where the two functions $\overline{\s}_+ (\la)$, $\s_- (\la)$ are
generalizations to excited states of the generalized dressed charge functions
introduced in \cite{BoGo10}. They are defined by the integral equations
\begin{subequations}
\begin{align}
     \overline{\s}_+ (\la) = 1 + \int_{{\cal C}_n} \rd m_+^\a (\m)
	K_\a (\la - \m) \overline{\s}_+ (\m) \epc \\
     \s_- (\la) = 1 + \int_{{\cal C}_n} \rd m_-^\a (\m) \s_- (\m) K_\a (\m - \la) \epp
\end{align}
\end{subequations}
Using the techniques developed in \cite{BoGo10} it is not difficult to show
that
\begin{equation}
     \overline{\s}_+^- \s_-^- = \ch^{-2} \biggl\{
        \int_{{\cal C}_n} \frac{\rd \la}{2 \p \i} \:
	\ln \biggl( \frac{1 + \fa_n (\la|\a)}{1 + \fa_0 (\la)} \biggr) \biggr\} \epp
\end{equation}

When both sets of rapidities in (\ref{slavnov}) satisfy the Bethe equation
we obtain two different formulae by using either of the two corresponding
auxiliary functions on the right hand side of the equation. In order to
derive (\ref{ancalform}) we used both possibilities in a symmetric fashion
and ended up with a particularly symmetric result. Using two times the same
formula and also (\ref{norm}) we obtain alternative expressions for $A_n (\a)$
containing the square of one of the two determinants that appear in the
numerator on the right hand side of (\ref{ancalform}) and a different prefactor.
Comparing these asymmetric formulae with (\ref{ancalform}) we obtain the
following interesting relation between the Fredholm determinants in the numerator,
\begin{multline} \label{fpm}
     \frac{\det_{\rd m^\a_-, {\cal C}_n} \bigl\{ 1 - \widehat{\cal K}_{\a} \bigr\}}
          {\det_{\rd m^\a_+, {\cal C}_n} \bigl\{ 1 - \widehat{\cal K}_{-\a} \bigr\}} =
	  \exp \biggl\{ \frac{\be \r_n' (0|\a)}{\r_n (0|\a)}
	       - \int_{{\cal C}_n} \frac{\rd \la}{\p \i} \:
	          \ln \biggl( \frac{1 + \fa_n (\la|\a)}{1 + \fa_0 (\la)} \biggr) \\
	       + \int_{{\cal C}_n} \frac{\rd \la}{2 \p \i} \biggl[
	         \frac{\La_n' (\la|\a)}{\La_n (\la|\a)} \ln \bigl( 1 + \fa_0 (\la)\bigr) -
	         \frac{\La_0' (\la)}{\La_0 (\la)} \ln \bigl( 1 + \fa_n (\la|\a)\bigr)
		 \biggr] \biggr\} \epp
\end{multline}

Equations (\ref{ancalform}) and (\ref{an}) are the main results of this section.
They are valid for finite Trotter number as well as in the Trotter limit. For
finite Trotter number the right hand sides of these equations are represented
by finitely many integrals (see (\ref{detalphan})). We will show below that the
high- and the low-temperature limits of the amplitudes can be extracted from these
formulae. Their suitability for numerical calculations will be explored in
future research.
\section{Amplitudes in the asymptotic expansion of the transversal
correlation functions} \label{sec:transamp}
In this section we consider a slight generalization of the amplitudes $A_n^{-+}$,
equation (\ref{amppm}), depending on a spectral parameter $\x$ and on a
twist parameter $\a$. We define
\begin{equation} \label{deffpm}
     F_- (\x) = \frac{\<\Ps_0|B(\x)|\Ps_n^\a\>}{\La_n (\x|\a) \<\Ps_0|\Ps_0\>} \epc \qd
     F_+ (\x) = \frac{\<\Ps_n^\a|C(\x)|\Ps_0\>}{\La_0 (\x) \<\Ps_n^\a|\Ps_n^\a\>}
\end{equation}
and
\begin{equation}
     A_n^{-+} (\x) = F_- (\x) F_+ (\x) \epp
\end{equation}

For $F_+ (\x)$ we obtain in appendix~\ref{app:pm}
\begin{equation} \label{fplus}
     F_+ (\x) = \frac{G_+^- (\x) \re^{- \x} \prod_{j=1}^M d(\la_j) \re^{\la_j}}
		     {\prod_{j=1}^{M-1} \r_n (\m_j|\a) q^{- \a} d(\m_j) \re^{\m_j}} \,
                \frac{\det_{M-1} \Bigl\{ \de^j_k - \frac{\r_n (\m_j|\a)}{\fa_n' (\m_j|\a)}
		      K_{1 - \a} (\m_j - \m_k) \Bigr\}}
		     {\det_{M-1} \Bigl\{\de^j_k - \frac{1}{\fa_n' (\m_j|\a)}
		      K (\m_j - \m_k)\Bigr\}} \epp
\end{equation}
Here
\begin{equation}
     G_+^- (\x) = \lim_{\Re \la \rightarrow - \infty} G_+ (\la, \x) \epc
\end{equation}
and $G_+ (\la, \x)$ is the solution of the linear integral equation
\begin{multline}
     G_+ (\la, \x) = 
        - \cth(\la - \x) \\ + q^{\a - 1} \r_n (\x|\a) \cth(\la - \x - \h)
	+ \int_{{\cal C}_n} \rd m_+^\a (\m) K_{1-\a} (\la - \m) G_+(\m, \x) \epp
\end{multline}

The simplest way to obtain a formula for the amplitudes from (\ref{fplus}) is to
combine (\ref{fplus}) with (\ref{slavnov}) and (\ref{norm}). Then
\begin{multline} \label{apmas}
     A^{-+}_n (\x) = \frac{F_+^2 (\x) \<\Ps_n^\a|\Ps_n^\a\>}
                          {\r_n (\x|\a) \<\Ps_0|\Ps_0\>} = \\[2ex]
                \frac{G_+^- (\x)^2
		      \re^{2 (\sum_{j=1}^M \la_j - \sum_{j=1}^{M-1} \m_j - \x)}
                      \prod_{j=1}^M d(\la_j)}
		     {\r_n (\x|\a) \prod_{j=1}^{M-1} q^{- \a} d(\m_j) \r_n (\m_j|\a)} \,
		\frac{\prod_{j=1}^{M-1} \La_0 (\m_j)}
		     {\prod_{j=1}^M \La_0 (\la_j)} \\[1ex] \times
                \frac{\det_{M-1}^2 \Bigl\{ \de^j_k
		                           - \frac{\r_n (\m_j|\a)}{\fa_n' (\m_j|\a)}
		      K_{1 - \a} (\m_j - \m_k) \Bigr\}}
		     {\det_{M-1} \Bigl\{\de^j_k - \frac{1}{\fa_n' (\m_j|\a)}
		      K (\m_j - \m_k)\Bigr\}
		      \det_M \Bigl\{\de^j_k - \frac{1}{\fa_0' (\la_j)}
		      K (\la_j - \la_k)\Bigr\}} \epp
\end{multline}
As before the determinants turn into Fredholm determinants in the Trotter limit.
Some more care is necessary when dealing with the prefactors. All in all we
obtain
\begin{multline} \label{apmcalform}
     A_n^{-+} (\x) =  C(\x) \exp \biggl\{
                   \int_{{\cal C}_n} \frac{\rd \la}{2 \p \i} \:
	           \frac{\r_n' (\la|\a)}{\r_n (\la|\a)}
		   \ln \biggl( \frac{1 + \fa_n (\la|\a)}
		                    {1 + \fa_0 (\la)} \biggr) \biggr\}\\ \times
		   \frac{\det_{\rd m^\a_+, {\cal C}_n}^2
		         \bigl\{ 1 - \widehat{K}_{1-\a} \bigr\}}
			{\det_{\rd m^\a_0, {\cal C}_n}
			 \bigl\{ 1 - \widehat{K} \bigr\}
			 \det_{\rd m, {\cal C}_n}
			 \bigl\{ 1 - \widehat{K} \bigr\}} \epc
\end{multline}
where
\begin{multline}
     C(\x) = \frac{G_+^- (\x)^2 \re^{- 2\x}}{\r_n(\x|\a)} \,
             \frac{q^{\a + \k - 1} + q^{- \a - \k + 1}}{(q^\k + q^{- \k})^2} \\[1ex]
	     \mspace{-90.mu} \times
	  \exp \biggl\{ \frac{\be \r_n' (0|\a)}{\r_n (0|\a)}
	       - \int_{{\cal C}_n} \frac{\rd \la}{\p \i} \: \la \6_\la
	          \ln \biggl( \frac{1 + \fa_n (\la|\a)}{1 + \fa_0 (\la)} \biggr) \\
	       + \int_{{\cal C}_n} \frac{\rd \la}{2 \p \i} \biggl[
	         \frac{\La_n' (\la|\a)}{\La_n (\la|\a)} \ln \bigl( 1 + \fa_0 (\la)\bigr) -
	         \frac{\La_0' (\la)}{\La_0 (\la)} \ln \bigl( 1 + \fa_n (\la|\a)\bigr)
		 \biggr] \biggr\} \epp
\end{multline}

Note the similarity between the exponential term on the right hand side of
this equation and the right hand side of equation (\ref{fpm}). This similarity
hints that a formula like (\ref{fpm}) might also exist for the present case
and that a simpler, more symmetric formula for $A_n^{-+} (\x)$ can be derived.

There is indeed an alternative, less obvious formula for the form factor
$F_- (\x)$,
\begin{multline} \label{fminus}
     F_- (\x) =
     \frac{\overline{G}_-^+ (\x)}{(q^{1 + \a} - q^{- 1 - \a})(q^\a - q^{-\a})} \\
     \frac{\re^\x \prod_{j=1}^{M-1} q^{- \a} d(\m_j) \re^{\m_j}}
     {\prod_{j=1}^M \r_n^{-1} (\la_j|\a) d(\la_j) \re^{\la_j}} \,
      \frac{\det_M \Bigl\{ \de^j_k - \frac{\r_n^{-1} (\la_j|\a)}{\fa_0' (\la_j|\a)}
            K_{1 + \a} (\la_j - \la_k) \Bigr\}}
	   {\det_M \Bigl\{\de^j_k - \frac{1}{\fa_0' (\la_j|\a)}
	    K (\la_j - \la_k)\Bigr\}} \epc
\end{multline}
where
\begin{equation}
     \overline{G}_-^+ (\x)
        = \lim_{\Re \la \rightarrow \infty} \overline{G}_- (\la, \x) \epc
\end{equation}
and $\overline{G}_- (\la, \x)$ is the solution of the linear integral equation
\begin{multline}
     \overline{G}_- (\la, \x) = 
        - \cth(\la - \x) \\ + q^{\a + 1} \r_n^{-1} (\x|\a) \cth(\la - \x - \h)
	+ \int_{{\cal C}_n} \rd m_-^\a (\m) \overline{G}_- (\m, \x) K_{1+\a} (\m - \la)
	  \epp
\end{multline}
The derivation is shown in appendix~\ref{app:pm}.

Combining (\ref{fplus}) and (\ref{fminus}) we obtain a more symmetric expression
for the amplitudes $A_n^{-+} (\x)$,
\begin{multline} \label{apmsym}
     A_n^{-+} (\x) = \frac{G_+^- (\x) \overline{G}_-^+ (\x)}
                          {(q^{1 + \a} - q^{- 1 - \a})(q^\a - q^{-\a})}
                     \frac{\prod_{j=1}^M \r_n (\la_j|\a)}
		          {\prod_{j=1}^{M-1} \r_n (\m_j|\a)} \\ \times
                \frac{\det_{M-1} \Bigl\{ \de^j_k - \frac{\r_n (\m_j|\a)}{\fa_n' (\m_j|\a)}
		      K_{1 - \a} (\m_j - \m_k) \Bigr\}}
		     {\det_{M-1} \Bigl\{\de^j_k - \frac{1}{\fa_n' (\m_j|\a)}
		      K (\m_j - \m_k)\Bigr\}}
                \frac{\det_M \Bigl\{ \de^j_k
		      - \frac{\r_n^{-1} (\la_j|\a)}{\fa_0' (\la_j|\a)}
		      K_{1 + \a} (\la_j - \la_k) \Bigr\}}
		     {\det_M \Bigl\{\de^j_k - \frac{1}{\fa_0' (\la_j|\a)}
		      K (\la_j - \la_k)\Bigr\}} \epp
\end{multline} In the Trotter limit it turns into
\begin{multline} \label{amp}
     A_n^{-+} (\x) = \frac{q^{\a + \k - 1} + q^{- \a - \k + 1}}
                          {(q^{1 + \a} - q^{- 1 - \a})(q^\a - q^{-\a})
			   (q^\k + q^{- \k})} \\ \times
		     G_+^- (\x) \overline{G}_-^+ (\x)
                     \exp \biggl\{
		        \int_{{\cal C}_n} \frac{\rd \la}{2 \p \i} \:
			\frac{\r_n' (\la|\a)}{\r_n (\la|\a)}
			\ln \biggl( \frac{1 + \fa_n (\la|\a)}
			    {1 + \fa_0 (\la)} \biggr) \biggr\}\\ \times
		     \frac{\det_{\rd m^\a_+, {\cal C}_n}
		           \bigl\{ 1 - \widehat{K}_{1-\a} \bigr\}
		           \det_{\rd m^\a_-, {\cal C}_n}
		           \bigl\{ 1 - \widehat{K}_{1+\a} \bigr\}}
			  {\det_{\rd m^\a_0, {\cal C}_n}
			   \bigl\{ 1 - \widehat{K} \bigr\}
			   \det_{\rd m, {\cal C}_n}
			   \bigl\{ 1 - \widehat{K} \bigr\}} \epp
\end{multline}

What we find remarkable about this expression is that it is basically
of the same form as the expression (\ref{an}) for the amplitudes
in the form factor expansion of the generating function of the
$zz$-correlation functions. The exponential term and the ratio of
the products of two determinants appear in both cases, and the prefactor
is a product of two functions satisfying `conjugate' linear integral
equations. All expressions are parameterized by the functions $\r_n$
and $\fa_n$ which depend upon the specific excitation under
consideration and are, of course, different in both cases. In
appendix~\ref{app:pm} we argue that $\lim_{\a \rightarrow 0}
\overline{G}_-^+ (\x) = 0$, such that the prefactor in (\ref{amp})
is non-singular in this limit.

Following \cite{BJMST08a,BJMS10} we can introduce `the spin' of 
a local operator as the eigenvalue of the $z$-component of the
operator of the total spin when acting adjointly on it. Then
$\s^-$, $\s^z$, $\s^+$ have eigenvalues $s = -, 0 , +$, and we can
interpret the parameterization of the kernels in the Fredholm determinants
in the numerators of (\ref{an}) and (\ref{amp}) as $\whk_{s - \a}$ and
$\whk_{\a - s}$. In this combination the spin also appears in the
$TQ$ equation.

\section{Generating function at high temperatures}
We are convinced that our formula for the amplitudes are not only
aesthetically appealing but that they will also prove to be
efficient for the future analytical and numerical analysis of the
long-distance asymptotic behaviour of the finite-temperature 
correlation functions of the XXZ chain. In this and in the
following section, we shall support this claim with two examples
concerning the high- and the low-temperature asymptotics of
the generating function of the longitudinal correlation function.

First of all, we use \eqref{an} to calculate the leading high-temperature
contribution to the generating function $\bigl\< q^{2 \a S(m)}\bigr\>$. Since
the next-to leading eigenvalues vanish in the high-temperature limit, we only
have to consider the first term $n=0$ of the expansion \eqref{genfunexp},
\begin{equation}
\bigl\<q^{2 \a S(m)}\bigr\>
        =  A_0 (\a) \r_0^m (0|\a) + \mathcal{O}(1/T^m) \epp
  \label{genfunexpHTE}
\end{equation}
%if we are interested in the leading behaviour.
As is well known (see e.g.\ \cite{TsSh05}), the dominant eigenvalue of the
(twisted) quantum transfer matrix and the corresponding auxiliary function
can be expanded in a power series in  $1/T$. Let us briefly review this
procedure.

It follows from (\ref{nlie}) with $M = N/2$ that
\begin{equation}
     \mathfrak{a}_0(\la|\alpha) = q^{-2\alpha}
\end{equation}
if $\beta=\kappa=0$. In order to calculate the next order, we assume the
asymptotic expansion
\begin{equation}
     \mathfrak{a}_0(\la|\alpha) = q^{-2\alpha}
        \left( 1+ \sum_{k=1}^\infty \frac{a_k(\la|\alpha)}{T^k} \right) \epp
\label{ansatzHTE}
\end{equation}
By substituting \eqref{ansatzHTE} into \eqref{nlie}
and expanding in powers of $1/T$, we see that $a_1(\la|\alpha)$ is the solution
of the linear integral equation
\begin{equation}
     a_1(\la|\alpha) = -h - 2J\sh(\eta) \operatorname{e}(\lambda)
        - \int_{\mathcal{C}_0} \frac{\mathrm{d}\mu}{2\pi i}
	  K(\la-\mu)   \frac{a_1(\mu|\alpha)}{1+ q^{2\alpha}} \epp
\label{lie1}
\end{equation}
Obviously, the function
\begin{equation}
     a_1(\la|\alpha) =  -h- 2J\sh(\eta)
        \left( \operatorname{e}(\la) - \frac{K(\la)}{1+q^{2\alpha}} \right)
\label{solutionlie1}
\end{equation}
is a solution of \eqref{lie1}. Since $\mathfrak{a}_0(\la) = \mathfrak{a}_0(\la|0)$,
the corresponding expression for $\mathfrak{a}_0(\la)$ can be obtained from
\eqref{solutionlie1} by setting $\alpha=0$. These results can be substituted
into \eqref{rhoint}, after evaluating the contour integrals we arrive at
\begin{equation}
     \rho_0(\la|\alpha) = \frac{q^\alpha + q^{-\alpha}}{2}
        \left( 1+ \frac{h}{2T} \frac{q^\a - q^{-\a}}{q^\a + q^{-\a}} + \frac{J\sh(\eta)}{2T}
	\left( \frac{q^\a - q^{-\a}}{q^\a + q^{-\a}} \right)^2
	   K(\la) \right) + \mathcal{O}(1/T^2) \epp
\label{rhoHTE}
\end{equation}
By the same reasoning, we obtain
\begin{equation}
     \overline{\s}_+^- \s_-^- = 1+ \mathcal{O}(1/T^2)~, \quad \exp \biggl\{
                   \int_{{\cal C}_0} \frac{\rd \la}{2 \p \i} \:
	           \frac{\r_0' (\la|\a)}{\r_0 (\la|\a)}
		   \ln \biggl( \frac{1 + \fa_0 (\la|\a)}{1 + \fa_0 (\la)} \biggr)
		   \biggr\} =  1+ \mathcal{O}(1/T^2) \epp
\end{equation}
Lastly, we have to consider the Fredholm determinants appearing in \eqref{an}.
Since every pole of the measure comes with a factor $1/T$, we observe that
the $k$-th term of the series \eqref{detalpha} is of the order $\mathcal{O}(1/T^k)$.
Hence, for our purpose, it is sufficient to calculate the first term of the
expansion \eqref{detalpha}. Using \eqref{solutionlie1} and \eqref{rhoHTE} we
obtain after some calculations
\begin{equation}
     A_0(\alpha) =  1 + \frac{J\Delta}{T}
        \left( \frac{q^\a - q^{-\a}}{q^\a + q^{-\a}} \right)^2 + \mathcal{O}(1/T^2) \epp
\label{amplitudeHTE}
\end{equation} 
Combining \eqref{rhoHTE} and \eqref{amplitudeHTE} with \eqref{genfunexpHTE},
we end up with
\begin{align} \label{genhight}
     \langle q^{2\alpha S(m)} \rangle &
        = \left( \frac{q^\alpha+q^{-\alpha}}{2} \right)^m 
	  + \frac{mh}{2T} \left(\frac{q^\alpha-q^{-\alpha}}{2} \right)
	    \left(\frac{q^\alpha+q^{-\alpha}}{2} \right)^{m-1} \nonumber \\
	    &~~~-(m-1)\frac{J \Delta}{T}
	     \left(\frac{q^\alpha-q^{-\alpha}}{2} \right)^2
	     \left(\frac{q^\alpha+q^{-\alpha}}{2} \right)^{m-2}
	     + \! \mathcal{O}(1/T^2 ) \epp
\end{align}
Note that this result was already obtained in \cite{Seel05} within 
the multiple-integral approach to correlation functions.

Using \eqref{genzz} we can calculate the longitudinal correlators,
\begin{subequations}
\label{onetwohight}
\begin{align}
     \langle \sigma^z_1 \sigma^z_2 \rangle &
        = -\frac{J \Delta}{T} + \mathcal{O}(1/T^2) \epc \\
     \langle \sigma^z_1 \sigma^z_{m+1} \rangle &
        = \mathcal{O}(1/T^2)~, \quad m \geq 2 \epp
\end{align}
\end{subequations}
The emptiness formation probability is given by
\begin{equation} \label{empthight}
     \bigl\< {e_1}_+^+ \dots {e_m}_+^+\bigr\> 
        = \left( \frac{1}{2} \right)^m
	  \left( 1 + \frac{mh}{2T} - (m-1)\frac{J\Delta}{T} \right)
	  + \mathcal{O}(1/T^2) \epp
\end{equation}

Note that (\ref{genhight}), (\ref{onetwohight}) and (\ref{empthight}) are
valid for all $\D > - 1$. It will be interesting to extend the high-temperature
expansion systematically to higher orders.

\section{Generating function at low temperatures}
\label{sec:lot}
The low-temperature limits of the massless and the massive regimes
($|\D| \le 1$ and $\D > 1$, respectively) have to be analyzed separately.
For $\D > 1$ the spectrum of the quantum transfer matrix remains
gapped as $T \rightarrow 0$, but for $|\D| < 1$ it becomes gapless, and
the long-distance asymptotics of the static correlation functions
is expected to be described by conformal field theory. Here we consider
the massless case at finite magnetic field, $h > 0$. The low-energy
(large-distance) physics of our spin chain has then much in common with
the physics of free Fermions. It is largely determined by particle-hole
like excitations around the Fermi edge. As we shall see, the low-temperature
analysis can be based on a generalization of the Sommerfeld expansion
\cite{Sommerfeld28} to the interacting case. Such an analysis was
partially carried out in \cite{KlSc03}, but here we have to go further
and include the quantum numbers of the excited states.
%In the following we
%shall set $\h = - \i \g$ and, for simplicity, impose the restriction
%$0 < \g < \p/2$. Then $0 < \D = \cos (\g) < 1$.

Since the spectrum becomes gapless for $T \rightarrow 0$, infinitely
many terms in the form factor expansion contribute to the large-distance
asymptotics of the correlation functions. As we shall see, each
individual contribution vanishes as a fractional power of the temperature.
The situation is very similar as in case of the so-called critical form
factors defined between eigenstates of the ordinary transfer matrix
\cite{KKMST11a} which vanish in the thermodynamic limit as
negative fractional powers of the system size. We shall also observe
a close analogy of our result with that obtained for the Bose gas
by one of the authors \cite{KMS11a,KMS11b} using a different method.
As in \cite{KMS11b} we will utilize a remarkable summation formula,
obtained in \cite{KOV93,Olshanski03,KKMST11b}, which allows us to sum
up the leading low-temperature terms. This is necessary since the limit
involved in the infinite series of the leading terms and the limit
$T \rightarrow 0$ do not commute.

The technique we are going to apply can be used for analyzing
both, the longitudinal as well as the transversal correlation
functions at small temperatures. Here we shall restrict ourselves
to the analysis of the generating function of the longitudinal
correlation functions and postpone the transversal case to a
future publication. We start from the form factor expansion
(\ref{genfunexp}). In order to determine the leading low-temperature
behaviour of the amplitudes and eigenvalue ratios we use
(\ref{ancalform}) and (\ref{rhoint}). The full calculation is
rather lengthy. It consists of the following steps.
\begin{enumerate}
\item
Straightening of the contours in the non-linear integral
equations (\ref{nlie}) and in the expressions for the
amplitudes (\ref{ancalform}) and eigenvalue ratios (\ref{rhoint}).
\item
Low-temperature analysis of the non-linear integral equations.
\item
Low-temperature analysis of the eigenvalue ratios.
\item
Low-temperature analysis of the amplitudes.
\item
Summation of the leading terms.
\end{enumerate}
Fortunately, the most sophisticated step (v) can be adopted 
from \cite{KKMST11b}, and essentially only steps (i)-(iv) remain
to be done.
\subsection{Straightening the contours}
The contour ${\cal C}_0$ consists of two straight lines parallel to
the real axis and intersecting the imaginary axis at $\pm \h/2$ such
that the real axis is `encircled' in counterclockwise direction.
The contours ${\cal C}_n$ are deformations of ${\cal C}_0$ such that
all Bethe roots are included and all holes are excluded. To straighten
${\cal C}_n$ means to deform it into ${\cal C}_0$. In the process of
this deformation the particle and hole parameters are crossed and
the driving terms in the non-linear integral equations are supplemented
by terms depending explicitly on these parameters.

We denote the number of hole parameters by $n_h$ and the number of
particle parameters by $n_p$, set $n' = n_h + n_p$ and define the functions
\begin{equation}
     E(\la) = \ln \biggl( \frac{\sh (\la)}{\sh(\h + \la)} \biggr) \epc \qd
     \th (\la) = \ln \biggl( \frac{\sh (\h - \la)}{\sh(\h + \la)} \biggr)
\end{equation}
which are antiderivatives of the bare energy $\re (\la)$ and of
the kernel $K(\la)$. We further define
\begin{equation} \label{defz}
     z(\la) = - \frac{1}{2 \p \i}
                \ln \biggl( \frac{1 + \fa_n (\la|\a)}{1 + \fa_0 (\la)} \biggr) \epp
\end{equation}

A straightening of the contours in (\ref{nlie}) leads to
\begin{multline} \label{nliec0}
     \ln (\fa_n (\la)) =  \i \p (n_h - n_p)
        - (\k + N/2 - M - n_h + n_p) 2 \h - \be \re (\la) \\
        + \sum_{j=1}^{n_h} \th (\la - \la_j^h)
        - \sum_{j=1}^{n_p} \th (\la - \la_j^p)
        - \int_{{\cal C}_0} \frac{\rd \m}{2\p \i} \:
	  K(\la - \m) \ln \bigl(1 + \fa_n (\m)\bigr) \epp
\end{multline}
This defines an $n'$-parametric family of functions, depending on $\{\la_j^h\}$
and $\{\la_j^p\}$. The individual functions $\fa_n$ are then determined
by the subsidiary conditions
\begin{equation} \label{subs}
     \fa_n (\la_j^h) = \fa_n (\la_k^p) = - 1 \epc \qd j = 1, \dots, n_h \epc \qd
                       k = 1, \dots, n_p
\end{equation}
fixing these parameters to a discrete set of values.

We assume that for sufficiently small temperature
\begin{equation} \label{phcond}
     N/2 - M - n_h + n_p = 0
\end{equation}
and use this condition in our low-temperature analysis without further
mentioning. Then the eigenvalue ratios take the form
\begin{equation} \label{evratc0}
     \r_n (0|\a) = q^\a \exp \biggl\{
        \sum_{j=1}^{n_h} E (\la_j^h) - \sum_{j=1}^{n_p} E (\la_j^p)
	- \int_{{\cal C}_0} \rd \la \: \re (\la) z(\la) \biggr\} \epp
\end{equation}

The amplitudes (\ref{ancalform}) consist of a determinant factor times
\begin{equation} \label{defazero}
     A_n^{(0)} (\a) = \exp \biggl\{
                   \int_{{\cal C}_n} \frac{\rd \la}{2 \p \i} \:
	           \frac{\r_n' (\la|\a)}{\r_n (\la|\a)}
		   \ln \biggl( \frac{1 + \fa_n (\la|\a)}{1 + \fa_0 (\la)} \biggr) \biggr\}
                   \epp
\end{equation}
The determinant factor is treated in appendix~\ref{app:detlim}. Inserting
(\ref{rhoint}) into (\ref{defazero}) and straightening the contours we obtain
\begin{align} \label{amps}
     A_n^{(0)} (& \a) = % (-1)^{n_p - n_h}
	    \exp \biggl\{ - \int_{{\cal C}_0} \rd \la \: \int_{{\cal C}_0'} \rd \m \:
	         z(\la) \re' (\la - \m) z(\m) \biggr\} \notag \\
        & \times 
	    \exp \biggl\{ 2 \int_{{\cal C}_0} \rd \la \: z(\la)
	         \biggl[ \sum_{j=1}^{n_p} \cth(\la - \la_j^p)
	                 - \sum_{j=1}^{n_h} \cth(\la - \la_j^h) \biggr] \biggr\} \notag \\
        & \times 
	    \exp \biggl\{ - \int_{{\cal C}_0} \rd \la \: z(\la)
	         \biggl[ \sum_{j=1}^{n_p} \bigl( \cth(\la - \la_j^p + \h)
	                                         + \cth(\la - \la_j^p - \h) \bigr)
						   \notag \\
        & \mspace{180.mu} - \sum_{j=1}^{n_h} \bigl( \cth(\la - \la_j^h + \h)
			                         + \cth(\la - \la_j^h - \h) \bigr) \biggr]
		 \biggr\} \notag \\
        & \times
	  \biggl[ \prod_{j=1}^{n_p}
	          \Bigl( \6_\la \re^{- 2 \p \i z(\la)} \Bigr)^{-1}_{\la = \la_j^p} \biggr]
	  \biggl[ \prod_{j=1}^{n_h}
	          \Bigl( \6_\la \re^{- 2 \p \i z(\la)} \Bigr)^{-1}_{\la = \la_j^h} \biggr]
		  \notag \displaybreak[0] \\
        & \times \frac{\biggl[ \prod_{\substack{j, k = 1 \\ j \ne k}}^{n_h}
	                      \sh(\la_j^h - \la_k^h) \biggr]
                       \biggl[ \prod_{\substack{j, k = 1 \\ j \ne k}}^{n_p}
	                      \sh(\la_j^p - \la_k^p) \biggr]}
                      {\Bigl[ \prod_{j=1}^{n_h} \prod_{k=1}^{n_p}
		              \sh(\la_j^h - \la_k^p) \Bigr]^2} \notag \displaybreak[0] \\
        & \times \frac{\prod_{j = 1}^{n_h} \prod_{k = 1}^{n_p}
	               \sh(\la_j^h - \la_k^p + \h) \sh(\la_j^h - \la_k^p - \h)}
                      {\Bigl[ \prod_{j, k = 1}^{n_h} \sh(\la_j^h - \la_k^h - \h) \Bigr]
                       \Bigl[ \prod_{j, k = 1}^{n_p} \sh(\la_j^p - \la_k^p + \h) \Bigr]}
		       \epp
\end{align}
Here ${\cal C}_0'$ is a contour infinitesimally close to ${\cal C}_0$
and inside ${\cal C}_0$. The equation holds for $N = 2M$. For $N \ne 2M$
additional phase factors may arise.
\subsection{Low-temperature analysis of the non-linear integral equations}
From now on we restrict ourselves to the computation of quantities related
to the generating function. Then $M = N/2$. We introduce
\begin{equation}
     \e_0 (\la) = h - \frac{4 J(1 - \D^2)}{\ch (2 \la) - \D}
\end{equation}
and
\begin{equation}
     u_0 (\la) = - T \ln \bigl( \fa_0 (\la + \i \g/2) \bigr) \epc \qd
     u(\la) = - T \ln \bigl( \fa_n (\la + \i \g/2 |\a) \bigr) \epc
\end{equation}
where we suppressed the index $n$ and the dependence on $\a$ in the
definition of $u$.

It follows from (\ref{nliec0}) that
\begin{multline} \label{nlieu}
     u (\la) = \e_0 (\la) + T \biggl[ 2 \p \i \a'
        + \sum_{j=1}^{n_p} \th (\la - \la_j^p + \i \g/2)
        - \sum_{j=1}^{n_h} \th (\la - \la_j^h + \i \g/2) \biggr] \\
        + T \int_{{\cal C}_0 - \i \g/2} \frac{\rd \m}{2\p \i} \:
	  K(\la - \m) \ln \Bigl(1 + \re^{- \frac{u(\m)}T} \Bigr) \epc
\end{multline}
where we introduced $\a' = \h \a/\i \p$. A similar equation without
the contribution proportional to $T$ in the driving term holds for
$u_0$,
\begin{equation} \label{nlieu0}
     u_0 (\la) = \e_0 (\la)
                 + T \int_{{\cal C}_0 - \i \g/2} \frac{\rd \m}{2\p \i} \:
		            K(\la - \m) \ln \Bigl(1 + \re^{- \frac{u_0(\m)}T} \Bigr) \epp
\end{equation}
Since $\th$ is bounded on the contour ${\cal C}_0 - \i \g$, the terms in
square brackets in (\ref{nlieu}) may be neglected compared to $\e_0 (\la)$
when $T$ becomes small. Thus, $u$ and $u_0$ have the same zero temperature
limit $\e$. The integrals in (\ref{nlieu}) and (\ref{nlieu0}) vanish on those
parts of the contour on which $\Re \e > 0$ and are nonzero on their complement.
We claim that this complement is an interval $[-Q, Q]$ on the real axis.
Indeed, if $\e$ satisfies the linear integral equation
\begin{equation} \label{lieeps}
     \e (\la) = \e_0 (\la) + \int_{-Q}^Q \frac{\rd \m}{2\p \i} \: K(\la - \m) \e (\m) \epc
\end{equation}
where $Q$ is determined by the condition $\e (\pm Q) = 0$, then $\e$ behaves
similarly as the driving term. It is real and even on $\mathbb R$, has at most the
two zeros $\pm Q$ and is negative on $[-Q, Q]$. One can prove that
$\Re \e > h/4 > 0$ for all $\g \in (0,\p/2)$ on ${\mathbb R} - \i \g$,
which is the lower part of the integration contour in (\ref{nlieu}), (\ref{nlieu0}).
%This is obvious in the special cases when we know $\e$ explicitly
%(e.g.\ for $\D = 0$ or for $h = 0$) and can at least be verified on a
%computer for general values of the parameters $\D$ and $h$.
Hence, $\e$ satisfies (\ref{nlieu}) and (\ref{nlieu0}) up to
corrections at most of the order of $T$. The function $\e$ is the well-known
dressed energy function.

Our argument becomes rigorous by employing the following `generalized Sommerfeld
lemma' which also allows us to obtain the first and second order temperature
corrections below.
\begin{lemma} \label{lem:sommerfeld}
Let let $u, f$ be holomorphic in an open set containing a contour ${\cal C}_u$,
and let f be bounded on ${\cal C}_u$. Let $v = \Re u$, $w = \Im u$. Assume that
$v$ has exactly two zeros $Q_\pm$ on ${\cal C}_u$ separating ${\cal C}_u$ into
a part ${\cal C}_u^-$ between $Q_-$ and $Q_+$ on which $v$ is negative and
a remainder ${\cal C}_u^+$ on which $v$ is positive. Assume that $\exists p \in
{\mathbb Z}$ such that $w(Q_\pm) = 2\p p T$. Assume that ${\cal C}_u$ is oriented
in such a way that $Q_-$ comes before $Q_+$ on ${\cal C}_u^-$. Then (for $T > 0$)
\begin{multline}
     T \int_{{\cal C}_u} \rd \la \: f(\la) \ln \Bigl(1 + \re^{- \frac{u(\la)}T} \Bigr)
        = - \int_{Q_-}^{Q_+} \rd \la \: f(\la) \bigl(u(\la) - 2 \p \i p T\bigr) \\
	  + \frac{T^2 \p^2}{6}
	    \biggl( \frac{f(Q_+)}{u'(Q_+)} - \frac{f(Q_-)}{u'(Q_-)} \biggr)
          + \CO (T^4) \epp
\end{multline}
\end{lemma}
We present a proof of this lemma in appendix~\ref{app:sommerfeld}. The lemma is
designed in such a way that it can be directly applied to the integral equations
(\ref{nlieu}), (\ref{nlieu0}). For small $T$ the functions $u$ and $u_0$ are
close to $\e$ and therefore satisfy the requirements of Lemma~\ref{lem:sommerfeld}.
Hence,
\begin{multline} \label{ubart}
     \uq (\la) = \e_0 (\la) + T r_1 (\la)
                 + \frac{\i \p T^2}{12} \biggl[ \frac{K(\la - Q_+)}{u'(Q_+)}
		                              - \frac{K(\la - Q_-)}{u'(Q_-)} \biggr] \\
                 + \int_{Q_-}^{Q_+} \frac{\rd \m}{2 \p \i} K(\la - \m) \uq (\m)
		 + \CO (T^4) \epc
\end{multline}
where $\uq (\la) = u(\la) - 2 \p \i p T$ and
\begin{equation} \label{defr1}
     r_1 (\la) = 2 \p \i (\a' - p) + \sum_{j=1}^{n_p} \th (\la - \la_j^p + \i \g/2)
                 - \sum_{j=1}^{n_h} \th (\la - \la_j^h + \i \g/2) \epp
\end{equation}
An equation similar to (\ref{ubart}) holds for $u_0$.

If we neglect the $\CO (T^4)$ terms, equation (\ref{ubart}) is already a linear
integral equation for $\uq$. In the next step we resolve the $T$-dependence of
$Q_\pm$. We make the ansatz
\begin{equation}
     u(\la) = \e (\la) + T u_1 (\la) + T^2 u_2 (\la) + \CO (T^3) \epc \qd
     Q_\pm = \pm Q + T Q^{(1)}_\pm + \CO (T^2) \epp
\end{equation}
Using that $u(Q_\pm) = 2 \p \i p T$ and $\e (\pm Q) = 0$ we immediately obtain
\begin{equation} \label{boundariesot}
     Q_\pm = \pm Q \mp \frac{\uq_1 (\pm Q)}{\e' (Q)} T + \CO (T^2) \epc
\end{equation}
where we agree upon $\uq_1 (\la) = u_1 (\la) - 2 \p \i p$.

Inserting (\ref{boundariesot}) into (\ref{ubart}) and expanding up to the second
order in $T$ we obtain
\begin{equation} \label{ubart2}
     \uq (\la) = \e_0 (\la) + T r_1 (\la) + T^2 r_2 (\la)
                 + \int_{-Q}^{Q} \frac{\rd \m}{2 \p \i} K(\la - \m) \uq (\m)
		 + \CO (T^3) \epc
\end{equation}
where
\begin{equation}
     r_2 (\la) = \frac{\i \p}{4 \e' (Q)}
                 \biggl[ K(\la - Q) \biggl(\frac13 + \frac{\uq_1^2 (Q)}{\p^2}\biggr)
		       + K(\la + Q) \biggl(\frac13 + \frac{\uq_1^2 (-Q)}{\p^2}\biggr)
		       \biggr] \epp
\end{equation}
Then the linearity of (\ref{ubart2}) implies that $\uq_1$ and $u_2$ satisfy
\begin{subequations}
\label{u1u2}
\begin{align}
     \uq_1 (\la) & = r_1 (\la)
                     + \int_{-Q}^{Q} \frac{\rd \m}{2 \p \i} K(\la - \m) \uq_1 (\m) \epc \\
     u_2 (\la) & = r_2 (\la)
                   + \int_{-Q}^{Q} \frac{\rd \m}{2 \p \i} K(\la - \m) u_2 (\m) \epp
\end{align}
\end{subequations}

We can express these functions by means of a set of standard functions that
appear in the description of the ground state properties of the model. We
introduce the dressed charge function $Z$, the root density function $\r$,
the dressed phase $\Ph$ and the resolvent $R$ as solutions of the linear
integral equations
\begin{subequations}
\label{zphr}
\begin{align} \label{liz}
     Z (\la) & = 1 + \int_{-Q}^{Q} \frac{\rd \m}{2 \p \i} K(\la - \m) Z (\m) \epc \\
     \r (\la) & = - \frac{\re (\la + \i \g/2)}{2 \p \i}
                       + \int_{-Q}^{Q} \frac{\rd \m}{2 \p \i} K(\la - \m) \r (\m)
		       \epc \\ \label{liph}
     \Ph (\la, \n) & = - \frac{\th (\la - \n)}{2 \p \i}
                       + \int_{-Q}^{Q} \frac{\rd \m}{2 \p \i} K(\la - \m) \Ph (\m, \n)
		       \epc \\
     R (\la, \n) & = K(\la - \n)
                     + \int_{-Q}^{Q} \frac{\rd \m}{2 \p \i} K(\la - \m) R (\m, \n) \epp
\end{align}
\end{subequations}
Comparing (\ref{u1u2}) and (\ref{zphr}) we conclude that
\begin{subequations}
\label{u1u2exp}
\begin{align} \label{u1exp}
     \uq_1 (\la) & = 2 \p \i \biggl[ (\a' - p) Z(\la)
                        + \sum_{j=1}^{n_h} \Ph (\la, \la_j^h - \i \g/2)
                        - \sum_{j=1}^{n_p} \Ph (\la, \la_j^p - \i \g/2) \biggr] \epc \\
			\label{u2exp}
     u_2 (\la) & = \frac{\i \p}{4 \e' (Q)}
                   \biggl[ R(\la,Q) \biggl(\frac13 + \frac{\uq_1^2 (Q)}{\p^2}\biggr)
		         + R(\la,-Q) \biggl(\frac13 + \frac{\uq_1^2 (-Q)}{\p^2}\biggr)
		         \biggr] \epp
\end{align}
\end{subequations}
The root density $\r$ will reappear later when we calculate the low-%
temperature asymptotics of the correlation lengths. Recall that
$\ln (\fa_n \bigl(\la + \i \g/2| \a)\bigr) = - \e(\la)/T - u_1(\la)
- u_2(\la) T + \CO (T^2)$. Thus, we have obtained the solutions of
(\ref{nliec0}) for $M = N/2$ and $n_p = n_h$ up to corrections of order
$T^2$.

It remains to determine the hole and particle parameters $\la_j^h$ and
$\la_k^p$ by means of the subsidiary condition (\ref{subs}). In terms
of the function $\uq$ the latter reads
\begin{equation} \label{subsu}
     \uq (\la_j^h - \i \g/2) = 2 \p \i T (h_j - 1/2) \epc \qd
     \uq (\la_k^p - \i \g/2) = 2 \p \i T (p_k - 1/2) \epc
\end{equation}
where $h_j, p_k \in {\mathbb Z}$ are such that the $\la_j^h$ are located 
below ${\cal C}_0$ while the $\la_k^p$ are located above ${\cal C}_0$.
Assuming that the quantum numbers $h_j$, $p_k$ of the particles and
holes are all different and that the $\la_j^h$ and $\la_k^p$ are
uniquely determined by (\ref{subsu}) we may write
\begin{equation}
     \la_j^h = \i \g/2 + x_{h_j} \epc \qd \la_k^p = \i \g/2 + y_{p_k} \epp
\end{equation}
We restrict ourselves to excitations close to the Fermi edges $\pm Q$
for which
\begin{equation} \label{defxy}
     x_{h_j}^\pm = \pm Q + \CO (T) \epc \qd y_{p_k}^\pm = \pm Q + \CO (T)
\end{equation}
and denote the number of holes and particles close to the left and
right Fermi edge by $n_h^-$, $n_p^-$ and $n_h^+$, $n_p^+$, respectively.
Since we assume that $n_h = n_p = n'$ we have
\begin{equation} \label{defl}
     n_h^- - n_p^- = n_p^+ - n_h^+ = \ell \epp
\end{equation}
Using (\ref{defxy}), (\ref{defl}) and the identity $\Ph (\la, Q) - \Ph (\la, -Q)
= Z(\la) - 1$, following from (\ref{liz}), (\ref{liph}), we infer that
\begin{equation} \label{defu1ell}
     \uq_1 (\la) = \uq_1^{(\ell)} (\la) + \CO (T) \epc \qd
     \uq_1^{(\ell)} (\la) = 2 \p \i \bigl( (\a' - \ell - p) Z(\la) + \ell \bigr) \epp
\end{equation}
Hence,
\begin{subequations}
\begin{align} \label{xpmdet}
     \uq (x_{h_j}^\pm) & = \mp 2 \p \i T (h_j^\pm - 1/2)
        = \pm \e' (Q) (x_{h_j}^\pm \mp Q) + \uq_1^{(\ell)} (Q) T + \CO (T^2) \epc \\
	  \label{ypmdet}
     \uq (y_{p_k}^\pm) & = \pm 2 \p \i T (p_k^\pm - 1/2)
        = \pm \e' (Q) (y_{p_k}^\pm \mp Q) + \uq_1^{(\ell)} (Q) T + \CO (T^2) \epc
\end{align}
\end{subequations}
where the signs in the first equations have to be understood as part of the
definitions of $h_j^\pm$ and $p_k^\pm$. It follows that
\begin{subequations}
\label{hppms}
\begin{align}
     x_{h_j}^\pm & = Q_\pm - \frac{2 \p \i T}{\e' (Q)} \bigl( h_j^\pm - 1/2 \bigr)
                     + \CO (T^2) \epc \\
     y_{p_k}^\pm & = Q_\pm + \frac{2 \p \i T}{\e' (Q)} \bigl( p_k^\pm - 1/2 \bigr)
                     + \CO (T^2) \epp
\end{align}
\end{subequations}
Our definition of particles and holes implies that
\begin{equation} \label{sepph}
     \Im x_{h_j}^\pm < 0 \epc \qd \Im y_{p_k}^\pm > 0 \epp
\end{equation}

Equation (\ref{hppms}) together with the constraint (\ref{sepph})
determines the roots of the function $1 + \fa_n (\la| \a)$ close to
the Fermi points to first order in~$T$. To this order they are
equidistantly located on two parallel lines intersecting the canonical
contour ${\cal C}_0$ at a right angle in $\i \g/2 \mp Q$. These roots
can be occupied by particles or holes. The picture is very much the
same as for free Fermions. The parameter $\ell$, introduced in
(\ref{defl}), is the excess of particles over holes at the right Fermi
edge which, in conventional language, is describing the `number of Umklapp
processes' involved in the excitation. The analogy with free Fermions
(the case $\D = 0$) also implies that $p$ is the number of roots below
$Q_+$ if $Q_+$ is above ${\cal C}_0$ (minus the number of roots between
$Q_+$ and ${\cal C}_0$ if $Q_+$ is below). For $p \ne 0$ we cannot naively
place particles or holes between $Q_\pm$ and ${\cal C}_0$, since
this would be inconsistent with our derivation. For $p \ne 0$ we have
to take into account that we might cross particle or hole parameters
in the course of deforming the contour ${\cal C}_0$ into the contour
${\cal C}_u$ required by Lemma~\ref{lem:sommerfeld}. In this case
the function $r_1$ in (\ref{defr1}) has to be modified accordingly.
In the following we shall avoid such complication by simply assuming
that $p = 0$. We shall see that this is already enough to derive the
leading large-distance asymptotics of the longitudinal correlation
functions for small temperatures and shall come back to the general
case in a forthcoming publication.

Since $p = 0$ by hypothesis, we suppress the bar in $\uq$, $\uq_1$ etc.\
in the following except in some general statements that are valid for any
$p \in {\mathbb Z}$. Setting $p = 0$ means that the points $Q_\pm + \i \g /2$
are located between the roots closest to ${\cal C}_0$, whence
$|\Im u_1^{(\ell)} (Q)| < \p$. Using (\ref{defu1ell}) with $p = 0$
and setting ${\cal Z} = Z(Q)$ this condition is equivalent to
\begin{equation} \label{pnulluneq}
     - \frac{\Re \a'}{{\cal Z}^{-1} - 1} - \frac{1}{2(1 - {\cal Z})} < \ell <
     - \frac{\Re \a'}{{\cal Z}^{-1} - 1} + \frac{1}{2(1 - {\cal Z})} \epp
\end{equation}
The latter inequality restricts the range of $\ell$ for given $\a'$,
${\cal Z}$. It is easy to see that $1/\sqrt{2} < {\cal Z} < 1$ for
$0 < \D < 1$. Thus, (\ref{pnulluneq}) is always satisfied for $\ell =
- 1, 0, 1$ as long as $\Re \a'$ is small enough. As we shall see,
these values determine the leading large-distance asymptotics of the
correlation functions. Obviously the inequality (\ref{pnulluneq}) can
be satisfied by any $\ell \in {\mathbb Z}$ if ${\cal Z}$ is close
enough to $1$ or if we choose $\a'$ appropriately. For this reason
we keep $\ell \in {\mathbb Z}$ general in our subsequent analysis.

For the low-temperature analysis of the eigenvalue ratios in the next
subsection we need $u_1^{(\ell)}$ up to the first order in $T$.
We insert (\ref{defu1ell}), (\ref{hppms}) into (\ref{u1exp}) and use
$\6_\n \Ph (\la, \n) = R(\la, \n)/2 \p \i$ to obtain
\begin{multline} \label{u1expexp}
     u_1 (\la) = u_1^{(\ell)} (\la) + \frac{2 \p \i T}{\e'(Q)}
                   \biggl\{ R(\la,Q) \biggl[ \frac{\ell u_1^{(\ell)} (Q)}{2 \p \i}
		      - \sum_{j=1}^{n_h^+} \Bigl(h_j^+ - \frac12\Bigr)
		      - \sum_{j=1}^{n_p^+} \Bigl(p_j^+ - \frac12\Bigr) \biggr] \\
                    + R(\la, -Q) \biggl[ \frac{\ell u_1^{(\ell)} (Q)}{2 \p \i}
		      - \sum_{j=1}^{n_h^-} \Bigl(h_j^- - \frac12\Bigr)
		      - \sum_{j=1}^{n_p^-} \Bigl(p_j^- - \frac12\Bigr) \biggr] \biggr\}
		      + \CO (T^2) \epp
\end{multline}
Equations (\ref{u2exp}) and (\ref{u1expexp}) together with the corresponding
linear integral equations then determine $\ln(\fa_n (\la|\a))$ up to the order $T$.
Using these equations we can readily calculate the leading order low-temperature
expansion of the eigenvalue ratios.

\subsection{Low-temperature analysis of the eigenvalue ratios}
We start from the representation (\ref{evratc0}) of the eigenvalue ratio. In
order to expand the sums in the exponent we use (\ref{hppms}). Then
\begin{multline}
     \sum_{j=1}^{n_h} E (\la_j^h) - \sum_{j=1}^{n_p} E (\la_j^p) =
        - 2 \ell E(Q + \i \g/2) \\
	+ \frac{2 \p \i T \re(Q + \i \g/2)}{\e' (Q)} \biggl[
	\frac{\ell u_1^{(\ell)} (Q)}{\i \p} - \sum_{j=1}^{n'} (h_j + p_j - 1) \biggr]
	+ \CO (T^2) \epp
\end{multline}
It remains to calculate the integral
\begin{equation}
	- \int_{{\cal C}_0} \rd \la \: \re (\la) z(\la)
	   = \int_{{\cal C}_0 - \i \g/2} \frac{\rd \la}{2 \p \i} \:
	     \re (\la + \i \g/2) \ln \Biggl( \frac{1 + \re^{- \frac{u(\la)}T}}
	                                          {1 + \re^{- \frac{u_0(\la)}T}} \Biggr)
             \epp
\end{equation}
For this purpose one can employ Lemma~\ref{lem:sommerfeld} as well as the
formulae of the previous section. Introducing the Fermi momentum
\begin{equation} \label{deffermimom}
     k_F = \p \int_{-Q}^Q \rd \la \: \r(\la) \epc
\end{equation}
and the sound velocity
\begin{equation}
     v_0 = \frac{\e'(Q)}{2 \p \r(Q)}
\end{equation}
we obtain altogether
\begin{multline}
     \r_n (0|\a) = q^\a \exp \biggl\{ - 2 \i (\a' - \ell) k_F \\
                   - \frac{2 \p T}{v_0} \biggl[ (\a' - \ell)^2 {\cal Z}^2 - \ell^2
		      + \sum_{j=1}^{n'} (h_j + p_j - 1) \biggr] \biggr\} + \CO(T^2) \epp
\end{multline}
\subsection{Low-temperature analysis of the amplitudes}
The low-temperature analysis of the amplitudes $A_n^{(0)} (\a)$, eqn.\ (\ref{amps}),
is the most tedious part of the calculation. Those terms in the expression
for the amplitudes that depend explicitly on the hole and particle parameters
can be expanded using (\ref{hppms}). We obtain for the respective leading terms
\begin{subequations}
\label{disc}
\begin{align}
     & \biggl[ \prod_{j=1}^{n'}
               \Bigl( \6_\la \re^{- 2 \p \i z(\la)} \Bigr)^{-1}_{\la = \la_j^p}
               \Bigl( \6_\la \re^{- 2 \p \i z(\la)} \Bigr)^{-1}_{\la = \la_j^h} \biggr]
       = (-1)^\ell \biggl( \frac{T}{\e'(Q)} \biggr)^{2 n'}
                   \bigl(1 - \re^{u_1^{(\ell)} (Q)}\bigr)^{2 n'} \epc \\[-2.ex]
    & \biggl[ \prod_{\substack{j, k = 1 \\ j \ne k}}^{n'}
                       \sh(\la_j^h - \la_k^h) \sh(\la_j^p - \la_k^p) \biggr]
              \biggl[ \prod_{j, k = 1}^{n'} \sh(\la_j^h - \la_k^p) \biggr]^{-2}
            \notag \displaybreak[0] \\[-2.ex] & \mspace{36.mu}
       = \biggl( \frac{2 \p \i T}{\e'(Q)} \biggr)^{2 \ell^2 - 2 n'} \sh^{- 2 \ell^2} (2Q)
         \prod_{\e = \pm} \frac{\prod_{1 \le j < k \le n_h^\e} (h_j^\e - h_k^\e)^2
	                        \prod_{1 \le j < k \le n_p^\e} (p_j^\e - p_k^\e)^2}
			       {\prod_{j=1}^{n_h^\e} \prod_{k=1}^{n_p^\e}
			             (h_j^\e + p_k^\e - 1)^2} \epc \\[-2.ex]
    & \prod_{j, k = 1}^{n'}
         \frac{\sh(\la_j^h - \la_k^p + \h) \sh(\la_j^h - \la_k^p - \h)}
	      {\sh(\la_j^h - \la_k^h - \h) \sh(\la_j^p - \la_k^p + \h)}
       = \biggl( \frac{\sh(\h + 2 Q) \sh(\h - 2Q)}{\sh^2 (\h)} \biggr)^{\ell^2} \epp
\end{align}
\end{subequations}

The integrals in the exponent in (\ref{amps}) which have a non-singular integrand
can be evaluated by means of Lemma~\ref{lem:sommerfeld} which together with
(\ref{boundariesot}) implies the following
\begin{corollary} \label{cor:away}
For $f$ holomorphic in a finite strip around ${\cal C}_0$ and $z$ defined in (\ref{defz})
\begin{equation}
     \int_{{\cal C}_0} \rd \la \: f(\la) z(\la)
        = - \int_{-Q}^Q \frac{\rd \la}{2 \p \i} f(\la + \i \g/2) \uq_1^{(\ell)} (\la)
	  + \CO (T) \epp
\end{equation}
\end{corollary}
This can be used to conclude that to leading order in $T$
\begin{multline} \label{nonsingint}
     - \int_{{\cal C}_0} \rd \la \: z(\la)
          \biggl[ \sum_{j=1}^{n'} \bigl( \cth(\la - \la_j^p + \h) 
	                                  + \cth(\la - \la_j^p - \h) \bigr) \\[-2.ex]
       \mspace{180.mu}
                - \sum_{j=1}^{n'} \bigl( \cth(\la - \la_j^h + \h)
		                         + \cth(\la - \la_j^h - \h) \bigr) \biggr] \\
     + \int_{{\cal C}_0} \rd \la \: \int_{{\cal C}_0'} \rd \m \:
              z(\la) \cth' (\la - \m + \h) z(\m) \displaybreak[0] \\
     = \ell^2 \ln \biggl( \frac{\sh^2 (\h)}{\sh(\h + 2 Q) \sh(\h - 2Q)} \biggr)
     - (\a' - \ell)^2 \int_{-Q}^Q \rd \la \int_{-Q}^Q \rd \m
          \frac{Z(\la) Z(\m)}{\sh^2 (\la - \m + \h)} \epp
\end{multline}

The remaining integrals involve singularities close to the contour
of integration and are harder to estimate. Fortunately these integrals
are similar to integrals that appeared previously in the calculation
of the low-temperature asymptotics of the generating function of
the density-density correlators of the Bose gas in \cite{KMS11b}.
The following lemma is slightly stronger than the corresponding
statement in case of the Bose gas in appendix B of \cite{KMS11b}
and can be obtained in the same spirit.
\begin{lemma}
Let $u$ and ${\cal C}_u$ be subject to the same assumptions as in
Lemma~\ref{lem:sommerfeld}. Let $\la_+$ be located above ${\cal C}_u$
and $\la_-$ below. Then the Cauchy-type integral
\begin{equation}
     I_u (\la_\pm) = \int_{{\cal C}_u} \rd \la \: \cth(\la - \la_\pm)
                        \ln \Bigl(1 + \re^{- \frac{u(\la)}T} \Bigr)
\end{equation}
admits a low-temperature expansion whose form depends on the distance
of $\la_\pm$ from the zeros $Q_\pm$ of the real part of $u$.

For $\de > 0$ define $V_\pm = \bigl\{ z \in {\mathbb C} \big| |u(z)| < \de/2,
\text{$z$ close to $Q_\pm$} \bigr\}$. If $\la_\pm$ are uniformly away from
$Q_\pm$, then Lemma~\ref{lem:sommerfeld} applies, and
\begin{equation}
     I_u (\la_\pm) =
        - \int_{Q_-}^{Q_+} \rd \la \: \cth(\la - \la_\pm) \frac{\uq (\la)}{T}
	+ \CO (T) \epp
\end{equation}
If $\la_\pm \in V_+$, then
\begin{multline}
    I_u (\la_\pm) =  - \int_{Q_-}^{Q_+} \rd \la \:
                       \cth(\la - \la_\pm) \frac{\uq (\la) - \uq (\la_\pm)}{T}
                     \ \mp 2 \p \i \ln \biggl\{ \G \biggl(
		        \2 \pm \frac{\uq (\la_\pm)}{2 \p \i T} \biggr) \biggr\} \\
      \pm \p \i \ln (2 \p) + \frac{\uq (\la_\pm)}T \biggl\{
                     \ln \biggl(\frac{\uq (\la_\pm)}{\pm 2 \p \i T} \biggr) - 1
                     - \ln \biggl(\frac{\sh (Q_+ - \la_\pm)}{\sh (Q_- - \la_\pm)} \biggr)
		     \biggr\} + \CO (T) \epp
\end{multline}
If $\la_\pm \in V_-$, then
\begin{multline}
    I_u (\la_\pm) =  - \int_{Q_-}^{Q_+} \rd \la \:
                       \cth(\la - \la_\pm) \frac{\uq (\la) - \uq (\la_\pm)}{T}
                     \ \mp 2 \p \i \ln \biggl\{ \G \biggl(
		        \2 \mp \frac{\uq (\la_\pm)}{2 \p \i T} \biggr) \biggr\} \\
      \pm \p \i \ln (2 \p) - \frac{\uq (\la_\pm)}T \biggl\{
                     \ln \biggl(\frac{\uq (\la_\pm)}{\mp 2 \p \i T} \biggr) - 1
                     + \ln \biggl(\frac{\sh (Q_+ - \la_\pm)}{\sh (Q_- - \la_\pm)} \biggr)
		     \biggr\} + \CO (T) \epp
\end{multline}
\end{lemma}

First of all this can be used to calculate the periodic Cauchy transform
\begin{equation} \label{cauchyz}
     L_{{\cal C}_0} [z] (\n) =  \int_{{\cal C}_0} \rd \m \: z(\m) \cth(\m - \n)
\end{equation}
when $\n$ is close to $\i \g/2 \pm Q$. Assuming this and setting $s = \sign \bigl(
\Im (\la) \bigr)$ we find
\begin{multline} \label{cthcauchy}
     L_{{\cal C}_0} [z] (\la + \i \g/2) =
        - \int_{-Q}^Q \frac{\rd \m}{2 \p \i} \:
	      \cth(\m - \la) \bigl(u_1 (\m) - u_1 (\la)\bigr) \\
	      + \frac{u_1 (\la)}{2 \p \i}
	        \ln \biggl( \frac{\e^{\pm 1} (\la) \sh(\la + Q)}{\sh(\la - Q)} \biggr)
	      \mp \frac{u_1 (\la)}{2 \p \i} \ln( \pm s 2 \p \i T) \\
	      - s \ln \biggl( \frac{\G \bigl(1/2 \pm s u (\la)/2 \p \i T\bigr)}
	                           {\G \bigl(1/2 \pm s u_0 (\la)/2 \p \i T\bigr)} \biggr)
				   + \CO (T) \epp
\end{multline}
This formula is useful for the evaluation of the remaining integrals. Using also 
(\ref{xpmdet}) and (\ref{hppms}) we obtain
\begin{align} \label{singint1}
     & 2 \int_{{\cal C}_0} \rd \la \: z(\la)
         \biggl[ \sum_{j=1}^{n'} \cth(\la - \la_j^p)
	        - \sum_{j=1}^{n'} \cth(\la - \la_j^h) \biggr]
		\notag \displaybreak[0] \\
     & \mspace{18.mu} = \frac{2 \ell u_1^{(\ell)} (Q)}{\p \i}
          \ln \biggl( \frac{\e'(Q) \sh(2Q)}{2 \p T} \biggr) - n' u_1^{(\ell)} (Q)
       \notag \\ & \mspace{54.mu}
	  - 4 \ell \int_{- Q}^Q \frac{\rd \la}{2 \p \i}
	              \bigl(u_1^{(\ell)} (\la) - u_1^{(\ell)} (Q) \bigr) \cth(\la - Q)
		      %\bigl(\cth(\la - Q) - \cth(\la + Q)\bigr)
       \notag \displaybreak[0] \\ & \mspace{54.mu}
          + \ln \Biggl\{ \Biggl[ \prod_{j=1}^{n_h^+}
	       \frac{\G^2 \Bigl( h_j^+ + \frac{u_1^{(\ell)} (Q)}{2 \p \i} \Bigr)}
	            {\G^2 ( h_j^+ )} \Biggr]
               \Biggl[ \prod_{j=1}^{n_p^+}
	       \frac{\G^2 \Bigl( p_j^+ - \frac{u_1^{(\ell)} (Q)}{2 \p \i} \Bigr)}
	            {\G^2 ( p_j^+ )} \Biggr] \Biggr. \notag \\ & \mspace{108.mu} \times
               \Biggl[ \prod_{j=1}^{n_h^-}
	       \frac{\G^2 \Bigl( h_j^- - \frac{u_1^{(\ell)} (Q)}{2 \p \i} \Bigr)}
	            {\G^2 ( h_j^- )} \Biggr]
               \Biggl[ \prod_{j=1}^{n_p^-}
	       \frac{\G^2 \Bigl( p_j^- + \frac{u_1^{(\ell)} (Q)}{2 \p \i} \Bigr)}
	            {\G^2 ( p_j^- )} \Biggr] \Biggr\} + \CO (T \ln (T)) \epp
\end{align}

Finally the most delicate term is the remaining double integral with
singular integrand. For this integral we can use again (\ref{cthcauchy})
and proceed analogously to appendix B of \cite{KMS11b}. Then to leading
order in $T$
\begin{multline} \label{singint2}
     - \int_{{\cal C}_0} \rd \la \: \int_{{\cal C}_0'} \rd \m \:
          z(\la) \cth' (\la - \m) z(\m) = 
	  2 \biggl( \frac{u_1^{(\ell)} (Q)}{2 \p \i} \biggr)^2
	  \ln \biggl( \frac{2 \p T}{\e' (Q) \sh(2Q)} \biggr) \\
	  + 2 \ln \biggl\{
	        G\biggl(1 - \frac{u_1^{(\ell)} (Q)}{2 \p \i} \biggr)
		G\biggl(1 + \frac{u_1^{(\ell)} (Q)}{2 \p \i} \biggr) \biggr\}
          + C_1 \biggl[ \frac{u_1^{(\ell)}}{2 \p \i} \biggr] \epc
\end{multline}
where $G$ is the Barnes $G$-function and $C_1$ is a functional defined by
\begin{multline}
     C_1 [v] = \2 \int_{- Q}^Q \rd \la \int_{- Q}^Q \rd \m \:
		  \frac{v' (\la) v (\m) - v (\la) v' (\m)}{\tgh (\la - \m)}
               + 2 v (Q) \int_{- Q}^Q \rd \la
	         \frac{v (\la) -  v (Q)}{\tgh (\la - Q)} \epp
%               - v (-Q) \int_{- Q}^Q \rd \la
%	         \frac{v (\la) -  v (-Q)}{\tgh (\la + Q)} \epp
\end{multline}

When we insert all the terms (\ref{disc}), (\ref{nonsingint}), (\ref{singint1})
and (\ref{singint2}) into (\ref{amps}) and also use (\ref{defu1ell}) they
neatly combine. The amplitude $A_n^{(0)}$ decomposes into two factors $A_n^{(\pm)}$
pertaining to the left and right Fermi edge:
\begin{equation}
     A_n^{(0)} (\i \p \a/\h) = A_n^{(-)} (\a) A_n^{(+)} (\a) \epc
\end{equation}
where
\begin{multline} \label{ampssmallt}
     A_n^{(\e)} (\a) = G^2 \bigr(1 - \e [\ell + (\a - \ell) {\cal Z}] \bigl)
	\biggl( \frac{2 \p T \re^{C [Z/{\cal Z}]}}
	             {\e' (Q) \sh(2Q)} \biggr)^{(\a - \ell)^2 {\cal Z}^2} \\[1ex] \times
        \biggl( \frac{\sin(\p (\a - \ell) {\cal Z})}{\p} \biggr)^{2 n_h^\e}
        {\cal R}_{n_h^\e, n_p^\e}
	\bigl( \{h_j^\e\}, \{p_j^\e\} \big| \e [(\a - \ell) {\cal Z} + \ell] \bigr) \epc
\end{multline}
for $\e = \pm 1$, and where we have introduced the shorthand notations
\begin{equation}
     C[v] = \2 \int_{- Q}^Q \rd \la \int_{- Q}^Q \rd \m \:
               \biggl[ \frac{v' (\la) (v (\m) + 1)
	                   - v' (\m) (v (\la) + 1)}{2 \tgh (\la - \m)}
		     - \frac{v (\la) v (\m)}{\sh^2 (\la - \m + \h)} \biggr]
\end{equation}
and
\begin{multline} \label{rcomb}
     {\cal R}_{n_1, n_2} \bigl( \{h_j\} , \{p_j\} \big| v \bigr)
        = \frac{\prod_{1 \le j < k \le n_1}
	        (h_j - h_k)^2 \prod_{1 \le j < k \le n_2} (p_j - p_k)^2}
	       {\prod_{j=1}^{n_1} \prod_{k=1}^{n_2} (h_j + p_k - 1)^2} \\ \times
          \Biggl[ \prod_{j=1}^{n_1}
	       \frac{\G^2 ( h_j + v)} {\G^2 (h_j)} \Biggr]
          \Biggl[ \prod_{j=1}^{n_2}
	       \frac{\G^2 ( p_j - v)} {\G^2 (p_j)} \Biggr] \epp
\end{multline}
Thus, in the low-temperature limit we have have obtained an entirely
explicit description of the `universal contributions' $A_n^{(0)}$ to
the amplitudes in terms of the dressed charge and the dressed energy.
Let us recall that the above formulae (\ref{ampssmallt})-(\ref{rcomb})
are valid for a certain class of excitations characterized by (i)
$n_p = n_h$, and (ii) $x_{h_j}^\pm, y_{p_j}^\pm = \pm Q + \CO(T)$.

The Fredholm determinant part of the amplitudes is a fraction
with two Fredholm determinants in the numerator and two Fredholm
determinants in the denominator. In order to calculate the zero
temperature limit of the determinants in the denominator we recall
that the weight functions $\bigl(1 + \fa^{-1}_0 (\la)\bigr)^{-1}$ and
$\bigl(1 + \fa^{-1}_n (\la|\a)\bigr)^{-1}$ play the role of the Fermi functions
for our system. In the zero temperature limit they turn into the
characteristic functions of the `interval' $\i \g/2 + [- Q, Q]$
on ${\cal C}_0$. Thus,
\begin{equation} \label{limden}
     \lim_{T \rightarrow 0+} \:
        \det_{\rd m^\a_0, {\cal C}_n} \bigl\{ 1 - \widehat{\cal K} \bigr\} =
     \lim_{T \rightarrow 0+} \:
        \det_{\rd m, {\cal C}_n} \bigl\{ 1 - \widehat{\cal K} \bigr\} =
        \det_{\frac{\rd \la}{2\p \i}, [-Q, Q]} \bigl\{ 1 - \widehat{\cal K} \bigr\}
	\epp
\end{equation}
The expression on the right hand side depends only on the anisotropy parameter
and on the magnetic field. For the proof we note the estimate \cite{ReSi78_13}
\begin{equation}
     | \det (1 - A) - \det (1 - B) |
        \le || A - B ||_1 \exp \bigl( ||A||_1 + ||B||_1 + 1 \bigr) \epc
\end{equation}
where $||A||_1 = \tr |A|$ and which is valid for trace class operators $A, B$
\cite{ReSi80_6}.

The zero temperature limit of the determinants in the numerator is more
technical. In order to obtain it we need to assume certain general
properties of the low-temperature behaviour of zeroes of certain functions
(see appendix~\ref{app:detlim}). Setting
\begin{equation} \label{measureatzero}
     \rd \widehat{M}_\pm^\a (\la) =
        \frac{\rd \la}{2 \p \i}
	\frac{\re^{\pm \i \p \a'
	      \pm (\a' - \ell) \int_{-Q}^Q \rd \m \: \re(\m - \la) Z(\m) }}
	     {1 - \re^{\pm 2 \p \i (\a' - \ell) Z(\la)}}
\end{equation}
and denoting by $\G [- Q, Q]$ a contour which encircles the interval $[- Q, Q]$
counterclockwise we conjecture that
\begin{equation} \label{limnum}
     \lim_{T \rightarrow 0+} \:
        \det_{\rd m^\a_\pm, {\cal C}_n}
	      \bigl\{ 1 - \widehat{\cal K}_{\mp \a} \bigr\} =
        \det_{\rd \widehat{M}^\a_\pm, \G [- Q, Q]}
	      \bigl\{ 1 - \widehat{\cal K}_{\mp \a} \bigr\} \epp
\end{equation}
With (\ref{limden}), (\ref{limnum}) the zero temperature limit of the
determinant part of the amplitude is
\begin{equation} \label{ratdetzero}
     {\cal D} (\ell) = 
        \frac{\det_{\rd \widehat{M}^\a_+, \G [- Q, Q]}
	      \bigl\{ 1 - \widehat{\cal K}_{- \a} \bigr\}
	      \det_{\rd \widehat{M}^\a_-, \G [- Q, Q]}
	      \bigl\{ 1 - \widehat{\cal K}_{\a} \bigr\}}
	     {\det^2_{\frac{\rd \la}{2\p \i}, [-Q, Q]}
	      \bigl\{ 1 - \widehat{\cal K} \bigr\}} \epp
\end{equation}
The important point about this expression is that it depends only on
$\ell$ but not on the `quantum numbers' $p_j$, $h_k$ of the
particle hole excitations.

In appendix~\ref{app:detlim} we justify the above conjecture. We also
provide alternative expressions for the ratios of Fredholm determinants
that allow us to compare our formulae with those obtained for
zero temperature in \cite{KKMST11a} and to extract the $\alpha$
dependence from the determinants. Moreover, we provide expressions
in which ratios of determinants are replaced by determinants of a
different type of operators.
\subsection{Summation of the leading terms}
As we already mentioned above it is possible to sum up the form factors
of particle-hole type considered in the previous subsections. Summation is
possible for each individual value of $\ell$ by means of the same
summation formula as employed in \cite{KKMST11b} in the context of the
so-called critical form factors pertaining to the eigenstates of the
ordinary transfer matrix. Adapted to our notation it reads
\begin{multline} \label{critsum}
     \sum_{\substack{n_p, n_h \ge 0\\ n_p - n_h = \ell}}
     \sum_{\substack{p_1 < \dots < p_{n_p}\\ p_j \in {\mathbb N}}}
     \sum_{\substack{h_1 < \dots < h_{n_h}\\ h_j \in {\mathbb N}}} \mspace{-6.mu}
     \re^{- \frac{2 \p m T}{v_0}
          \bigl[ \sum_{j=1}^{n_p} (p_j - 1) + \sum_{j=1}^{n_h} h_j \bigr]}
	  \biggl( \frac{\sin(\p v)}{\p} \biggr)^{2 n_h} \mspace{-12.mu}
	  R_{n_h, n_p} \bigl( \{h_j\}, \{p_j\} \big| v \bigr) \\[-2.ex]
        = \frac{G^2 (1 + \ell - v)}{G^2 (1 - v)} \,
          \frac{\re^{- \frac{\p m T \ell (\ell -1)}{v_0}}}
	       {\Bigl( 1 - \re^{- \frac{2 \p m T}{v_0}} \Bigr)^{(\ell - v)^2}} \epp
\end{multline}

Summing the contributions to the form factors series of the form
considered in the previous subsections amounts to summing over all sets
of mutually distinct quantum numbers $\{h_j^+\}$, $\{p_k^+\}$ and
$\{h_j^-\}$, $\{p_k^- \}$ under the constraint $n_p^+ - n_h^+ = n_h^-
- n_p^- = \ell$ and then summing over the allowed values of
$\ell \in {\mathbb Z}$. Using (\ref{critsum}) we end up with
\begin{equation} \label{leadoscsum}
     \bigl\< \re^{2 \p \i \a S(m)} \bigr\> \sim
          \sum_\ell \re^{\i \p \a m + 2\i k_F (\ell - \a) m}
	  {\cal D} (\ell) {\cal A} (\ell - \a)
	  \biggl( \frac{\p T/v_0}{\sh( m \p T/v_0)} \biggr)^{2 (\ell - \a)^2 {\cal Z}^2}
	  \epc
\end{equation}
where
\begin{equation} \label{defa}
     {\cal A} (x) = \frac{\re^{2 x^2 C [Z/{\cal Z}] {\cal Z}^2} \;
        G^2 \bigr(1 + x {\cal Z} \bigl)
        G^2 \bigr(1 - x {\cal Z} \bigl)}
           {\bigl(2 \p \r (Q) \sh(2Q) \bigr)^{2 x^2 {\cal Z}^2}} \epp
\end{equation}
%Here we have also assumed that $p$ is independent of $\ell$.

For every finite temperature (\ref{leadoscsum}) is a sum over terms
which decay exponentially with the distance $m$.  For $T \rightarrow 0+$
the nature of the asymptotics changes from exponential to algebraic
decay, $(\p T/v_0)/\sh(m \p T/v_0) \rightarrow 1/m$, as is expected at
a critical point. Up to an adjustment of conventions and notation
(\ref{leadoscsum}) turns into the sum obtained in \cite{KKMST11b}.
In particular, the amplitude ${\cal D} (\ell) {\cal A} (\ell - \a)$
is the same as in the zero temperature case.

Looking at it the other way round, the lowest order effect of
switching on the temperature is a deformation, $1/m
\rightarrow (\p T/v_0)/\sh(m \p T/v_0)$, exactly as expected from
conformal field theory by mapping the complex plane to a cylinder of
finite circumference \cite{Cardy84}.  Here we have obtained this
result directly from an expansion of the generating function in
terms of form factors of the quantum transfer matrix.

The series (\ref{leadoscsum}) is neither an asymptotic series for small
$T$ nor for large $m$. In \cite{KMS11b} it was characterized as a sum
gathering the `leading orders for each oscillatory term.' This means
that for each order of exponential decay we have neglected algebraic
corrections in $T$ which would contribute lower order terms than the
next-order exponentials. In addition, we have neglected higher order
correction to the inverse correlation lengths that would amount to
contributions of the form $\exp \bigl( {\cal O} (m T^2) \bigr)$.
Thus, (\ref{leadoscsum}) provides only the leading order asymptotics. On
the other hand, each oscillating term appears with its coefficient of
leading order in $T$.

Recall that we have assumed in our above analysis that $p = 0$. This
assumption restricts the allowed range of summation over $\ell$ for
fixed $\a$ to the values determined by (\ref{pnulluneq}). For
any $\D \in (0,1)$ and any value of the magnetic field this range
includes $\ell = - 1, 0, 1$ if $\a$ is small enough. It becomes large
for magnetic fields close to saturation and for anisotropies close
to zero. We believe that an extension of our analysis to $p \ne 0$
will show that the sum may be extended to all $\ell \in {\mathbb Z}$.
This would also restore the periodicity in $\a$ of the generating
function. We shall come back to this in future work.

\subsection{Large-distance asymptotics of the longitudinal correlation
functions}
The large-distance asymptotics of the longitudinal correlation functions
can be obtained from (\ref{leadoscsum}) by means of (\ref{genzz}),
(\ref{genonep}). From (\ref{defa}) and from appendix~\ref{app:detlim}
we know that
\begin{subequations}
\begin{align}
     & {\cal D} (0) = 1 + \CO (\a^2) \epc
     &  {\cal D} (\ell) = \CO (\a^2)\ \text{for $\ell \ne 0$} \epc \\ 
     & {\cal A} (0) = 1 \epc
     &  {\cal A} (\ell - \a) = {\cal A} (\ell) + \CO (\a)\ \text{for $\ell \ne 0$} \epp
\end{align}
\end{subequations}
Then (\ref{genonep}) implies
\begin{equation}
     \<\s_1^z\> = 1 - \frac{2 k_F}{\p}
\end{equation}
which is the known relation between magnetization and Fermi momentum. Using
this formula and (\ref{genzz}) we obtain to leading order (i.e.\ by taking into
account only $\ell = -1, 0 , 1$ in (\ref{leadoscsum}))
\begin{multline} \label{longassy}
     \<\s_1^z \s_{m+1}^z\> - \<\s_1^z\> \<\s_{m+1}^z\> =
        - \frac{2 {\cal Z}^2}{\p^2} \biggl( \frac{\p T/v_0}{\sh( m \p T/v_0)} \biggr)^2 \\
        + \frac{4 \sin^2 (k_F)}{\p^2}
	  {\cal A} (1) {\cal D}'' (1) \cos(2m k_F)
	  \biggl( \frac{\p T/v_0}{\sh( m \p T/v_0)} \biggr)^{2 {\cal Z}^2} \epp
\end{multline}
Here we have introduced the shorthand notation ${\cal D}'' (1) = \6_\a^2
{\cal D} (1)|_{\a = 0}$ and used that ${\cal D}'' (-1) = {\cal D}'' (1)$
and ${\cal A} (-1) = {\cal A} (1)$.

Depending on whether ${\cal Z} > 1$ or ${\cal Z} < 1$ either the first or
the second term determines the large-distance asymptotics of the correlation
function. It is easy to see from the integral equation (\ref{liz}) for
the dressed charge function that ${\cal Z} > 1$ for $- 1 < \Delta < 0$
and ${\cal Z} < 1$ for $0 < \Delta < 1$ provided that $Q > 0$. Thus, as
is well known \cite{LuPe75,BIK86}, the low-temperature correlation
functions are asymptotically always negative if $- 1 < \Delta < 0$,
while they exhibit $2 k_F$ oscillations if $0 < \Delta < 1$. At the free
Fermion point $\Delta = 0$ both types of behaviour coexist. The
amplitude in the first term is proportional to the square of the
dressed charge and is numerically easily accessible. It was obtained
using a perturbative scheme around $\D = 0$ in \cite{BIK86}. The
amplitude in the second term was apparently not known until recently
\cite{KKMST09a}. It is not yet clear how well the Fredholm determinant
part can be computed numerically. This question deserves further attention.

Note that we have derived (\ref{longassy}) assuming that $0 < \D < 1$.
It remains valid for $- 1 < \D < 0$, since the correlation functions
in this regime of the anisotropy can be obtained from those for
$0 < \D < 1$ by a simple transformation \cite{YaYa66a} consisting of
a sign reversal of the coupling $J$ and a conjugation of all local
operators on even sites by $\s^z$.

\section{Ground state correlation functions of the finite chain}
The description of the spectral properties of integrable finite-length
systems can be based on similar auxiliary functions as introduced above
in order to deal with the distributions of Bethe roots of the
quantum transfer matrix. In fact it is exactly this context in which 
auxiliary functions and non-linear integral equations first became relevant
for integrable quantum systems \cite{KluBatPea91}. As we have pointed
out elsewhere \cite{DGHK07,BoGo09} the form of many formulae for
correlation functions is the same for the infinite chain at finite
temperature and for the finite chain in the ground state with no
magnetic field applied. Here we claim that this also holds for the
formulae for the form factors and amplitudes obtained above.
Moreover, if one is willing to accept slightly more complicated
contours, then the finite-length case with non-zero magnetic field
can be treated in a similar manner. The resulting formulae may be
problematic for a numerical treatment, but we expect that the large-$L$
asymptotics may be analyzed in analogy with the low-temperature case in the
previous section. As compared to the approach pursued in \cite{KKMST09b,
KKMST11a} we would keep control of the finite-size dependence at
every stage of the calculation which might allow us, in principle,
to obtain higher order finite-size corrections. Yet, since we have not
carried out the calculation, we shall leave the realm of speculation at
this point and simply describe how equations (\ref{an}) and (\ref{amp})
for the amplitudes have to be modified.

In the finite-length case we work with the monodromy matrix
\begin{equation} \label{monol}
     T_a (\la) = q^{\k \s_a^z} R_{a, L} (\la, \h/2) \dots
                 R_{a, - L + 1} (\la, \h/2)
\end{equation}
acting on the tensor product of an auxiliary space with index $a$ and $2L$
quantum spaces. Here
\begin{equation}
     \k = \i \Ph/\h
\end{equation}
and $\Ph$ is a real flux parameter.

The corresponding transfer matrix $t(\la) = \Tr_a \{T_a (\la)\}$ generates the
Hamiltonian of the XXZ chain,
\begin{equation}
     H = 2 J \sh(\h) \, t^{-1} (\h/2) t' (\h/2)
       = J \sum_{j = - L + 1}^L \bigl( \s_{j-1}^x \s_j^x + \s_{j-1}^y \s_j^y
                               + \D (\s_{j-1}^z \s_j^z - 1) \bigr) \epc
\end{equation}
with twisted periodic boundary conditions,
\begin{equation}
     \s_{-L}^b = q^{- \k \s_L^z} \s_L^b q^{\k \s_L^z} \epc \qd b = x, y, z \epp
\end{equation}

Since $T_a (\la)$ satisfies the Yang-Baxter algebra relations (\ref{yba}),
the eigenstates and eigenvalues of $t(\la)$ can again be constructed by
means of the algebraic Bethe ansatz. The only essential difference is in 
the pseudo vacuum, which in this case is $|0\> = \binom{1}{0}^{\otimes 2L}$,
and, consequently, in the vacuum expectation values $a(\la)$, $d(\la)$ of the
diagonal elements of the monodromy matrix which are now given by
\begin{equation}
     a(\la) = q^\k \epc \qd d(\la) = q^{- \k} b^{2L} (\la, \h/2) \epp
\end{equation}

Inserting these vacuum expectation values into the Bethe equations (\ref{bae})
we obtain rather different patterns of roots in the complex plane. Still,
the definitions (\ref{auxa}) and (\ref{rho}) remain meaningful, and a non-linear
integral equation similar to (\ref{nlie}) can be derived for the auxiliary
functions,
\begin{equation} \label{nliefinisize}
     \ln (\fa_n (\la)) =  - 2 (\k - M) \h + 2 \i L \rp (\la)
        - \int_{{\cal C}_n} \frac{\rd \m}{2\p \i} \:
	  K(\la - \m) \ln \bigl(1 + \fa_n (\m)\bigr) \epp
\end{equation}
Here
\begin{equation}
     \rp (\la) = \i \ln \biggl( \frac{\sh(\la + \h/2)}{\sh(\la - \h/2)} \biggr)
\end{equation}
is the bare momentum, whose appearance seems rather naturally in view of the
fact that we have interchanged `space and time direction' in the underlying
six-vertex model. With the auxiliary functions (\ref{nliefinisize}) the formula
(\ref{rhoint}) for the ratio of two transfer matrix eigenvalues remains valid in
the finite-length case if we replace $N$ by $2L$.

The general formula (\ref{corn}) for the correlation functions remains valid
as well. Due our conventions for the monodromy matrix (\ref{monol}) the
operators and eigenvalues on the right hand side have to be evaluated at $\h/2$
instead of $0$,
\begin{equation} \label{corl}
     \bigl\< \CO_1^{(1)} \dots \CO_m^{(m)} \bigr\>
        = \frac{\<\Ps_0| \Tr \{\CO^{(1)} T(\h/2)\} \dots \Tr \{\CO^{(m)} T(\h/2)\}|\Ps_0\>}
               {\<\Ps_0|\Ps_0\> \La_0^m (\h/2)} \epp
\end{equation}
The corresponding form factor expansions of the generating function of the
longitudinal correlation functions and of the transversal two-point function
follow as
\begin{equation} \label{genfunexpl}
     \bigl\<q^{2 \a S(m)}\bigr\>_N
        = \sum_{n=0}^{N_M - 1} A_n (\a) \r_n^m \bigl(\tst{\frac{\h}{2}}|\a\bigr) \epc \qd
     \<\s_1^- \s_{m+1}^+\>_N = \sum_{n=1}^{N_M} A_n^{-+} \bigl(\tst{\frac{\h}{2}} \bigr)
                               \r_n^m \bigl(\tst{\frac{\h}{2}|\a}\bigr) \epp
\end{equation}
Here the number of overturned spins $M$ must be $L$ for the generating function
and $L - 1$ for the transversal correlation function. Not only the structure of
these asymptotic expansions is the same as in the finite-temperature case, 
but equations (\ref{an}) and (\ref{amp}) for the amplitudes remain valid as well.
Note, however, that in the derivation of the integral formula (\ref{rho}) for
the ratio of the eigenvalues as well as in the derivation of the formula for the
amplitudes in the asymptotic expansion of the transversal correlation functions
we have assumed that the spectral parameter is located inside the integration
contour. This must be kept in mind when sending it to $\h/2$. In order to deal
with a finite magnetic field one has to switch to a sector of non-zero eigenvalue
of the $z$-component of the total spin. $\La_0$ must be replaced by the
largest transfer matrix eigenvalue in this sector, and the contours have to
be changed accordingly.

\section{Conclusions}
In this work we have initiated a study of the finite-temperature correlation
functions of the XXZ chain by means of the form factors of the quantum
transfer matrix. Products of two form factors combine into the amplitudes
in the long-distance asymptotics expansions of the two-point correlation
functions. These amplitudes seem to be more natural than the form factors
themselves. There is no issue about their proper normalization, whereas
the normalization we chose e.g.\ for $F_- (\x)$ and $F_+ (\x)$ in
(\ref{deffpm}) seems to be somewhat arbitrary.
\enlargethispage{3ex}

For the amplitudes we were able to perform the Trotter limit analytically.
The resulting formulae (\ref{an}) and (\ref{amp}) for the longitudinal and
transversal cases have a remarkably similar structure. They consist of a
`Fredholm determinant part', a `universal amplitude part', which seems to carry
the critical behaviour in the zero temperature limit, and a `factorizing part'
by which we mean the product of $\s$-functions in (\ref{an}) or the product
of $G$-functions in (\ref{amp}), respectively. We believe that such type
of structure persists for more general form factors such as
$\<\Ps_n^\a|C(\x_1) C(\x_2)|\Ps_0\>$. This conjecture is supported by
\cite{JMS08,BJMS10,Boos11}, where factorization of correlation functions was
proved in a rather general context.

We have demonstrated that our formulae can be efficiently analyzed at least
for low and high temperatures. This analysis will be continued in future
work. We hope that the formulae for the amplitudes can also be evaluated
numerically. At fixed finite temperature a few (typically one or two)
amplitudes and the corresponding correlation lengths fix the leading
asymptotics. Due to the efficiency of asymptotic expansions a combination
with the exact short-distance results of \cite{BDGKSW08,TGK10a,SABGKTT11}
would give us numerical access to the full correlation functions and
would allow us to calculate their Fourier transforms which appear in
many applications in condensed matter physics.

An interesting feature of our Fredholm determinants
%that we would like to emphasize
is that they truncate exactly for every finite
Trotter number (every finite length in the finite-length case).
For finite Trotter number they are represented by finitely many
integrals (see (\ref{detalphan})). The Fredholm determinants as
well as the other factors in the expressions for the amplitudes
are parameterized by the auxiliary functions $\fa_n$ whose
low-temperature behaviour is rather well known. In particular
it can be also analyzed for vanishing magnetic field \cite{Kluemper98}.
We hope that this fact will help us to analyze the low-temperature
asymptotics of the full thermal correlation functions for zero
magnetic field which is one of the many interesting open problems
in our approach.

A further interesting issue to be studied in future work are finite-%
temperature singularities of the amplitudes which may occur
e.g.\ at the cross-over temperatures \cite{FKM99} where the
asymptotic behaviour of the correlation functions changes from
decaying and oscillating to monotonically decaying. Finally, we
would like to mention the possibility to study dynamical correlation
functions at finite temperature by means of the form factors of
the quantum transfer matrix. Results for zero temperature, based
on the form factor approach, were recently obtained in \cite{KKMST12}.
\\[.5ex]
{\bf Acknowledgment.}
The authors are grateful to H. Boos, N. Kitanine, A. Kl\"umper and
A. Wei{\ss}e for helpful and encouraging discussions. MD and FG
acknowledge financial support by the Volkswagen Foundation. KKK is
supported by the CNRS. His work has been partly financed by the grant
PEPS-PTI `Asymptotique d'int\'egrales multiples' and by a Burgundy
region PARI 2013 FABER grant `Structures et asymptotiques d'int\'egrales
multiples'.

\clearpage

{\appendix
\Appendix{Amplitudes of the longitudinal correlation functions}
\label{app:zz}
\noindent
In this appendix we sketch the derivation of the expression (\ref{aalphan})
for the amplitudes $A_n (\a)$ in the form factor expansion of the generating
function of the longitudinal two-point functions defined in (\ref{genfunexp}).

Recall that for the longitudinal correlation functions only excited states
of the twisted transfer matrix with $M = N/2$ Bethe roots contribute to the
form factor expansion. We denote these Bethe roots by $\m_j$, $j = 1, \dots, M$,
while the Bethe roots of the dominant state will be denoted $\la_j$, $j = 1,
\dots, M$. Using the scalar product formula (\ref{slavnov}) we obtain
\begin{multline} \label{twoslavnovz}
     \biggl[ \prod_{j=1}^M \frac{d(\m_j)}{d(\la_j)} \biggr]
        \frac{\<\Ps_n^\a|\Ps_0\>}{\<\Ps_n^\a|\Ps_n^\a\>} = 
        \biggl[ \prod_{j=1}^M - \Ph^{-1} (\m_k - \h) \biggr] \\ \times
	\biggl[ \prod_{1 \le j < k \le M}
	\frac{\sh(\m_j - \m_k)}{\sh(\la_j - \la_k)} \biggr]
     \frac{\det_M \bigl\{ \re(\la_j - \m_k) - \re(\m_k - \la_j) \fa_0 (\m_k) \bigr\}}
          {\det_M \bigl\{\de^j_k \fa_n' (\m_j|\a) - K(\m_j - \m_k)\bigr\}} \epc
\end{multline}
where we have introduced the function
\begin{equation} \label{defphi}
     \Ph (\la) = \prod_{j=1}^M \frac{\sh(\la - \m_j)}{\sh(\la - \la_j)} \epp
\end{equation}
Employing now the Cauchy determinant formula written in the form
\begin{equation}
	\prod_{1 \le j < k \le M} \frac{\sh(\m_j - \m_k)}{\sh(\la_j - \la_k)} =
        \det_M \biggl\{ \frac{\res \Ph (\la_k)}{\sh(\la_k - \m_j)} \biggr\}
\end{equation}
and the identity
\begin{equation}
     \Ph (\m_k - \h) \Ph^{-1} (\m_k + \h) = - q^{- 2\a} \fa_0 (\m_k) \epc
\end{equation}
which follows from the Bethe ansatz equations for the $\m_k$, we can simplify 
the right hand side of (\ref{twoslavnovz}). We find that
\begin{align} \label{calcf1z}
     & \biggl[ \prod_{1 \le j < k \le M}
             \frac{\sh(\m_j - \m_k)}{\sh(\la_j - \la_k)} \biggr]
	     \notag \\ & \mspace{54.mu} \times
        \biggl[ \prod_{j=1}^M - \re^{\la_j - \m_j} q^{- \a} \Ph^{-1} (\m_k - \h) \biggr]
           \det_M \bigl\{ \re(\la_j - \m_k) - \re(\m_k - \la_j) \fa_0 (\m_k) \bigr\}
	   \notag \\[1ex]
	& = \det_M \biggl\{ \frac{ \re^{\la_\ell - \m_j}
	                  \res \Ph (\la_\ell)}{\sh(\m_j - \la_\ell)} \biggr\}
           \det_M \bigl\{ q^{- \a} \Ph^{-1} (\m_k - \h) \re(\la_\ell - \m_k)
	                  + q^\a \Ph^{-1} (\m_k + \h) \re(\m_k - \la_\ell) \bigr\}
			  \notag \\[1ex]
	& = \det_M \biggl\{ \int_{\cal C} \frac{\rd \la}{2 \p \i} \,
	                  \frac{\re^{\la - \m_j} \Ph (\la)}{\sh(\m_j - \la)}
                 \bigl( q^{- \a} \Ph^{-1} (\m_k - \h) \re(\la - \m_k)
	                + q^\a \Ph^{-1} (\m_k + \h) \re(\m_k - \la) \bigr) \biggr\}
			\notag \\[1ex]
        & = \det_M \biggl\{ \de^j_k \r_n^{-1} (\m_j|\a) \fa_n' (\m_j|\a)
	  - \frac{q^{1 + \a}}{\sh(\m_j - \m_k - \h)}
	  + \frac{q^{-1 - \a}}{\sh(\m_j - \m_k + \h)} \biggr\} \epp
\end{align}
The contour $\cal C$ in the second equation is chosen in such a way that all
Bethe roots $\la_j$, $j = 1, \dots, M$, are included, but $\m_j, \m_k$ and $\m_k \pm \h$
are excluded. The integral can be calculated, since the integrand is an $\i \p$-periodic
function of $\la$ which decreases to zero as $\Re \la \rightarrow \pm \infty$.%
\footnote{A very similar calculation, inspired by \cite{IKMT99},
was performed in the appendix of \cite{BoGo09}.} In the last equation we have
inserted the identity
\begin{equation}
     \Ph' (\m_j) \bigl( q^{- \a} \Ph^{-1} (\m_j - \h) - q^\a \Ph^{-1} (\m_j + \h) \bigr)
        = \r_n^{-1} (\m_j|\a) \fa_n' (\m_j|\a)
\end{equation}
which follows from Baxter's $TQ$-equation. Comparing (\ref{twoslavnovz}) and
(\ref{calcf1z}) and taking into account the definition (\ref{kernela}) of the kernel
${\cal K}_\a$ we obtain
\begin{equation} \label{f1z}
     \biggl[ \prod_{j=1}^M \frac{q^{-\a} d(\m_j) \re^{\la_j}}{d(\la_j) \re^{\m_j}} \biggr]
        \frac{\<\Ps_n^\a|\Ps_0\>}{\<\Ps_n^\a|\Ps_n^\a\>} = 
	   \biggl[ \prod_{j=1}^M \r_n^{-1} (\m_j|\a) \biggr]
                   \frac{\det_M \Bigl\{ \de^j_k - \frac{\r_n (\m_j|\a)}{\fa_n' (\m_j|\a)}
		         {\cal K}_{- \a} (\m_j - \m_k) \Bigr\}}
		        {\det_M \Bigl\{\de^j_k - \frac{1}{\fa_n' (\m_j|\a)}
			 {\cal K} (\m_j - \m_k)\Bigr\}} \epp
\end{equation}
A very similar calculation with the roles of $\la_j$ and $\m_k$ interchanged yields
\begin{equation} \label{f2z}
     \biggl[ \prod_{j=1}^M \frac{q^\a d(\la_j) \re^{\m_j}}{d(\m_j) \re^{\la_j}} \biggr]
        \frac{\<\Ps_0|\Ps_n^\a \>}{\<\Ps_0|\Ps_0\>} = 
	   \biggl[ \prod_{j=1}^M \r_n (\la_j|\a) \biggr]
	           \frac{\det_M \Bigl\{ \de^j_k - \frac{\r_n^{-1} (\la_j|\a)}{\fa_0' (\la_j)}
		         {\cal K}_\a (\la_j - \la_k) \Bigr\}}
		        {\det_M \Bigl\{\de^j_k - \frac{1}{\fa_0' (\la_j)}
			 {\cal K} (\la_j - \la_k)\Bigr\}} \epp
\end{equation}
Then the equation (\ref{aalphan}) for the amplitudes follows by multiplying
(\ref{f1z}) and (\ref{f2z}).

\Appendix{Amplitudes of the transversal correlation functions}
\label{app:pm}
\noindent
In this appendix we derive the expressions (\ref{fplus}) and (\ref{fminus})
for the form factors $F_+ (\x)$ and $F_- (\x)$ defined in (\ref{fpm}).
\subsection{The form factor $F_+ (\x)$}
Now the excited state $|\Ps_n^\a\>$ is parameterized by $M - 1$ Bethe roots
still denoted $\m_j$. Again a function $\Ph$, now defined by
\begin{equation} \label{defphipm}
     \Ph (\la) = \frac{\prod_{j=1}^{M - 1} \sh(\la - \m_j)}
                      {\prod_{j=1}^M \sh(\la - \la_j)} \epc
\end{equation}
will be a useful tool in our calculation. We further use a degenerate
variant of the Cauchy determinant formula obtained by sending the real
part of one of the variables ($\m_M$ in our case) to $- \infty$ and
picking up the asymptotically leading term,
\begin{equation}
     \frac{\dst{\prod_{1 \le j < k \le M - 1}} \mspace{-14.mu} \sh(\m_j - \m_k)}
          {\dst{\prod_{1 \le j < k \le M}} \sh(\la_j - \la_k)}
        = \biggl[ \prod_{j=1}^M \res \Ph (\la_j) \biggr]
	  \det_M %\left|
	     %\begin{array}{@{}ccc@{}} 
	     \begin{pmatrix}
	     \frac{\re^{\la_1 - \m_1}}{\sh(\la_1 - \m_1)} & \dots &
	     \frac{\re^{\la_M - \m_1}}{\sh(\la_M - \m_1)} \\
	     \vdots & & \vdots \\
	     \frac{\re^{\la_1 - \m_{M-1}}}{\sh(\la_1 - \m_{M-1})} & \dots &
	     \frac{\re^{\la_M - \m_{M-1}}}{\sh(\la_M - \m_{M-1})} \\
	     1 & \dots & 1
	     \end{pmatrix}
	     %\end{array}
	  %\right|
	  \epp
\end{equation}

Inserting the determinant formula (\ref{slavnov}) and the expression
(\ref{abaeval}) for the eigenvalue into the definition (\ref{fpm})
of $F_- (\x)$ and using the above formulae we obtain
\begin{equation}  \label{fpnotsimple}
     F_+ (\x) = A \det_M \left( \binom{C}{E} \uv_1, \dots, \binom{C}{E} \uv_M \right) \epc
\end{equation}
where
\begin{equation} \label{defap}
     A = \frac{\prod_{j=1}^M d(\la_j) \re^{\la_j}}
              {\prod_{j=1}^{M-1} q^{- \a} d(\m_j) \re^{\m_j}} \,
         \frac{1}{\det_{M-1} \bigl\{ \de^j_k \fa_n' (\m_j|\a) - K(\m_j - \m_k)\bigr\}}
	 \epc
\end{equation}
where
\begin{subequations}
\begin{align}
     C^j_\ell & = \frac{1}{\sh(\la_\ell - \m_j)} \epc \qd
                  j = 1, \dots, M - 1 \, ;\ \ell = 1, \dots, M \epc \\
     E^1_\ell & = \re^{- \la_\ell} \epc \qd \ell = 1, \dots, M \epc
\end{align}
\end{subequations}
and where
\begin{subequations}
\begin{align}
     (\uv_k)^\ell & =
        - \res \Ph (\la_\ell) \bigl\{q^{-\a} \Ph^{-1} (\m_k - \h) \re(\la_\ell - \m_k)
            + q^\a \Ph^{-1} (\m_k + \h) \re(\m_k - \la_\ell) \bigr\},  \\
            & \mspace{284.mu} k = 1, \dots, M - 1 \, ;\ \ell = 1, \dots, M \epc \notag \\
     (\uv_M)^\ell & = - \frac{\Ph^{-1} (\x)}{1 + \fa_0 (\x)} \res \Ph (\la_\ell)
        \bigl\{\re(\la_\ell - \x) - \re(\x - \la_\ell) \fa_0 (\x) \bigr\} \epc \qd
	\ell = 1, \dots, M \epp
\end{align}
\end{subequations}

The determinant on the right hand side of (\ref{fpnotsimple}) can be simplified
in a similar way as shown in (\ref{calcf1z}) for one of the form factors
of the generating function. The four cases $C \uv_k$, $k = 1, \dots, M - 1$,
$C \uv_M$, $E \uv_k$, $k = 1, \dots, M - 1$, and $E \uv_M$ have to be treated
separately. We obtain
\begin{subequations}
\begin{align} \label{cfplus}
     (C \uv_k)^j & = \de^j_k \fa_n' (\m_j|\a) \r_n^{-1} (\m_j|\a) 
                     - \frac{q^\a}{\sh(\m_j - \m_k - \h)}
                     + \frac{q^{- \a}}{\sh(\m_j - \m_k + \h)} \epc \\ \label{efplus}
     E \uv_k & = (q^{\a - 1} - q^{1 - \a}) \re^{- \m_k} \epc \\ \label{cmfplus}
     (C \uv_M)^j & = \frac{1}{\sh(\x - \m_j)}
                     - \frac{q^{-\a} \r_n (\x|\a)}{1 + \fa_n (\x|\a)}
                       \frac{1}{\sh(\x - \m_j - \h)}
                     - \frac{q^{\a} \r_n (\x|\a)}{1 + \faq_n (\x|\a)}
                       \frac{1}{\sh(\x - \m_j + \h)} \epc \\ \label{emfplus}
     E \uv_M & = \re^{- \x} \biggl( 1 - \frac{q^{1 - \a} \r_n (\x|\a)}{1 + \fa_n (\x|\a)} 
                      - \frac{q^{\a - 1} \r_n (\x|\a)}{1 + \faq_n (\x|\a)} \biggr) \epc
\end{align}
\end{subequations}
where $\faq_n = 1/\fa_n$ by definition.

If we directly continue with these matrices we end up with Fredholm
determinants with kernel ${\cal K}$. Having in mind the more
involved case of the form factor $F_- (\x)$ we prefer to switch at
this point to the kernel $K$. For this purpose we subtract $\re^{\m_j}$
times (\ref{efplus}) from (\ref{cfplus}) and $\re^{\m_j}$ times
(\ref{emfplus}) from (\ref{cmfplus}) which does not change the
determinant in (\ref{fpnotsimple}). Then
\begin{subequations}
\label{fpdetfull}
\begin{align}
     (C \uv_k)^j - \re^{\m_j} E \uv_k & = \de^j_k \fa_n' (\m_j|\a) \r_n^{-1} (\m_j|\a)
          - \re^{\m_j - \m_k} K_{1 - \a} (\m_j - \m_k) \notag \\
	& = \re^{\m_j - \m_k} (\vv_k)^j \epc \\
     (C \uv_M)^j - \re^{\m_j} E \uv_M
        & = \re^{\m_j - \x} \biggl[ \cth(\x - \m_j)
	  - \frac{q^{1-\a} \r_n (\x|\a)}{1 + \fa_n (\x|\a)} \cth(\x - \m_j - \h) \notag \\
	& \mspace{72.mu} - \frac{q^{\a-1} \r_n (\x|\a)}{1 + \faq_n (\x|\a)}
	                   \cth(\x - \m_j + \h) \biggr] \notag \\
	& = \re^{\m_j - \x} (\vv_M)^j \epc
\end{align}
\end{subequations}
where $j = 1, \dots, M - 1$. These equations define an $(M - 1) \times (M - 1)$
matrix $V$ with matrix elements $V^j_k = (\vv_k)^j$ and an $(M - 1)$-component
column vector $\vv_M$.

Using (\ref{fpdetfull}) in (\ref{fpnotsimple}) and expanding the determinant
with respect to its last row we see that
\begin{equation}  \label{fpsimpler}
     F_+ (\x) = A \re^{- \x} \det_{M - 1} (V)
        \bigl\{ E \uv_M \re^{\x}
	        - (E \uv_1 \re^{\m_1} , \dots, E \uv_{M-1} \re^{\m_{M-1}} )
		   V^{-1} \vv_M \bigr\} \epp
\end{equation}
Here we have assumed that $V$ is invertible, i.e.\ that
\begin{equation} \label{fpv}
     \det_{M - 1} (V)
        = \biggl[ \prod_{j=1}^{M-1} \frac{\fa_n' (\m_j|\a)}{\r_n (\m_j|\a)} \biggr]
          \det_{M - 1} \biggl\{ \de^j_k - \frac{\r_n (\m_j|\a)}{\fa_n' (\m_j|\a)}
	                        K_{1-\a} (\m_j - \m_k) \biggr\}
\end{equation}
is non-zero. We do not have a general proof of this assumption, but we checked
it with several examples for small $M$.

It remains to calculate the terms in curly brackets on the right hand
side of (\ref{fpsimpler}). Setting $\xv = V^{-1} \vv_M$ we find that the
coordinates $x^j$ of this vector satisfy the set of linear equations
\begin{multline}
     \fa_n' (\m_j|\a) \r_n^{-1} (\m_j|\a) x^j = - \cth(\m_j - \x)
                     + \frac{q^{1 - \a} \r_n (\x|\a)}{1 + \fa_n (\x|\a)}
                       \cth(\m_j - \x + \h) \\
                     + \frac{q^{\a - 1} \r_n (\x|\a)}{1 + \faq_n (\x|\a)}
                       \cth(\m_j - \x - \h)
                     + \sum_{k=1}^{M-1} K_{1-\a} (\m_j - \m_k) x^k \epp
\end{multline}
This suggests to define a function
\begin{multline}
     G_+ (\la, \x) = - \cth(\la - \x)
                     + \frac{q^{1 - \a} \r_n (\x|\a)}{1 + \fa_n (\x|\a)}
                       \cth(\la - \x + \h) \\
                     + \frac{q^{\a - 1} \r_n (\x|\a)}{1 + \faq_n (\x|\a)}
                       \cth(\la - \x - \h)
                     + \sum_{k=1}^{M-1} K_{1-\a} (\la - \m_k) x^k
\end{multline}
which has the following properties:
\begin{equation} \label{xg}
     G_+ (\m_k, \x) = \fa_n' (\m_k|\a) \r_n^{-1} (\m_k|\a) x^k \epc
\end{equation}
and $G_+ (\la, \x)$ as a function of $\la$ is meromorphic inside the
contour ${\cal C}_n$ with a single simple pole at $\la = \x$ with residue $-1$.
It follows that $G_+ $ satisfies the linear integral equation
\begin{multline}
     G_+ (\la, \x) = 
        - \cth(\la - \x) \\ + q^{\a - 1} \r_n (\x|\a) \cth(\la - \x - \h)
	+ \int_{{\cal C}_n} \rd m_+^\a (\m) K_{1-\a} (\la - \m) G_+(\m, \x) \epc
\end{multline}
where
\begin{equation}
     \rd m_+^\a (\la)
        = \frac{\rd \la \: \r_n (\la|\a)}{2 \p \i (1 + \fa_n (\la|\a))} \epp
\end{equation}
Finally
\begin{align} \label{fpintpart}
     & E \uv_M \re^\x
          - (E \uv_1 \re^{\m_1}, \dots, E \uv_{M-1} \re^{\m_{M-1}}) V^{-1} \vv_M
        = E \uv_M \re^\x - \sum_{j=1}^{M-1} E \uv_j \re^{\m_j} x^j \notag \\
     & \mspace{54.mu}
	= 1 - q^{\a-1} \r_n (\x|\a) - (q^{\a-1} -q^{1-\a})
	      \int_{{\cal C}_n} \rd m_+^\a (\la) G_+ (\la, \x) \notag \\
     & \mspace{54.mu}
        = \lim_{\Re \la \rightarrow - \infty} G_+ (\la, \x) = G_+^- (\x) \epp
\end{align}
Here we have used (\ref{xg}) in the second equation.  Inserting (\ref{defap}),
(\ref{fpv}) and (\ref{fpintpart}) into (\ref{fpsimpler}) we arrive at equation
(\ref{fplus}) of the main text.
\subsection{The form factor $F_- (\xi)$}
Here we can proceed as in the previous section by using in (\ref{deffpm})
the second equation (\ref{slavnov}) with $\fa_0$ on the right hand side.
We obtain a formula similar to (\ref{fplus}), but with a different determinant
in the denominator. Multiplying this expression for $F_- (\xi)$ with
(\ref{fplus}) we obtain equation (\ref{apmas}) for the amplitude $A_n^{-+} (\x)$.
In order to obtain the simpler and more symmetric formula (\ref{apmsym}) of
the main text, however, we have to perform a more cumbersome calculation which
we shall sketch below.

Instead of using (\ref{slavnov}) directly in (\ref{deffpm}) we first of all move
$B(\x)$ through the product of $C$ operators in $\<\Ps_0|$ to the left, using the
well-known Yang-Baxter algebra relations (cf.\ e.g.\ \cite{KBIBo,GKS05}). This
produces a double sum $F_- (\x) = \sum_{\substack{\ell, m \\ \ell \ne m}}^{M+1}
t_{\ell, m}$, where
\begin{multline}
     t_{\ell, m} = d(\la_\ell) a(\la_m) c(\la_\ell, \x) c(\x, \la_m) \\
        \biggl[
        \prod_{\substack{k=1 \\ k \ne \ell}}^{M+1} \frac{1}{b(\la_\ell, \la_k)} \biggr]
	\biggl[
        \prod_{\substack{k=1 \\ k \ne \ell, m}}^{M+1} \frac{1}{b(\la_k, \la_m)} \biggr]
	\frac{\<0| \prod_{\substack{k=1 \\ k \ne \ell, m}}^{M+1} C(\la_k) |\Ps_n^\a\>}
	     {\La_n (\x|\a) \<\Ps_0|\Ps_0\>}
\end{multline}
and $\la_{M+1} = \x$ in the various products above. Here we insert the scalar
product formula (\ref{slavnov}) on the right hand side. The result can be
written in the following form,
\begin{multline}
     t_{\ell, m} = \frac{B \r^{-1}_n (\x|\a)}{1 + \fa_0 (\x)} \det_M (D)
	\det_{M-1} \bigl\{ \Ph (\la_{\ell_k} - \h) [\re(\m_j - \la_{\ell_k})
	                   - \re(\la_{\ell_k} - \m_j) \fa_n (\la_{\ell_k}|\a)] \bigr\} \\
        \times \frac{(- \fa_0 (\la_\ell)) \sign (m - \ell) (-1)^{\ell - m}}
	            {\sh( \la_\ell - \la_m + \h)} \epc
\end{multline}
where $\ell_k \in \{1, \dots, M + 1\} \setminus \{\ell, m\}$, $k = 1, \dots, M - 1$,
$\ell_1 < \ell_2 < \dots < \ell_{M-1}$, and
\begin{subequations}
\begin{align}
     B & = \frac{\prod_{j=1}^{M-1} q^{- \a} d(\m_j) \re^{\m_j}}
              {\prod_{j=1}^M d(\la_j) \re^{\la_j}} \,
         \frac{1}{\det_M \bigl\{ \de^j_k \fa_0' (\la_j) - K(\la_j - \la_k)\bigr\}}
	 \epc \\[1ex]
     D & = \begin{pmatrix}
             \frac{q^\a \res \Ph^{-1} (\m_1)}{\sh(\la_1 - \m_1)} & \dots &
	     \frac{q^\a \res \Ph^{-1} (\m_{M-1})}{\sh(\la_1 - \m_{M-1})} & \re^{\la_1} \\
	     \vdots & & \vdots & \vdots \\
	     \frac{q^\a \res \Ph^{-1} (\m_1)}{\sh(\la_M - \m_1)} & \dots &
	     \frac{q^\a \res \Ph^{-1} (\m_{M-1})}{\sh(\la_M - \m_{M-1})} & \re^{\la_M}
	   \end{pmatrix} \epp
\end{align}
\end{subequations}
The function $\Ph$ was defined in equation (\ref{defphipm}) above. At this point
we see already that the prefactor in $B$ which would be somewhat inconvenient in
the Trotter limit will cancel the corresponding prefactor in A, equation
(\ref{defap}). What remains to be done is to sum up the $t_{\ell, m}$ into a
single determinant. We shall divide this task into several steps.

Let us define
\begin{subequations}
\begin{align} \label{defav}
     (\av_k)^j & = \Ph (\la_k - \h)
        \bigl( \re(\m_j - \la_k) - \re(\la_k - \m_j) \fa_n (\la_k|\a)\bigr) \epc \\
      s_{k j} & = \frac{1}{\sh(\la_j - \la_k + \h)} \epc
\end{align}
\end{subequations}
where $j = 1, \dots, M - 1$ in (\ref{defav}) and where the remaining indices
run from $1$ to $M + 1$. Then
\begin{multline} \label{sumone}
     \sum_{\ell = 1}^M t_{\ell, M + 1} = 
        \frac{B \r^{-1}_n (\x|\a)}{1 + \fa_0 (\x)} \det_M (D)
        \sum_{\ell = 1}^M (-1)^{M + \ell -1} s_{M+1, \ell}
           \det_{M-1} \bigl( \av_1, \dots, \av_{\ell - 1}, \av_{\ell + 1},
                             \dots, \av_M \bigr) \\
        = - \frac{B \r^{-1}_n (\x|\a)}{1 + \fa_0 (\x)}
	    \det_M \biggl( D \binom{\av_1}{s_{M+1, 1}},
	                   \dots, D \binom{\av_M}{s_{M+1, M}} \biggr) \epp
\end{multline}
We denote the matrix in the determinant on the right hand side of this equation
by $X$. The entries of this matrix can be simplified the same way as in the
previous section, and
\begin{equation}
     X^j_k = \de^j_k \r_n(\la_j|\a) \fa_0' (\la_j)
             - \frac{q^{-\a}}{\sh(\la_j - \la_k - \h)}
             + \frac{q^\a}{\sh(\la_j - \la_k + \h)} + \re^{\la_j} s_{M+1, k} \epp
\end{equation}
This expression looks appealing as it is a sum of a matrix and a one-dimensional
projector. One might be tempted to pull the matrix $X^j_k - \re^{\la_j}
s_{M+1, k}$ out of the determinant. This is impossible, however, since the matrix
is not invertible. The latter fact can be easily seen by replacing $s_{M+1, k}$
on the right hand side of (\ref{sumone}) by zero. A left eigenvector with zero
eigenvalue has the components $\res \Ph (\la_j)$.

In order to resolve the problem we introduce the kernel $K$ at this point and write
\begin{subequations}
\begin{align} \label{intoldk}
     \det_M (X) & = \det_M \bigl\{ \de^j_k \r_n(\la_j|\a) \fa_0' (\la_j) -
                    K_{1 + \a} (\la_j - \la_k) + \wts_{M+1, k} \bigr\} \epc \\
     \wts_{j, k} & = \re^{\la_k} s_{j, k} + q^{- 1 - \a} - q^{1 + \a} \epp
\end{align}
\end{subequations}
We further define
\begin{equation}
     \whk^j_k = \frac{K_{1 + \a} (\la_j - \la_k)}{\r_n(\la_j|\a) \fa_0' (\la_j)} \epc \qd
     \vv^j = \frac{1}{\r_n(\la_j|\a) \fa_0' (\la_j)}
\end{equation}
and assume henceforth that $1 - \whk$ is invertible. This seems to be true for
general $M$. It is easy to prove for $M = 1$, and for larger $M$ we verified
it numerically for a number of examples. Setting
\begin{equation} \label{defwm}
     \wv = (1 - \whk)^{-1} \vv
\end{equation}
we conclude that
\begin{equation} \label{detxtimessum}
     \det_M (X) =
        \biggl[ \prod_{j=1}^M \frac{\fa_0' (\la_j)}{\r_n^{-1} (\la_j|\a)} \biggr]
	  \det_M (1 - \whk) \biggl[ 1 + \sum_{j=1}^M \wts_{M+1, j} \wv^j \biggr] \epp
\end{equation}

As before we convert (\ref{defwm}) into an integral equation. It is equivalent
to the linear equation
\begin{equation}
     \wv^j \r_n(\la_j|\a) \fa_0' (\la_j)
        = 1 + \sum_{k=1}^M K_{1+\a} (\la_j - \la_k) \wv^k \epp
\end{equation}
Defining
\begin{equation}
     \s (\la) = 1 + \sum_{k=1}^M K_{1+\a} (\la - \la_k) \wv^k
\end{equation}
we conclude that
\begin{equation} \label{wsigma}
     \wv^j = \frac{\s(\la_j)}{\r_n(\la_j|\a) \fa_0' (\la_j)}
\end{equation}
and
\begin{equation} \label{dressedapp}
     \s (\la) = 1 + \int_{{\cal C}_n} \rd m_-^\a (\m) K_{1+\a} (\la - \m) \s (\m) \epp
\end{equation}
The function $\s(\la)$ is a generalized dressed charge function of the type
introduced in \cite{BoGo10}. It allows us to write
\begin{equation} \label{sumwintsigma}
     1 + \sum_{j=1}^M \wts_{M+1, j} \wv^j = \s_\infty
        - \int_{{\cal C}_n} \rd \overline{m}_-^\a (\la)
	  \frac{\re^\la \s (\la)}{\sh(\la - \x + \h)} \epc
\end{equation}
where
\begin{equation}
     \rd \overline{m}_-^\a (\la)
        = \frac{\rd \la \, \r_n^{-1} (\la|\a)}{2 \p \i (1 + \faq_0 (\la))} \epp
\end{equation}

We finally argue that
\begin{equation} \label{limsigma}
     \s_\infty = \lim_{\Re \la \rightarrow + \infty} \s (\la) = 0 \epp
\end{equation}
This can be seen as follows. We could have introduced a parameter, $\g$ say,
into the last column of the matrix $D$ above and at the same time could have
multiplied the whole sum (\ref{sumone}) we are considering by $1/\g$. This trivial
modification cannot change our result, but we would have a term $\s_\infty/\gamma$
instead of $\s_\infty$ in the final formula, whence (\ref{limsigma}) must hold.
Inserting (\ref{detxtimessum}), (\ref{sumwintsigma}) and (\ref{limsigma}) into
(\ref{sumone}) we obtain
\begin{equation} \label{sumonefin}
     \sum_{\ell = 1}^M t_{\ell, M + 1} =
        \frac{B \r^{-1}_n (\x|\a)}{1 + \fa_0 (\x)} 
        \biggl[ \prod_{j=1}^M \frac{\fa_0' (\la_j)}{\r_n^{-1} (\la_j|\a)} \biggr]
	\det_M (1 - \whk)
        \int_{{\cal C}_n} \rd \overline{m}_-^\a (\la)
	\frac{\re^\la \s (\la)}{\sh(\la - \x + \h)} \epp
\end{equation}
In a similar way
\begin{equation} \label{sumtwofin}
     \sum_{m= 1}^M t_{M + 1, m} =
        - \frac{B \r^{-1}_n (\x|\a)}{1 + \faq_0 (\x)} 
        \biggl[ \prod_{j=1}^M \frac{\fa_0' (\la_j)}{\r_n^{-1} (\la_j|\a)} \biggr]
	\det_M (1 - \whk)
        \int_{{\cal C}_n} \rd m_-^\a (\la)
	\frac{\re^\la \s (\la)}{\sh(\x - \la + \h)} \epp
\end{equation}

To prepare for the calculation of the remaining double sum with $\ell, m < M + 1$
we introduce the $M \times M$ matrix
\begin{equation}
     U_m = \begin{pmatrix}
              \av_1 & \dots & \av_{m-1} & \av_{M+1} & \av_{m+1} & \dots & \av_M \\
	      s_{m, 1} & \dots & s_{m, m - 1} & s_{m , m} & s_{m, m + 1} & \dots
	      & s_{m, M}
           \end{pmatrix} \epp
\end{equation}
Then
\begin{multline} \label{sumellmstart}
     \sum_{\substack{\ell = 1\\ \ell \ne m}}^M t_{\ell, m} = 
        \frac{B \r^{-1}_n (\x|\a)}{1 + \fa_0 (\x)} \det_M (D) \\ \times
	\Bigl\{ \det_M (U_m) - (-1)^{M - m} s_{m, m}
	        \det_{M-1} (\av_1, \dots, \av_{m-1}, \av_{m+1}, \dots, \av_M) \Bigr\}
\end{multline}
which still has to be summed over $m$.

For the second term in curly brackets we can proceed as above and obtain
\begin{multline} \label{sumthreefin}
     \frac{B \r^{-1}_n (\x|\a)}{1 + \fa_0 (\x)} \det_M (D)
	\sum_{m=1}^M (-1)^{M - m + 1} s_{m, m}
	\det_{M-1} (\av_1, \dots, \av_{m-1}, \av_{m+1}, \dots, \av_M) \\ =
        - \frac{B \r^{-1}_n (\x|\a)}{1 + \fa_0 (\x)} 
        \biggl[ \prod_{j=1}^M \frac{\fa_0' (\la_j)}{\r_n^{-1} (\la_j|\a)} \biggr]
	\det_M (1 - \whk)
        \int_{{\cal C}_n} \rd m_-^\a (\la)
	\frac{\re^\la \s (\la)}{\sh(\h)} \epp
\end{multline}
For the first term we first simplify $D U_m$ and perform the summation afterwards.
Thus,
\begin{subequations}
\begin{align}
     (D U_m)^j_k & = \de^j_k \r_n(\la_j|\a) \fa_0' (\la_j) -
                     \re^{\la_j - \la_k} \bigl( K_{1 + \a} (\la_j - \la_k)
		     - \wts_{m, k} \bigr) \epc \qd \text{for $k \ne m$,} \\[1ex]
     (D U_m)^j_m & = \re^{\la_j - \x} \bigl( \r_n (\la_j|\a) \fa_0' (\la_j) \yv^j
                                             + \wts \bigr) \epc
\end{align}
\end{subequations}
where
\begin{subequations}
\begin{align}
     & \r_n (\la_j|\a) \fa_0' (\la_j) \yv^j =
        - \r_n (\x|\a) \bigl(1 + \fa_0 (\x)\bigr) \cth(\la_j - \x) \notag \\
	  & \mspace{154.mu}
	+ q^{\a + 1} \cth(\la_j - \x + \h) 
	+ \fa_0 (\x) q^{- \a - 1} \cth(\la_j - \x - \h) \epc \\[1ex]
     & \wts = \r_n (\x|\a) \bigl(1 + \fa_0 (\x)\bigr)
              - q^{\a + 1} - \fa_0 (\x) q^{- \a - 1} + \re^\x s_{m, m} \epc
\end{align}
\end{subequations}
and hence
\begin{multline} \label{dum}
     \det_M (D U_m) =
        \re^{\la_m - \x}
	\biggl[ \prod_{j=1}^M \frac{\fa_0' (\la_j)}{\r_n^{-1} (\la_j|\a)} \biggr]
	\det_M (1 - \whk) \\ \times
	\Biggl[ \zv^m + \wts \wv^m + \sum_{\ell = 1}^M \wts_{m, \ell}
	       \det \begin{pmatrix} \wv^\ell & \zv^\ell \\ \wv_m & \zv^m \end{pmatrix}
	       \biggr] \epc
\end{multline}
where we defined
\begin{equation} \label{defzm}
     \zv = (1 - \whk)^{-1} \yv \epp
\end{equation}

Once again we convert (\ref{defzm}) into an integral equation. Setting
\begin{multline}
     G_-(\la, \x) = - \cth(\la - \x)
                    + \frac{q^{\a + 1} \r_n^{-1} (\x|\a)}{1 + \fa_0 (\x)}
		      \cth(\la - \x + \h) \\
                    + \frac{q^{- \a - 1} \r_n^{-1} (\x|\a)}{1 + \faq_0 (\x)}
		      \cth(\la - \x - \h)
                    + \sum_{k=1}^M K_{1+\a} (\la - \la_k)
		      \frac{\r_n^{-1} (\x|\a) \zv^k}{1 + \fa_0 (\x)}
\end{multline}
implies that
\begin{equation} \label{zgm}
     \zv^j = \frac{\r_n (\x|\a) (1 + \fa_0 (\x))G_-(\la_j, \x)}
                  {\r_n (\la_j|\a) \fa_0'(\la_j)}
\end{equation}
and
\begin{multline} \label{gmapp}
     G_-(\la, \x) = - \cth(\la - \x) \\
                    + q^{- \a - 1} \r_n^{-1} (\x|\a) \cth(\la - \x - \h)
                    + \int_{{\cal C}_n} \rd m_-^\a (\m) K_{1+\a} (\la - \m) G_- (\m, \x)
		      \epp
\end{multline}

Using (\ref{wsigma}) and (\ref{zgm}) we can calculate the sum on the right
hand side of (\ref{dum}) and can convert it into an integral,
\begin{multline} \label{sumdum}
     \frac{\r_n^{-1} (\x|\a)}{1 + \fa_0 (\x)}
     \sum_{m=1}^M \re^{\la_m}
	\Biggl[ \zv^m + \wts \wv^m + \sum_{\ell = 1}^M \wts_{m, \ell}
	       \det \begin{pmatrix} \wv^\ell & \zv^\ell \\ \wv_m & \zv^m \end{pmatrix}
	       \biggr] \\
        = \int_{{\cal C}_n} \rd m_-^\a (\la) \re^\la
	  \int_{{\cal C}_n} \rd \overline{m}_-^\a (\m) \re^\m \:
	  \frac{G_- (\la, \x) \s(\m) - G_- (\m, \x) \s(\la)}{\sh (\la - \m - \h)} \\
          + \frac{\re^\x \r_n^{-1} (\x|\a)}{1 + \fa_0 (\x)}
	    \int_{{\cal C}_n} \rd m_-^\a (\la) \frac{\re^\la \s(\la)}{\sh(\h)}
          + \frac{\re^\x \r_n^{-1} (\x|\a)}{1 + \faq_0 (\x)}
	    \int_{{\cal C}_n} \rd m_-^\a (\la)
	    \frac{\re^\la \s(\la)}{\sh(\x - \la + \h)} \\
          - \frac{\re^\x \r_n^{-1} (\x|\a)}{1 + \fa_0 (\x)}
	    \int_{{\cal C}_n} \rd \overline{m}_-^\a (\la)
	    \frac{\re^\la \s(\la)}{\sh(\la - \x + \h)}
          - G_-^+ (\x) \int_{{\cal C}_n} \rd \overline{m}_-^\a (\la) \re^\la \s(\la) \epc
\end{multline}
where by definition
\begin{equation}
     G_-^+ (\x) = \lim_{\Re \la \rightarrow + \infty} G_-(\la, \x) \epp
\end{equation}
Notice that in (\ref{sumdum}) we have used the measure $\rd \overline{m}_-^\a$
when rewriting the finite sums as integrals and that we employed (\ref{limsigma})
to simplify the right hand side of (\ref{sumdum}). Using a very similar
argument as in the derivation of (\ref{limsigma}) we find that
\begin{equation} \label{limgm}
     G_-^+ (\x) \int_{{\cal C}_n} \rd \overline{m}_-^\a (\la) \re^\la \s(\la) = 0 \epp
\end{equation}

With this we can summarize what we have obtained so far. Combining (\ref{sumonefin}),
(\ref{sumtwofin}), (\ref{sumellmstart}), (\ref{sumthreefin}), (\ref{dum}),
(\ref{sumdum}) and (\ref{limgm}) we conclude that
\begin{multline}
     F_-(\x) = \frac{\re^{- \x} \prod_{j=1}^{M-1} q^{- \a} d(\m_j) \re^{\m_j}}
                    {\prod_{j=1}^M d(\la_j) \re^{\la_j} \r_n^{-1} (\la_j|\a)} \,
               \frac{\det_M
	             \Bigl\{ \de^j_k - \frac{\r_n^{-1} (\la_j|\a)}{\fa_0' (\la_j|\a)}
		      K_{1 + \a} (\la_j - \la_k) \Bigr\}}
	            {\det_M \Bigl\{\de^j_k - \frac{1}{\fa_0' (\la_j|\a)}
		      K (\la_j - \la_k)\Bigr\}} \\ \times
               \int_{{\cal C}_n} \rd m_-^\a (\la) \re^\la
	       \int_{{\cal C}_n} \rd \overline{m}_-^\a (\m) \re^\m \:
	       \frac{G_- (\la, \x) \s(\m) - G_- (\m, \x) \s(\la)}{\sh (\la - \m - \h)}
	       \epp
\end{multline}
This is already a quite remarkable formula. The prefactor nicely combines with
the corresponding prefactor in our expression (\ref{fplus}) for $F_+ (\x)$. The
double integral is much reminiscent of the double integral representing the
inhomogeneous neighbour correlation function $\<\s^-_1 \s^+_2\>$ derived in
\cite{BoGo09}. In particular, its structure is such that it can be factorized
by means of the technique suggested in \cite{BGKS06,BoGo09}.

For this purpose we first note that
\begin{equation}
     \rd \overline{m}_-^\a (\la) =
        \frac{\rd \la}{2 \p \i \r_n (\la|\a)} - \rd m_-^\a (\la) \epp
\end{equation}
Then
\begin{multline}
   I = \int_{{\cal C}_n} \rd m_-^\a (\la) \re^\la
      \int_{{\cal C}_n} \rd \overline{m}_-^\a (\m) \re^\m \:
      \frac{G_- (\la, \x) \s(\m) - G_- (\m, \x) \s(\la)}{\sh (\la - \m - \h)} \\
   = \re^\x \r_n^{-1} (\x|\a) \int_{{\cal C}_n} \rd m_-^\a (\la)
	    \frac{\re^\la \s(\la)}{\sh(\la - \x - \h)} \\ 
     - \int_{{\cal C}_n} \rd m_-^\a (\la)
       \int_{{\cal C}_n} \rd m_-^\a (\m) \: \frac{\re^{\la + \m}}{\sh (\la - \m - \h)}
       \det \begin{pmatrix} G_- (\la, \x) & G_- (\m, \x) \\ \s(\la) & \s(\m) \end{pmatrix}
       \epp
\end{multline}
Define
\begin{equation}
     r(\la, \m) = \frac{\re^{\la + \m}}{\sh (\la - \m - \h)} \epc \qd
     s(\la, \m) = - \frac{q^{1+\a} - q^{- 1 - \a} + K_{1+\a} (\la - \m)}
                           {q^\a - q^{-\a}} \epp
\end{equation}
We observe that $r(\la, \m) - \re^{2 \la} s(\la, \m)$ is a symmetric function.
Hence, we can replace $r(\la, \m)$ by $\re^{2 \la} s(\la, \m)$ under the double
integral. Then, further using the integral equations (\ref{dressedapp}) and
(\ref{gmapp}) as well as the asymptotics (\ref{limsigma}) and the elementary identity
\begin{multline} \label{decsh}
     \frac{q^\a - q^{-\a}}{\sh (\la - \x - \h)} \\ = 
        q^{\a+1} \re^{- \la + \x} \bigl( \cth(\la - \x - \h) + 1 \bigr) -
	q^{- \a - 1} \re^{\la - \x} \bigl( \cth(\la - \x - \h) - 1 \bigr) \epc
\end{multline}
we obtain
\begin{multline} \label{factorpmone}
     I = \frac{G_-^+ (\x)}{q^\a - q^ {- \a}}
                \int_{{\cal C}_n} \rd m_-^\a (\la) \: \re^{2 \la} \s(\la) \\
                + \frac{\re^{2 \x}}{q^\a - q^ {- \a}}
		\int_{{\cal C}_n} \rd m_-^\a (\la) \: \s(\la)
		\bigl( - \cth(\la - \x) + q^{\a+1} \r_n^{-1} (\x|\a) \cth(\la - \x - \h)\\
		       - 1 + q^{\a+1} \r_n^{-1} (\x|\a) \bigr) \epp
\end{multline}
Equation (\ref{limgm}) states that $G_-^+ (\x) = 0$ or the integral is zero.
For $M = 1$ it is easy to see that the integral in (\ref{limgm}) is non-zero.
Hence, $G_-^+ (\x) = 0$. We have verified this numerically for more examples and
in the following assume it to be generally true. Then $I$ is reduced to
the second term in (\ref{factorpmone}).

Introducing a function $\overline{G}_-$ as the solution of the linear
integral equation
\begin{multline}
     \overline{G}_- (\la, \x) =
        - \cth(\la - \x) \\
        + q^{\a + 1} \r_n^{-1} (\x|\a) \cth(\la - \x - \h)
	+ \int_{{\cal C}_n} \rd m_-^\a (\m) \:
	\overline{G}_- (\m, \x) K_{1+\a} (\m - \la)
\end{multline}
and using the dressed function trick\footnote{Schematically: $f_1 = g_1 + \int K f_1$
and $f_2 = g_2 + \int K^t f_2$ implies $\int f_1 g_2 = \int f_2 g_1$.} we end up with
\begin{equation}
     I = \frac{\re^{2 \x} \overline{G}_-^+ (\x)}
                     {(q^\a - q^ {- \a})(q^{\a + 1} - q^{-\a - 1})} \epc
\end{equation}
where
\begin{equation}
     \overline{G}_-^+ (\x)
        = \lim_{\Re \la \rightarrow + \infty} \overline{G}_- (\la, \x) \epp
\end{equation}
This completes our derivation of equation (\ref{fminus}) of the main text.

We expect that
\begin{equation} \label{limgqa}
     \lim_{\a \rightarrow 0} \overline{G}_-^+ (\x) = 0
\end{equation}
for $I$ to be finite. For the verification note that $\a = 0$ in (\ref{decsh})
implies the identity
\begin{equation}
     K_1 (\la) = \re^{- 2 \la} \bigl[ K_{-1} (\la) - q^{-1} + q \bigr] + q^{-1} - q
\end{equation}
which allows us to replace the kernel $K_1$ by $K_{-1}$ in (\ref{intoldk}).
Proceeding with the calculation the function $\overline{G}^+_-$ (for $\a = 0$)
along with a modified function $\overline{\s}$ will appear instead of
$G^+_-$ and $\s$. For these functions an identity similar to (\ref{limgm})
can be derived which implies (\ref{limgqa}).

\Appendix{A generalized Sommerfeld lemma}
\label{app:sommerfeld}
\noindent
In this appendix we provide a proof of Lemma~\ref{lem:sommerfeld}. Let
$F$ be an antiderivative of $f$. Using integration by parts we obtain
\begin{align}
     & T \int_{{\cal C}_u^-} \rd \la \: f(\la) \ln \Bigl(1 + \re^{- \frac{u(\la)}T} \Bigr)
          \notag \\ & \mspace{45.mu}
        = T \ln (2) \bigr(F(Q_+) - F(Q_-)\bigl)
	  + \int_{Q_-}^{Q_+} \rd \la \: \frac{F(\la) u'(\la)}{1 + \re^{\frac{u(\la)}T}}
          \notag \displaybreak[0] \\ & \mspace{45.mu}
        = T \int_{{\cal C}_u^-} \rd \la \:
	             f(\la) \ln \Bigl(1 + \re^{\frac{u(\la)}T} \Bigr)
	  + \int_{Q_-}^{Q_+} \rd \la \: F(\la) u'(\la)
          \notag \\ & \mspace{45.mu}
        = T \int_{{\cal C}_u^-} \rd \la \:
	           f(\la) \ln \Bigl(1 + \re^{\frac{u(\la)}T} \Bigr)
	  - \int_{Q_-}^{Q_+} \rd \la \: f(\la) \bigl( u(\la) - 2 \p \i p T\bigr) \epp
\end{align}
Setting $\uq (\la) = u(\la) - 2 \p \i p T$ it follows that
\begin{multline}
     \D I = T \int_{{\cal C}_u} \rd \la \: f(\la) \ln \Bigl(1 + \re^{- \frac{u(\la)}T} \Bigr)
	    + \int_{Q_-}^{Q_+} \rd \la \: f(\la) \uq (\la) \\
          = T \int_{{\cal C}_u} \rd \la \:
	             f(\la) \ln \Bigl(1 + \re^{- \frac{u(\la) \sign (v(\la))}T} \Bigr) \epp
\end{multline}
Then the real part of the exponent on the right hand side is always negative except
at $Q_\pm$. This means that the leading contribution to the integral for
$T \rightarrow 0$ comes from the (infinitesimally small) vicinities of these two
points. In order to quantify the leading contribution we fix $\de > 0$ small enough.
Since $u$ and $f$ are holomorphic we can deform the contour locally in a small
vicinity of $Q_\pm$ into contours $J_\pm^\de$ such that $w(\la) = 2 \p p T$ for
$\la \in J_\pm^\de$ and $v (\la) = \pm \de$ at the boundaries of $J_\pm^\de$.
Note that $v$ is monotonic on $J_\pm^\de$, since it has simple zeros at $Q_\pm$.
It follows that
\begin{equation}
     \D I = T \int_{J_-^\de \cup J_+^\de} \rd \la \:
                    f(\la) \ln \Bigl(1 + \re^{- \frac{|v(\la)|}T} \Bigr)
            + \CO (T^\infty) \epp
\end{equation}
We parameterize $J_\pm^\de$ by $x = v(\la) \Leftrightarrow \la = v^{-1} (x)$.
Then
\begin{multline}
     T \int_{J_-^\de} \rd \la \: f(\la) \ln \Bigl(1 + \re^{- \frac{|v(\la)|}T} \Bigr)
        = T \int_\de^{-\de} \rd x \: \frac{f(v^{-1} (x))}{v'(v^{-1} (x))}
	                    \ln \Bigl(1 + \re^{- \frac{|x|}T} \Bigr) \\
        = - \frac{T^2 \p^2}{6} \frac{f(Q_-)}{u'(Q_-)} + \CO (T^4) \epp
\end{multline}
Here we have used that $v^{-1} (0) = Q_-$ and that $\ln \bigl(1 + \re^{- |x|}\bigr)$
is even. At $Q_+$ we can perform a similar calculation with the only difference
that $v(\la)$ is ascending in the direction of the contour, whence the sign will
be positive. This completes the proof of the lemma.

\Appendix{Properties of the Fredholm determinants}
\label{app:detlim}
\noindent
In this appendix we obtain alternative forms of the Fredholm determinants
or their ratios and discuss the zero-temperature limit of the determinants
appearing in the numerator of the determinant part of the amplitudes.
For the sake of brevity we restrict ourselves to the longitudinal case.
\subsection{On the zero-temperature limit}
In order to calculate the $T \rightarrow 0+$ limit, we note that the 
$TQ$-equation implies (see \cite{BoGo10})
\begin{multline} \label{niceid}
     \frac{1}{\r_n (\la|\a) (1 + \fa_0 (\la))} =
        - \frac{q^{- \a} \Ph (\la + \h)}
	     {\r_n (\la|\a) \bigl(q^\a \Ph (\la - \h) - q^{- \a} \Ph (\la + \h)\bigr)} \\
        + \frac{\Ph(\la)}{q^\a \Ph (\la - \h) - q^{- \a} \Ph (\la + \h)} \epc
\end{multline}
where $\Ph$ is the ratio of $Q$-functions defined in (\ref{defphi}). The function
on the left hand side is the weight function in the measure $\rd m_-^\a$.
The function $\Ph (\la + \h)/\r_n (\la|\a)$ has no poles inside ${\cal C}_n$.
We shall assume that the contour of the Fredholm determinant with measure
$\rd m_-^\a$ can be deformed into a contour $\G_n^{(-)}$, containing no zero of
$q^\a \Ph (\la - \h) - q^{- \a} \Ph (\la + \h)$ and no pole of the kernel,
without changing the value of the determinant. In the Fredholm determinant
with this deformed contour we can use (\ref{niceid}) to replace the measure
$\rd m_-^\a$ by
\begin{equation}
     \rd M_-^\a (\la) = \frac{\rd \la}{2 \p \i}\
        \frac{\Ph(\la)}{q^\a \Ph (\la - \h) - q^{- \a} \Ph (\la + \h)} \epp
\end{equation}
In fact, we expect most of the zeros of $q^\a \Ph (\la - \h) -
q^{- \a} \Ph (\la + \h)$ to be located close to $\pm \h$, where the
essential singularities of the functions $\Ph (\la \pm \h)$ appear in the
Trotter limit. Thus, $\G_n^{(-)}$ might be similar to ${\cal C}_n$.

The weight function in the new measure is easily re-expressed in terms of
the auxiliary functions,
\begin{equation}
     \frac{\Ph(\la)}{q^\a \Ph (\la - \h) - q^{- \a} \Ph (\la + \h)}
        = \frac{1 + \fa_n (\la|\a)}{\r_n (\la|\a) (1 + \fa_0 (\la))}
	  \frac{1}{1 - \fa_n (\la| \a)/\fa_0 (\la)} \epp
\end{equation}
We rewrite the factors on the right hand side in a form that is appropriate
for performing the zero temperature limit. Using the non-linear integral
equation (\ref{nlie}) we obtain
\begin{equation} \label{rata}
     \frac{\fa_n (\la|\a)}{\fa_0 (\la)} = q^{- 2 \a}
        \exp \biggl\{\int_{{\cal C}_n} \rd \m \: K (\m - \la) z(\m) \biggr\} \epc
\end{equation}
while (\ref{rhoint}) implies that
\begin{equation} \label{rhoout}
     \frac{1 + \fa_n (\la|\a)}{\r_n (\la|\a) (1 + \fa_0 (\la))} = q^{- \a}
        \exp \biggl\{\int_{{\cal C}_n} \rd \m \: \re (\m - \la) z(\m) \biggr\} \epc
\end{equation}
where $\la$ is now outside ${\cal C}_n$.

We now assume that we may shift the contour $\G_n^{(-)}$, such that
its upper part is slightly above ${\cal C}_0$, uniformly for all $T$.
Clearly the poles of the function $\Ph$ do not prevent us from
shifting the contour, but we must further assume that $q^\a \Ph (\la - \h) -
q^{- \a} \Ph (\la + \h)$ has no zeros close to ${\cal C}_0$ for
$T \rightarrow 0+$. For the zero temperature limit of the Fredholm
determinant with measure $\rd M_-^\a$ it is then enough to perform
the limit in (\ref{rata}) and (\ref{rhoout}) for $\la$ away from
${\cal C}_0$. First straightening the contours in (\ref{rata}) and
(\ref{rhoout}) and then using corollary~\ref{cor:away} as well as
(\ref{defu1ell}) and (\ref{liz}) we obtain
\begin{subequations}
\label{measureszero}
\begin{align}
     & \int_{{\cal C}_n} \rd \m \: K (\m - \la) z(\m) = 2 \p \i (\a' - \ell)
        - 2 \p \i (\a' - \ell) Z(\la - \i \g/2) + {\cal O} (T) \epc \\
     & \int_{{\cal C}_n} \rd \m \: \re (\m - \la) z(\m) =
        - (\a' - \ell) \int_{-Q}^Q \rd \m \: \re (\m + \i \g/2 - \la) Z(\m)
	+ {\cal O} (T)
\end{align}
\end{subequations}
which fixes $\lim_{T \rightarrow 0+} \rd M_-^\a (\la)$ for $\la \in \G_n^{(-)}$.

A similar reasoning works for $\rd m_+^\a$. The identity
\begin{multline} \label{twiceid}
     \frac{\r_n (\la|\a)}{1 + \fa_n (\la|\a)} =
        - \frac{q^\a \Ph^{-1} (\la + \h) \r_n (\la|\a)}
	       {q^{-\a} \Ph^{-1} (\la - \h) - q^\a \Ph^{-1} (\la + \h)} \\
        + \frac{\Ph^{-1} (\la)}{q^{-\a} \Ph^{-1} (\la - \h) - q^\a \Ph^{-1} (\la + \h)}
\end{multline}
inspires the definition of the measure
\begin{multline}
     \rd M_+^\a (\la) = \frac{\rd \la}{2 \p \i}
        \frac{\Ph^{-1} (\la)}{q^{-\a} \Ph^{-1} (\la - \h) - q^\a \Ph^{-1} (\la + \h)} \\
	= \frac{\rd \la}{2 \p \i} 
          \frac{\r_n (\la|\a) (1 + \fa_0 (\la))}{1 + \fa_n (\la|\a)}
	  \frac{1}{1 - \fa_0 (\la)/\fa_n (\la|\a)}
\end{multline}
which replaces $\rd m_+^\a (\la)$ on an appropriate contour $\G_n^{(+)}$.
We assume it can be kept away from ${\cal C}_0$. Then eqs.\ (\ref{measureszero})
determine the zero temperature limit of $\rd M_+^\a$ on $\G_n^{(+)}$.

Our claims can be summarized as follows. Define
\begin{equation} \label{mhatapp}
     \rd \widehat{M}_\pm^\a (\la) =
        \frac{\rd \la}{2 \p \i}
	\frac{\re^{\pm \i \p \a'
	      \pm (\a' - \ell) \int_{-Q}^Q \rd \m \: \re(\m - \la) Z(\m) }}
	     {1 - \re^{\pm 2 \p \i (\a' - \ell) Z(\la)}} \epp
\end{equation}
Then
\begin{equation}
     \lim_{T \rightarrow 0+} \rd M_\pm^\a (\la) =
        \rd \widehat{M}_\pm^\a (\la - \i \g/2) \epp
\end{equation}
In order to obtain the formulae in the main text it remains to shrink
the contours $\G_n^{(\pm)}$ to a contour $\G [-Q, Q] + \i \g/2$, encircling
$[-Q, Q] + \i \g/2$ in counterclockwise manner and to shift this down to
the real axis.

Note that $Z(\la)$ is real on the real axis. Therefore, in accordance with
our assumptions, the denominator in (\ref{mhatapp}) is nonzero on the
real axis, as long as $\a'$ has a non-vanishing imaginary part.
\subsection{Relation to previous determinant formulae}
Using some of the results of the previous subsection we can rewrite the
determinants in the numerator and establish a relation with the determinants
obtained in \cite{KMS11a,KMS11b,KKMST11a}. For this purpose we first note
that by construction
\begin{equation} \label{detwethey}
     \det_{\rd m^\a_\pm, {\cal C}_n} \bigl\{ 1 - \widehat{\cal K}_{\mp \a} \bigr\} =
        \det_{\rd M^\a_\pm, \G_n^{(\pm)}}
	      \bigl\{ 1 - \widehat{\cal K}_{\mp \a} \bigr\} \epp
\end{equation}

Using eqs.\ (\ref{rata}) and (\ref{rhoout}) and inserting the definition
(\ref{cauchyz}) of the periodic Cauchy transform, we rewrite $\rd M_\pm^\a$ as
\begin{equation}
     \rd M_\pm^\a (\la) = \mp \frac{\rd \la}{2 \p \i}
        \frac{q^{- \a} \re^{\mp L_{{\cal C}_n} [z] (\la)}}
	     {\re^{\mp L_{{\cal C}_n} [z] (\la \pm \h)}
	      - q^{- 2\a} \re^{\mp L_{{\cal C}_n} [z] (\la \mp \h)}} \epp
\end{equation}
Due to (\ref{detalphan}) and (\ref{kernrel}) we may replace ${\cal K}_{- \a} (\la)$
by $K_{-\a} (\la) - q^\a + q^{-\a}$ and ${\cal K}_{\a} (\la)$ by
$K_{\a} (- \la) - q^{- \a} + q^\a$ in the determinants in (\ref{detwethey}).

We define $\be = q^{-2 \a}$,
\begin{equation}
     \widetilde{K}_\be (\la) = \cth (\la - \h) - \be \cth (\la + \h)
\end{equation}
and two new kernels
\begin{subequations}
\begin{align}
     & U_\th^{(-)} (\la, \m) = - \frac{1}{2 \p \i}
        \frac{\re^{L_{{\cal C}_n} [z] (\la)} \bigl(\widetilde{K}_\be (\la - \m)
	      - \widetilde{K}_\be (\th - \m) \bigr)}
	     {\re^{L_{{\cal C}_n} [z] (\la - \h)}
	      - \be \re^{L_{{\cal C}_n} [z] (\la + \h)}} \epc
        \\[1ex]
     & U_\th^{(+)} (\la, \m) = \frac{1}{2 \p \i}
        \frac{\re^{- L_{{\cal C}_n} [z] (\m)} \bigl(\widetilde{K}_\be (\la - \m)
	      - \widetilde{K}_\be (\la - \th)\bigr)}
	     {\re^{- L_{{\cal C}_n} [z] (\m + \h)}
	      - \be \re^{- L_{{\cal C}_n} [z] (\m - \h)}} \epp
\end{align}
\end{subequations}
Then clearly
\begin{equation}
     \det_{\rd m^\a_\pm, {\cal C}_n} \bigl\{ 1 - \widehat{\cal K}_{\mp \a} \bigr\} =
        \lim_{\th \rightarrow - \infty} \det_{\rd \la, \G_n^{(\pm)}}
	\bigl\{ 1 + \widehat{U}_\th^{(\pm)} \bigr\} \epp
\end{equation}

In order to compare with \cite{KMS11a,KMS11b} we note that
\begin{equation}
     \lim_{\th \rightarrow - \infty} \re^{L_{{\cal C}_n} [z] (\th \pm \h)} = 
        \re^{\int_{{\cal C}_n} \rd \la \: z(\la)} =
	\lim_{\Re \la \rightarrow - \infty} \Ph (\la) = b^{-1} \epc
\end{equation}
whence
\begin{equation}
     \lim_{\th \rightarrow - \infty} \:
        \frac{b^{\pm 1} (1 - \be)}
	     {\re^{\mp L_{{\cal C}_n} [z] (\th \pm \h)}
	      - \be \re^{\mp L_{{\cal C}_n} [z] (\th \mp \h)}} = 1 \epc
\end{equation}
and therefore
\begin{equation}
     \det_{\rd m^\a_\pm, {\cal C}_n} \bigl\{ 1 - \widehat{\cal K}_{\mp \a} \bigr\} =
        \lim_{\th \rightarrow - \infty} \biggl[
        \frac{b^{\pm 1} (1 - \be)}
	     {\re^{\mp L_{{\cal C}_n} [z] (\th \pm \h)}
	      - \be \re^{\mp L_{{\cal C}_n} [z] (\th \mp \h)}}
	\det_{\rd \la, \G_n^{(\pm)}}
	\bigl\{ 1 + \widehat{U}_\th^{(\pm)} \bigr\} \biggr] \epp
\end{equation}

Using exactly the same reasoning as in appendix A.3 of \cite{KKMST09a}
one can show that the expression in square brackets on the right hand
side is independent of $\th$. Hence,
\begin{equation} \label{uform}
     \det_{\rd m^\a_\pm, {\cal C}_n} \bigl\{ 1 - \widehat{\cal K}_{\mp \a} \bigr\} =
        \frac{b^{\pm 1} (1 - \be)}
	     {\re^{\mp L_{{\cal C}_n} [z] (\th \pm \h)}
	      - \be \re^{\mp L_{{\cal C}_n} [z] (\th \mp \h)}}
	\det_{\rd \la, \G_n^{(\pm)}}
	\bigl\{ 1 + \widehat{U}_\th^{(\pm)} \bigr\} \epp
\end{equation}
Introducing two independent parameters $\th_1$ and $\th_2$ for the
two cases, the determinants take the same form as in \cite{KMS11a,KMS11b}
(see e.g.\ eqs.\ (2.14)-(2.18) in \cite{KMS11b}). Taking the limit
$T \rightarrow 0+$ and setting $\th_1 = - Q$, $\th_2 = Q$ we reproduce
the expressions obtained in \cite{KKMST09a,KKMST11a}.

\subsection{Dependence on the twist parameter}
Equation (\ref{uform}) is important, because it allows us to understand
the $\a$ dependence of the determinants. Recalling that $\be = q^{- 2\a}$ we
infer that $\det_{\rd m^\a_\pm, {\cal C}_n} \{ 1 - \widehat{\cal K}_{\mp \a} \}
= {\cal O} (\a)$ unless $n = 0$.

This statement remains true for $T \rightarrow 0+$ unless $\ell = 0$. The
case $\ell = 0$ for $T \rightarrow 0+$ needs a more careful treatment
(compare \cite{KKMST09a,KMS11a}). Consider (\ref{measureatzero}) and (\ref{limnum})
for $\ell = 0$. Recall that the contour $\G [-Q, Q]$ encircles the interval
$[-Q, Q]$ counterclockwise. Thus, we may replace the contour by $[-Q, Q]$
if we replace at the same time the weight by
\begin{equation} \label{lzero}
     \mspace{-3.mu}
     \rd \breve{M}_\pm^\a (\la) =
        \frac{\rd \la \: \re^{\pm \i \p \a'}}
	     {2 \p \i \bigl(1 - \re^{\pm 2 \p \i \a' Z(\la)}\bigr)}
	\Bigl( \re^{\pm \a' \int_{-Q}^Q \rd \m \: \re(\m - \la_-) Z(\m) }
	     - \re^{\pm \a' \int_{-Q}^Q \rd \m \: \re(\m - \la_+) Z(\m) } \Bigr).
\end{equation}
Here $\la_+$ denotes the limit from above and $\la_-$ the limit from below.
This replacement is justified since the right hand side of (\ref{lzero})
can be recast as
\begin{equation} \label{lzerotwo}
     \rd \breve{M}_\pm^\a (\la) =
        \frac{\rd \la}{2 \p \i}
	\biggl( \frac{\sh(\la_- - Q) \sh(\la + Q - \h)}
	             {\sh(\la_- + Q) \sh(\la - Q - \h)} \biggr)^{\pm \a' {\cal Z}}
	\re^{\pm \i \p \a'
	     \pm \a' \int_{-Q}^Q \rd \m \: \re(\m - \la_-) (Z(\m) - {\cal Z})}
\end{equation}
and since for small $\a'$ the singularity of the latter expression can be
integrated over. We further conclude that $\lim_{\a \rightarrow 0}
\rd \breve{M}_\pm^\a (\la) = \rd \la/2 \p \i$.

For the ratio of determinants ${\cal D} (\ell)$ defined in (\ref{ratdetzero})
eq.\ (\ref{lzerotwo}) implies that
\begin{equation}
     {\cal D} (0) = 1 + {\cal O} (\a^2) \epc
\end{equation}
which is needed for calculating the longitudinal correlation functions
from the generating function.
\subsection{A quotient formula}
This subsection is of auxiliary character. We would like to point out
that the ratios of Fredholm determinants in the main text can be
combined into single Fredholm determinants. This may turn out to
be useful e.g.\ in the low-temperature analysis of the model for
vanishing magnetic field.

Consider, for instance, the ratio
\begin{equation} \label{defdm}
     {\cal D}_- =
        \frac{\det_M \Bigl\{ \de^j_k - \frac{\r_n^{-1} (\la_j|\a)}{\fa_0' (\la_j)}
	      {\cal K}_\a (\la_j - \la_k) \Bigr\}}
	     {\det_M \Bigl\{\de^j_k - \frac{1}{\fa_0' (\la_j)}
	      {\cal K} (\la_j - \la_k)\Bigr\}}
\end{equation}
occurring in eq.\ (\ref{aalphan}). We would like to write it as a single
Fredholm-type determinant of the form
\begin{equation} \label{rewdm}
     {\cal D}_- = {\det_M}^{-1} (\ev_1 + \wv_1, \dots, \ev_M + \wv_M) \epc
\end{equation}
where the $\ev_k$ denote the canonical unit vectors, $\ev_k^j = \de^j_k$ and
where the $\wv_k$ are to be determined. Comparing (\ref{defdm}) and (\ref{rewdm})
we see that they must satisfy the equations
\begin{multline}
     \wv_k^j \fa_0' (\la_j) = {\cal K}_\a (\la_j - \la_k) \r_n^{-1} (\la_k|\a)
        - {\cal K} (\la_j - \la_k) \\
	+ \sum_{\ell = 1}^M {\cal K}_\a (\la_j - \la_\ell)
	                    \r_n^{-1} (\la_\ell|\a) \wv_\ell^j \epp
\end{multline}
These can be turned into an integral equation. Defining
\begin{equation}
     F(\la, \n) = {\cal K}_\a (\la - \n) \r_n^{-1} (\n|\a) - {\cal K} (\la - \n)
	+ \int_{{\cal C}_n} \rd m_-^\a (\m) {\cal K}_\a (\la - \m) F(\m, \n)
\end{equation}
we conclude that $\wv_k^j = F(\la_j, \la_k)/\fa_0' (\la_j)$. Thus,
\begin{equation}
     {\cal D}_- =
        {\det_M}^{-1} \biggl\{ \de^j_k + \frac{F(\la_j, \la_k)}{\fa_0' (\la_j)} \biggr\}\
	\longrightarrow\
        {\det_{\rd m, {\cal C}_n}}^{-1} \bigl\{ 1 + \widehat F \bigr\} \epp
\end{equation}

In a similar way we obtain the alternative representation
\begin{equation}
     {\cal D}_- =
        \det_M \biggl\{ \de^j_k
	                + \frac{\overline F(\la_j, \la_k)}{\fa_0' (\la_j)} \biggr\}\
			  \longrightarrow\
        \det_{\rd m, {\cal C}_n} \bigl\{ 1 + \widehat{\overline{F}} \bigr\} \epc
\end{equation}
where
\begin{equation}
     \overline F(\la, \n) = {\cal K} (\la - \n)
        - {\cal K}_\a (\la - \n) \r_n^{-1} (\n|\a)
	+ \int_{{\cal C}_n} \rd m (\m) {\cal K} (\la - \m) \overline F(\m, \n) \epp
\end{equation}
Similar expressions can also be obtained for the other ratio of determinants
in (\ref{aalphan}).

}

%\bibliographystyle{amsplain}
%\bibliography{hub,fab,andreas}

\providecommand{\bysame}{\leavevmode\hbox to3em{\hrulefill}\thinspace}
\providecommand{\MR}{\relax\ifhmode\unskip\space\fi MR }
\providecommand{\MRhref}[2]{%
  \href{http://www.ams.org/mathscinet-getitem?mr=#1}{#2}
}
\providecommand{\href}[2]{#2}

\end{document}